\documentclass[11pt, draftclsnofoot, onecolumn]{IEEEtran}
\usepackage{graphicx,forloop,caption,subcaption,float,listings,color,booktabs,mathtools}
\usepackage[T1]{fontenc}
\usepackage{hyperref}
\usepackage{graphics}
\usepackage{amsmath}
\usepackage{amssymb}
\usepackage{listings}
\usepackage{overpic,color}
\usepackage{psfrag}
\usepackage{multirow}
\usepackage[absolute,showboxes]{textpos} %temporary
\usepackage{xspace}
\usepackage{bbm}
\input{mathlig}
\usepackage{mathrsfs}
\newcommand{\muspace}{\mspace{1mu}}

\DeclareRobustCommand{\scond}{\mathchoice{\muspace\vert\muspace}{\vert}{\vert}{\vert}}
\mathlig{|}{\scond}
\newcommand{\cond}{\mathchoice{\,\vert\,}{\mspace{2mu}\vert\mspace{2mu}}{\vert}{\vert}}

\DeclareRobustCommand{\discint}{\mathchoice{\mspace{-1.5mu}:\mspace{-1.5mu}}{\mspace{-1.5mu}:\mspace{-1.5mu}}{:}{:}}
\mathlig{::}{\discint}

\def\tr{\mathop{\rm tr}\nolimits}%
\def\Var{\mathop{\rm Var}\nolimits}%
\def\rank{\mathop{\rm rank}\nolimits}%
\def\P{{\mathsf P}}
\newcommand\norm[1]{\left\lVert#1\right\rVert}

\newcommand{\Ac}{\mathcal{A}}

\newcommand{\Cc}{\mathcal{C}}

\newcommand{\Ec}{\mathcal{E}}

\newcommand{\Sc}{\mathcal{S}}
\newcommand{\Tc}{\mathcal{T}}

\newcommand{\Xc}{\mathcal{X}}

\renewcommand{\Pr}{\mathscr{P}}
\newcommand{\Rr}{\mathscr{R}}

\newcommand{\mv}{{\bf m}}

\newcommand{\aep}{{\mathcal{T}_{\epsilon}^{(n)}}}
\newcommand{\aepvar}{{\mathcal{T}_{\epsilon'}^{(n)}}}

\newcommand{\mh}{{\hat{m}}}

\newcommand{\Rt}{{\tilde{R}}}

\newcommand{\yt}{{\tilde{y}}}

\def\l{\lambda}

\DeclareMathOperator\E{\textsf{E}}
\let\P\relax
\DeclareMathOperator\P{\textsf{P}}

\newcommand{\N}{\mathrm{N}}

\def\textiid{i.i.d.\@\xspace}
\newcommand\iid{\ifmmode\text{ i.i.d. } \else \textiid \fi}

\newcommand{\Real}{\mathbb{R}}

\def\mathllap{\mathpalette\mathllapinternal}
\def\mathllapinternal#1#2{%
  \llap{$\mathsurround=0pt#1{#2}$}}

\def\clap#1{\hbox to 0pt{\hss#1\hss}}
\def\mathclap{\mathpalette\mathclapinternal}
\def\mathclapinternal#1#2{%
  \clap{$\mathsurround=0pt#1{#2}$}}

\let\oldstackrel\stackrel
\renewcommand{\stackrel}[2]{\oldstackrel{\mathclap{#1}}{#2}}

\renewcommand{\hbar}{h\mathllap{\overline{\vphantom{h}\hphantom{\rule{4.6pt}{0pt}}}\mspace{0.77mu}}}

\catcode`~=11 % make LaTeX treat tilde (~) like a normal character
\newcommand{\urltilde}{\kern -.06em\lower -.06em\hbox{~}\kern .02em}
\catcode`~=13 % revert back to treating tilde (~) as an active character

\newcommand\numberthis{\addtocounter{equation}{1}\tag{\theequation}}
\usepackage{bbm}
\usepackage{enumitem}

\usepackage{amsthm}

\theoremstyle{plain}
\newtheorem{theorem}{Theorem}
\newtheorem*{theorem*}{Theorem}
\newtheorem{prop}{Proposition}
\newtheorem{lemma}{Lemma}
\newtheorem{corollary}{Corollary}
\theoremstyle{definition}

\newtheorem*{contribution*}{Our contribution}
\newtheorem{example}{Example}

\newtheorem*{observation*}{Observation}

\newtheorem{remark}{Remark}
\def\yb{{\mathbf y}}
\def\yt{{\tilde y}}
\def\ybt{\tilde{\mathbf{y}}}

\title{\huge{On the Capacity Regions of Cloud Radio Access\\ Networks with Limited Orthogonal Fronthaul}}
\author{
    \IEEEauthorblockN{Shouvik Ganguly\IEEEauthorrefmark{1},\ Seung-Eun Hong\IEEEauthorrefmark{2},\ Young-Han Kim\IEEEauthorrefmark{1}}\\
    \IEEEauthorblockA{\IEEEauthorrefmark{1}Department of Electrical and Computer Engineering\\ 
     University of California, San Diego}\\
     \IEEEauthorblockA{\IEEEauthorrefmark{1}Email:  shgangul@eng.ucsd.edu,\ yhk@ucsd.edu}\\
    \IEEEauthorblockA{\IEEEauthorrefmark{2}Telecommunications \& Media Research Lab, ETRI, Korea}\\
    \IEEEauthorblockA{\IEEEauthorrefmark{2}Email:  iptvguru@etri.re.kr}%
     \thanks{This work was supported in part by Institute of Information \& communications Technology Planning \& Evaluation (IITP) grant  2018-0-01410.}%
\thanks{The material in this manuscript was presented in part at the IEEE International Symposium on Information Theory 2017 \cite{gangulykimisit17} and at the IEEE International Symposium on Information Theory 2019 \cite{gangulykimisit19}.}%
}
\begin{document}
\maketitle

\begin{abstract}
Uplink and downlink cloud radio access networks are modeled as two-hop $K$-user $L$-relay networks, whereby small base-stations act as relays for end-to-end communications and are connected to a central processor via orthogonal fronthaul links of finite capacities. Simplified versions of network compress--forward (or noisy network coding) and distributed decode--forward are presented to establish inner bounds on the capacity region for uplink and downlink communications, that match the respective cutset bounds to within a finite gap independent of the channel gains and signal to noise ratios. These approximate capacity regions are then compared with the capacity regions for networks with no capacity limit on the fronthaul. Although it takes infinite fronthaul link capacities to achieve these ``fronthaul-unlimited'' capacity regions exactly, these capacity regions can be approached \emph{approximately} with finite-capacity fronthaul. The total fronthaul link capacities required to approach the fronthaul-unlimited sum-rates (for uplink and downlink) are characterized. Based on these results, the capacity scaling law in the large network size limit is established under certain uplink and downlink network models, both theoretically and via simulations.
\end{abstract} 

\section{Introduction}
\begin{figure}[h]
\begin{subfigure}{.5\textwidth}
\centering
\begin{overpic}[width=0.8\textwidth]{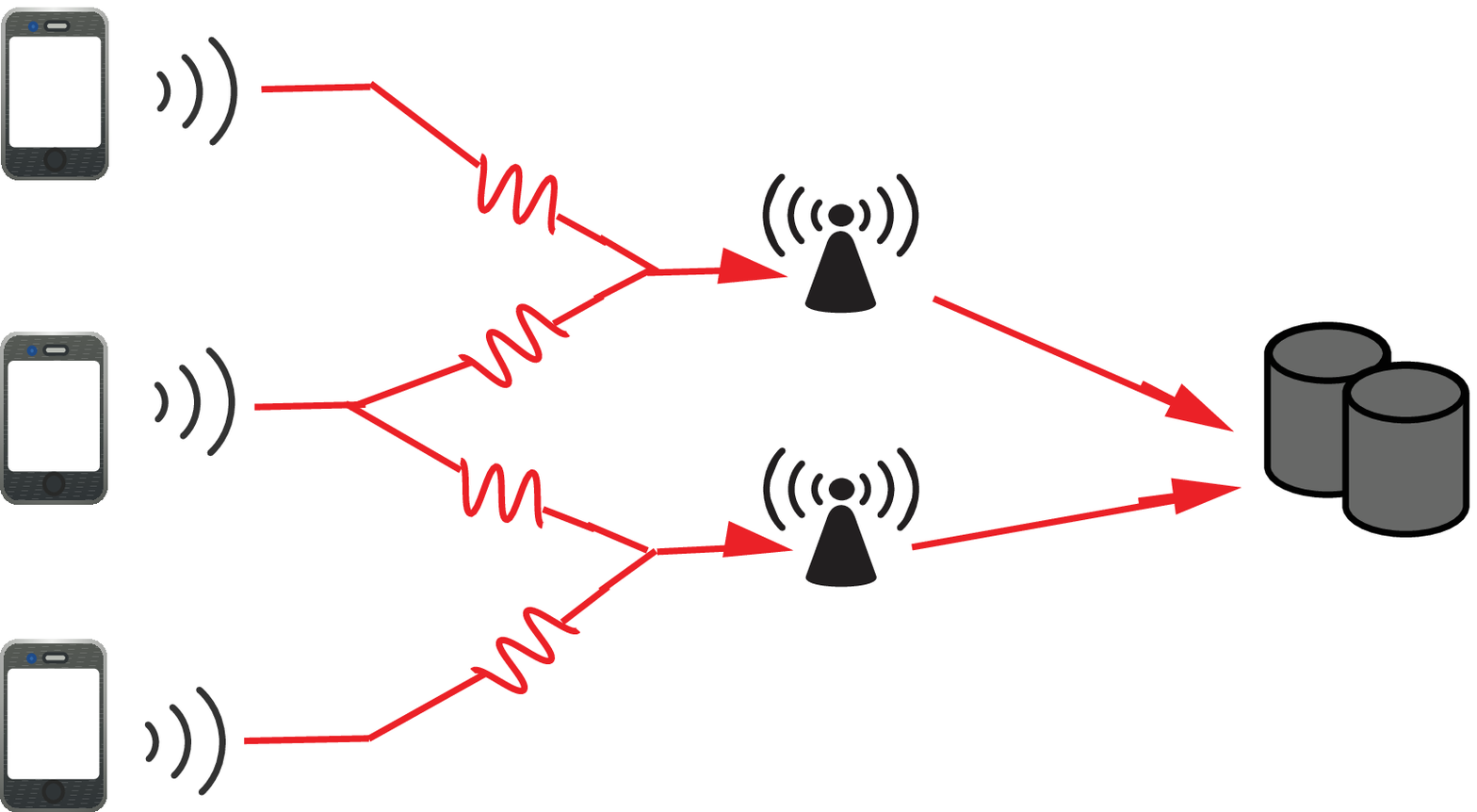}
\put(-10,-6){User devices}
\put(45,9){Radio heads}
\put(85,13){Central}
\put(83,8){processor}
\end{overpic}
\end{subfigure}
\begin{subfigure}{.5\textwidth}
\centering
\begin{overpic}[width=0.8\textwidth]{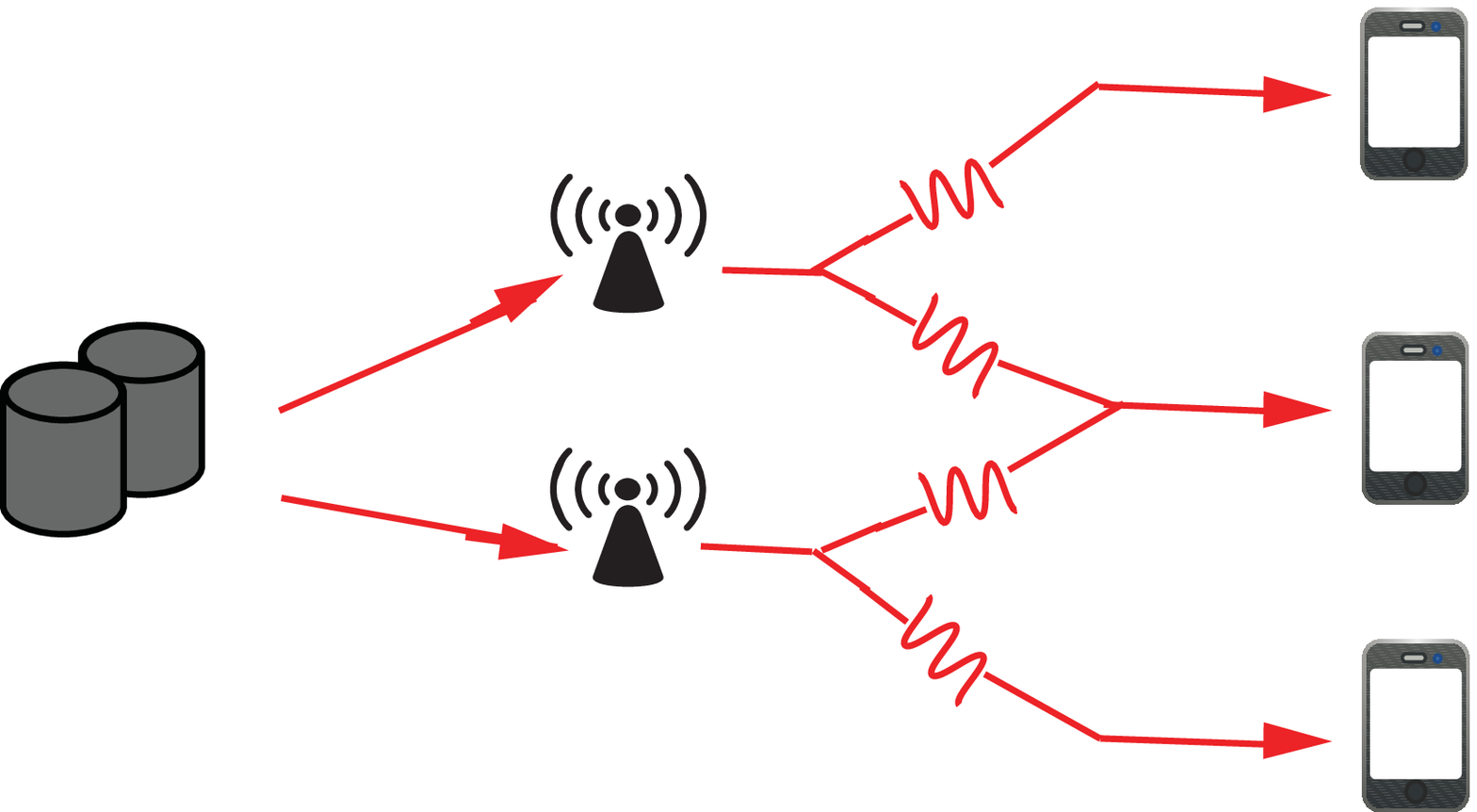}
\put(79,-6){User devices}
\put(30,9){Radio heads}
\put(-2,13){Central}
\put(-4,8){processor}
\end{overpic}
\end{subfigure}
\caption{(a) Uplink and (b) downlink cloud radio access networks.}\label{fig1}
\end{figure}

With ever-increasing demands for higher data rates, better coverage, more reliable connectivity for a large number of devices, new network architectures and protocols are expected to play an important role in future communication systems. The cloud radio access network (C-RAN) architecture \cite{quekpengsimoneyucranbook17, pengwanglaupoorwc15} is one of the promising candidates, in which communication over a group of cells is coordinated by a cloud-based central processor. Fig.~\ref{fig1} depicts uplink and downlink C-RAN systems schematically. 

Base stations in a C-RAN, unlike in conventional cellular networks, do not perform all network functionalities locally, but instead delegate most of them to a central processor by communicating with it over wired or wireless fronthaul links. If these links have unbounded capacities, the base stations act as spatially distributed antennas of a conventional multiple-input multiple-output (MIMO) system, that use beamforming to coordinate transmission and mitigate interference among multiple cells. For the purposes of this paper, these models with unlimited coordination are referred to as ``fronthaul-unlimited uplink'' or ``fronthaul-unlimited downlink'', as the case may be. For the more realistic situation of limited fronthaul link capacities, beamforming in a downlink C-RAN is typically performed at the central processor assuming no capacity constraints, and the corresponding baseband signals are digitized individually for each base-station and transmitted through the fronthaul links. For uplink, the received signal at each base-station is similarly digitized individually according to the corresponding link capacity, and transmitted to the base-station. These approaches often lead to high fronthaul capacity requirements. 

As an alternative to this greedy beamforming--digitization approach, this paper investigates near-optimal coding schemes for the C-RAN architecture and their achievable throughput tradeoffs by modeling the entire system as a two-hop relay network. In this model, which was studied, for example, in \cite{quekpengsimoneyucranbook17, zhouyujsa14, zhouxuyuchentit16}, the base stations act as relays that summarize the received signals from user devices to the central processor (uplink) and transmit the prescribed signals from the central processor to user devices (downlink). Communication-theoretic results on this model were presented in a recent volume edited by Quek, Peng, Simeone, and Yu \cite{quekpengsimoneyucranbook17}. 
 
\subsection{Uplink C-RAN}
Several coding schemes have been proposed in the literature for the uplink C-RAN with $K$ users (senders) and $L$ relays. %The two-hop relay network model for the uplink C-RAN with $K$ senders and $L$ relays was studied by 
Zhou and Yu \cite{zhouyujsa14} applied the network compress--forward relaying scheme \cite{kramergastparguptatit05} to this model and showed, by optimizing over quantizers, that under some symmetry assumptions, this scheme achieves the optimal sum-rate within $L/2$ bits per real dimension uniformly over all $K$ and all channel parameters. Sanderovich, Someskh, Poor, and Shamai \cite{sanderovichsomekhpoorshitztit09} used the same scheme and analyzed the large-user asymptotics (i.e., the scaling law) of symmetric achievable rates when all fronthaul links have equal capacities. Zhou, Xu, Yu, and Chen \cite{zhouxuyuchentit16} subsequently showed that under a sum-capacity constraint on the fronthaul links, the coding scheme in \cite{zhouyujsa14} and \cite{sanderovichsomekhpoorshitztit09} can be simplified through successive cancellation decoding, generalizing an earlier result for the single-sender multiple relay network \cite{sanderovichshamaisteinbergkramertit08}. Aguerri and Zaidi \cite{aguerrizaidi16} proposed a hybrid coding scheme of network compress--forward and compute--forward \cite{nazergastpar11}, and demonstrated that it outperforms the better of the two in general. Aguerri, Zaidi, Caire, and Shamai \cite{aguerrizaidicaireshitzarxiv17} specialized the noisy network coding scheme \cite{nnc} to the uplink C-RAN, the achievable rate region of which coincides with that of network compress--forward \cite{zhouyujsa14, sanderovichsomekhpoorshitztit09}. 

The most general outer bound on the capacity region of the uplink C-RAN can be obtained by specializing the cutset bound \cite{gamalcutset}; see, for example, \cite{zhouyujsa14} and the references therein, as well as Proposition~\ref{upgenouter} in this paper. The cutset bound has been further tightened under additional assumptions. %Simeone, Somekh, Erkip, Poor, and Shamai \cite{simeonesomeskherkippoorshitz11} studied the uplink single-user C-RAN with symmetric channel, \textit{oblivious relaying with enabled time sharing}, and unreliable fronthaul, where the fronthaul links are assumed to be liable to failure with some assigned probabilities, and derived upper and lower bounds on achievable rates.%Mathematically, oblivious relaying implies that encoders and decoders can choose different codebooks at random and the relays are unaware (and unable to deduce) the specific choice of codebook. If in addition, time sharing is enabled, the encoders and decoders can switch among different codebooks and the relays are assumed to be aware of the time sharing sequence. Upper and lower bounds on achievable rates were derived in terms of the link failure probabilities and under the oblivious relaying assumption, and were numerically shown to be close. 
Aguerri, Zaidi, Caire, and Shamai \cite{aguerrizaidicaireshitzarxiv17} studied the uplink C-RAN in which the relays are oblivious of the codebooks of the senders, and demonstrated that network compress--forward (or noisy network coding) achieves the capacity region. Simeone, Somekh, Erkip, Poor, and Shamai \cite{simeonesomeskherkippoorshitz11} studied the uplink C-RAN with one sender, $L$ oblivious relays, and unreliable fronthaul links, and derived an upper bound on the capacity, which was numerically shown to be close to the network compress--forward lower bound under certain network parameters.

\begin{contribution*}
In this paper, we apply network compress--forward (or equivalently, noisy network coding) to the uplink C-RAN and show that the scheme achieves the capacity region approximately within $(1/2)\log(eL)$ bits per user per real dimension, regardless of the channel gain matrix, power constraint, and the number of users $K$. When the fronthaul link capacities are unbounded, the approximation is precise and the network compress--forward inner bound (as well as the cutset outer bound) coincides with the fronthaul-unlimited uplink capacity region. We quantify the minimum fronthaul capacity required to achieve the latter approximately. We then use this result to characterize the scaling behavior of the uplink C-RAN sum-capacity for large network size under various channel models and demonstrate that the C-RAN sum-rate exhibits similar large-user asymptotics as the fronthaul-unlimited uplink sum-capacity for a range of channel models. 
\end{contribution*}

\subsection{Downlink C-RAN}
For the downlink C-RAN with $L$ relays and $K$ receivers, a variety of coding schemes have been proposed. Hong and Caire \cite{hongcairetit13} studied a low-complexity reverse compute--forward scheme for symmetric rates. Liu and Yu \cite{liuyuglobecom16} applied network coding and beamforming to the downlink model with a noiseless multi-hop fronthaul. Motivated by the MAC--BC duality, Liu, Patil, and Yu \cite{liupatilyuisit16} proposed compression-based schemes and established a duality between achievable rate regions for the uplink and downlink C-RANs. El Bakouri and Nazer \cite{bakourinazerallerton17, bakourinazerallerton18} applied integer-forcing based joint beamforming and compression strategies and demonstrated a duality between uplink and downlink C-RANs under this framework. Bidokhti and Kramer \cite{bidokhtikramertit16} studied the 2-relay, single-user downlink C-RAN and used rate-splitting across relays and Marton coding with common message to derive capacity lower bounds. Bidokhti, Kramer, and Shamai \cite{bidokhtikramershitzisit17} studied the $L$-relay, single-user downlink C-RAN and used Marton coding and rate-splitting across relays, but this time with no common message (due to the complexity for $L>2$). Wang, Wigger, and Zaidi \cite{wangwiggerzaiditit18} studied the three-hop, 2-relay, 2-user downlink C-RAN with relay cooperation, where the relays communicate with each other once, simultaneously, per network use. They applied the \textit{generalized data-sharing (G-DS)} and distributed decode--forward (DDF) \cite{ddf} coding schemes to this network, numerically showing that G-DS outperforms DDF in the low-power regime with a Gaussian second hop. More recently, Patil and Yu \cite{patilyuarxiv18} have shown that under fronthaul sum-capacity constraints, a successive encoding scheme achieves the same rate region as DDF.     

The most general outer bound on the capacity region of the downlink C-RAN can be obtained by specializing the cutset bound \cite{gamalcutset}; see, for example, \cite{bidokhtikramershitzisit17} and the references therein, as well as Proposition~\ref{downgenouter} in this paper. The cutset bound has been further tightened for specific network configurations. Bidokhti and Kramer \cite{bidokhtikramertit16} derived capacity upper bounds for the $2$-relay, single-user downlink C-RAN by tightening the cutset bound through channel enhancement techniques \cite{ozarowbstj80, kangliuisit11}. These bounds are tight for the single-user symmetric Gaussian C-RAN under certain parameters, establishing the capacity for those cases. These upper bounds were further generalized to $L\ge 3$ relays and a single user by Bidokhti, Kramer, and Shamai \cite{bidokhtikramershitzisit17}. 

\begin{contribution*}
In this paper, we specialize and simplify the distributed decode--forward coding scheme \cite{ddf} to the downlink C-RAN with capacity-limited single-hop fronthaul. In this scheme, multicoding at the encoder (as in Marton coding for broadcast channels \cite{nit}) is coupled with coding for fronthaul links, which allows more efficient coordination among the transmitted signals at the base-stations. We show that our rate region achieves a per-user gap of $(1/2)\log(eKL)$ bits per real dimension from the cutset bound. This result refines the best-known linear gap from capacity for this model (implicit in \cite{liupatilyuisit16}). Similar to the uplink case, we also quantify the minimum fronthaul link capacity required to achieve the performance of fronthaul-unlimited downlink approximately. Using this, we characterize the scaling behavior of the downlink C-RAN sum-capacity for large network size under various channel models and demonstrate that the C-RAN sum-rate exhibits the same large-user asymptotics as the fronthaul-unlimited downlink sum-capacity for a range of channel models.  
\end{contribution*}
  
\subsection{Organization of the Paper}
The rest of the paper is organized as follows. Section~\ref{upload} studies the uplink model. Section~\ref{upgen} describes the general inner and outer bounds on the capacity region; Section~\ref{upgauss} specializes the network compress--forward inner bound to the Gaussian network model and establishes the capacity gap; Section~\ref{updiscuss} establishes the fronthaul capacity needed to approach the performance of fronthaul-unlimited uplink; and Section~\ref{upscale} applies the results of Section~\ref{updiscuss} to examine the sum-capacity scaling for uplink C-RAN under various network models. Section~\ref{download} parallels the same flow for the downlink C-RAN. Section~\ref{mimocran} extends the results of Sections~\ref{upload} and \ref{download} to MIMO C-RANs, in which users and relays have multiple local antennas. Detailed coding schemes as well as the proofs of technical results stated in Sections~\ref{upload}, \ref{download}, and \ref{mimocran} are relegated to Appendices~\ref{upproofs}, \ref{downproofs}, and \ref{mimoproofs}, respectively. Section~\ref{conclude} concludes the paper. Each of Sections~\ref{upload}--\ref{mimocran} is accompanied by simulations of sum-rate scaling laws under stochastic geometry network models \cite{baccelligamaltsetit11} with LOS as well as NLOS channels. The simulation codes (in {MATLAB}) are available at \href{https://github.com/shouvik1234/Crancode}{https://github.com/shouvik1234/Crancode}. 

Throughout the paper, we follow the notation in \cite{nit}. In addition, we use the following. We use $||A||_F:=\sqrt{\tr(AA^T)}=\sqrt{\tr(A^TA)}$ to denote the Frobenius norm of a matrix $A$. For a natural number $n,$ we denote by $[n]$ the set $\{1, \ldots, n\}.$ We denote a finite tuple of objects $(x_l, l\in \Sc)$ by the shorthand notation $x(\Sc),$ for $\Sc\subseteq\mathbb{N}.$ For example, $x([n]) = x^n = (x_1, \ldots, x_n).$ For a tuple of random variables $X(\Sc):=(X_l, l\in\Sc)$ and a random variable $Y,$ we define the \emph{total correlation}
\[
I^*(X(\Sc)\cond Y) := \sum_{x(\Sc), y}p(x(\Sc),y)\log\frac{p(x(\Sc)\cond y)}{\prod_{l\in\Sc}p(x_l\cond y)} = \sum_{l\in\Sc}I(X_l; X([l-1]\cap\Sc)\cond Y)
\]
as a multivariate generalization of conditional mutual information  \cite{watanabeibm60}.
 For functions $f$ and $g$ from $\mathbb{N}$ to $\Real,$ we say $f\sim g$ if $f(n)/g(n)\to 1$ as $n\to\infty.$ Further, $\log(\cdot)$ and $\ln(\cdot)$ denote logarithms to base $2$ and base $e,$ respectively. All information measures are in bits.

\section{Uplink Communication}\label{upload}
\subsection{General Model}\label{upgen}
\begin{figure}[h]
\center
\includegraphics[scale=1.0]{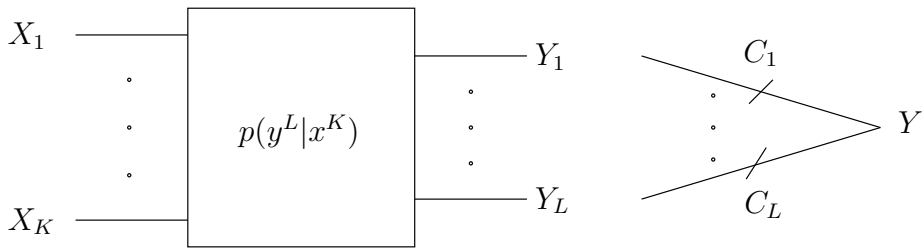}
\caption{Uplink network model.}\label{upfig}
\end{figure}

We model the uplink C-RAN as a two-hop relay network in Fig.~\ref{upfig}, where the first hop, namely, the (wireless) channel from the user devices (senders) to the radio heads (relays), is modeled as a discrete memoryless network $p(y^L|x^K)$, and the second hop, namely, the channel from the radio heads to the central processor, consists of orthogonal links of capacities $C_1,\ldots,C_L$ bits per real dimension, decoupled from the first hop. To be more precise, the channel output at the central processor (receiver) is $(W_1,\ldots,W_L)$, where $W_l\in[2^{nC_l}]$ is a reliable estimate of what relay $l$ communicates to the receiver over $n$ transmissions. We assume without loss of generality that these communication links are noiseless.

A $(2^{nR_1},\ldots,2^{nR_K},n)$ code for this network consists of $K$ message sets $[2^{nR_1}],\ldots,[2^{nR_K}]$; $K$ encoders, where encoder $k\in[K]$ assigns a codeword $x_k^n$ to each $m_k\in[2^{nR_k}]$; $L$ relay encoders, where relay encoder $l\in[L]$ assigns an index $w_l\in[2^{nC_l}]$ to each received sequence $y_l^n$; and a decoder that assigns message estimates $(\hat{m}_1,\ldots,\hat{m}_K)$ to each index tuple $w^L:=(w_1, \ldots, w_L)$. We assume that the messages $M_1,\ldots,M_K$ are uniformly distributed and independent of each other. The average probability of error is defined as $P_e^{(n)}=\P(\cup_{k=1}^K\{\hat{M}_k\ne M_k\})$. A rate tuple $(R_1,\ldots,R_K)$ is achievable if there is a sequence of $(2^{nR_1},\ldots,2^{nR_K},n)$ codes with $\lim_{n\to\infty}P_e^{(n)}=0$. The capacity region is defined as the closure of the set of all achievable rate tuples.

Sanderovich et al.~\cite{sanderovichsomekhpoorshitztit09} specialized the network compress--forward scheme \cite{sanderovichshamaisteinbergkramertit08} to this network. Roughly speaking, each relay compresses its received sequence and sends the bin index of the compression index. More specifically, to communicate message $M_k,$ sender $k\in[K]$ transmits a codeword $x_k^n(M_k).$ Upon receiving $Y_l^n,$ relay $l\in[L]$ compresses the sequence into a compression sequence $\hat{y}_l^n(W_l, T_l)$ and forwards $W_l.$ For a randomly generated compression codebook, the compression at relay $l$ succeeds w.h.p.\@ if
\[
C_l + \hat{R}_l > I(Y_l; \hat{Y}_l).\numberthis\label{ulnncinformal1}
\]
The receiver then recovers $(M_1, \ldots, M_K)$ based on $(W_1, \ldots, W_L),$ which succeeds w.h.p.\@ for a randomly generated codebook if 
\[
\sum_{k\in \Sc_1}R_k + \sum_{l\in \Sc_2}\hat{R}_l < I(X(\Sc_1);\hat{Y}(\Sc_2^c)|X(\Sc_1^c))+\sum_{l\in \Sc_2}I(Y_l;\hat{Y}_l)-\sum_{l\in \Sc_2}I\bigl(Y_l;\hat{Y}_l\cond\hat{Y}([l-1]\cap \Sc_2),\hat{Y}(\Sc_2^c),X^K\bigr)\numberthis\label{ulnncinformal2}
\]    
for every $\Sc_1\subseteq[K], \Sc_2\subseteq[L]$ such that $\Sc_1\ne\emptyset.$ Combining \eqref{ulnncinformal1} and \eqref{ulnncinformal2} to eliminate the auxiliary rates $\hat{R}_l$ and introducing a time-sharing random variable $Q$ leads to the following inner bound on the capacity region of this network. (See Appendix~\ref{upproofs} for a complete proof.) 
\begin{prop}[Network compress--forward inner bound for the uplink C-RAN]\label{propo1}  
A rate tuple $(R_1,\ldots,R_K)$ is achievable if
\[ 
\sum_{k\in \Sc_1}R_k  \le I(X(\Sc_1);\hat{Y}(\Sc_2^c)\cond X(\Sc_1^c), Q)+\sum_{l\in \Sc_2}C_l-\sum_{l\in \Sc_2}I\left(Y_l;\hat{Y}_l\cond\hat{Y}\left([l-1]\cap \Sc_2\right),\hat{Y}(\Sc_2^c),X^K, Q\right)\numberthis\label{upinngen}
\]
for all $\Sc_1\subseteq[K]$ and $\Sc_2\subseteq[L]$ for some pmf $p(q)\prod_{k=1}^Kp(x_k\cond q)\prod_{l=1}^Lp(\hat{y}_l\cond y_l, q)$. 
\end{prop}
%The inner bound can also be achieved by the following coding scheme, which is a specialization of the noisy network coding scheme \cite{nnc} to the uplink CRAN network model. 
The same inner bound can also be achieved by specializing \cite{aguerrizaidicaireshitzarxiv17} the noisy network coding scheme \cite{nnc}. 

Specializing the cutset bound \cite{gamalcutset} to the uplink C-RAN model leads to the following.
\begin{prop}[Cutset outer bound for the uplink C-RAN]\label{upgenouter}
If a rate tuple $(R_1,\ldots,R_K)$ is achievable for the uplink C-RAN, then
\begin{align*}
\sum_{k\in \Sc_1}R_k & \le I(X(\Sc_1);Y(\Sc_2^c)\cond X(\Sc_1^c),Q)+\sum_{l\in \Sc_2}C_l\numberthis\label{cutsetupbet}
\end{align*}
for all $\Sc_1\subseteq[K]$ and $\Sc_2\subseteq[L]$ for some pmf $p(q)\prod_{k=1}^Kp(x_k\cond q).$ 
\end{prop}
For completeness, we provide a proof of Proposition \ref{upgenouter} in Appendix \ref{upproofs}.

\begin{remark}\label{inffronthaullimit}
As the fronthaul capacities $C_1, \ldots, C_L$ tend to infinity, this uplink C-RAN channel model becomes identical to the ``fronthaul-unlimited'' uplink channel from $K$ senders to a single receiver with $L$ receive antennas, i.e., the multiple access channel $p(y^L|x^K)$ with $K$ senders $X_1, \ldots, X_K$ and one receiver $Y^L.$ In this regime, both the inner and outer bounds can be shown to converge to the capacity region of the fronthaul-unlimited uplink channel, characterized by rate tuples $(R_1, \ldots, R_K)$ satisfying
\[
\sum_{k\in \Sc_1}R_k \le I(X(\Sc_1); Y^L\cond X(\Sc_1^c), Q)
\]
for every $\Sc_1\subseteq[K]$ for some pmf $p(q)\prod_{k\in[K]}p(x_k\cond q).$
%Intuitively, having infinite fronthaul is equivalent to the relays being co-located with the receiver. 
In contrast, for \emph{finite} fronthaul link capacities $C_1, \ldots, C_L,$ no matter how large, we can always find networks for which the capacity region of the uplink C-RAN is strictly smaller than the fronthaul-unlimited uplink capacity region, as demonstrated in Section \ref{updiscuss}.
\end{remark} 

%To see this, rewrite \eqref{upinngen} and \eqref{cutsetupbet} as
%\begin{align*}
%\sum_{k\in S_1}R_k  & < \min_{S_2\subseteq[L]}\Bigl(I(X(S_1);\hat{Y}(S_2^c)|X(S_1^c), Q)+\sum_{l\in S_2}C_l\\
%& \quad\quad\quad\quad\quad -\sum_{l\in S_2}I\bigl(Y_l;\hat{Y}_l\cond\hat{Y}\left([1:l-1]\cap S_2\right),\hat{Y}(S_2^c),X^K, Q\bigr)\Bigr)\numberthis\label{upinngentimeshare_modified} 
%\end{align*} 
%and
%\[
%\sum_{k\in S_1}R_k \le \min_{S_2\subseteq[L]}\Bigl(I(X(S_1);Y(S_2^c)|X(S_1^c),Q)+\sum_{l\in S_2}C_l\Bigr)\numberthis\label{cutsetupbet_modified}
%\]
%respectively. Then, as $C_1, \ldots, C_L\to\infty,$ we observe that the minima in both \eqref{upinngentimeshare_modified} and \eqref{cutsetupbet_modified} are achieved by $S_2 = \emptyset,$ and both the regions reduce to
%\[
%\sum_{k\in S_1}R_k \le I(X(S_1);Y^L|X(S_1^c),Q)\numberthis\label{uplink_mac}
%\] 
%for some pmf $p(q)\prod_{k=1}^Kp(x_k|q),$ which is the capacity region of the $K$-sender centralized MIMO uplink with $L$ antennas at the receiver. 

\subsection{Gaussian Model}\label{upgauss}       
We now assume that the first hop of the network is Gaussian, i.e.,  
\begin{align*}Y^L & =GX^K+Z^L,\end{align*} 
where $G\in\Real^{L\times K}$ is a (deterministic) channel gain matrix and $Z^L$ is a vector of independent $\N(0,1)$ noise components. We also assume the average power constraint $P$ on each sender, i.e.,  
\begin{align*}\sum_{i=1}^nx_{ki}^2(m_k) & \le nP,\quad m_k\in[2^{nR_k}],\quad k\in[K].\end{align*}

The network compress--forward inner bound in Proposition~\ref{propo1} can be specialized to this Gaussian network model to show the achievability of all rate tuples $(R_1, \ldots, R_K)$ such that
\[
\sum_{k\in \Sc_1}R_k \le \frac{1}{2}\log\Big\vert\frac{P}{\sigma^2+1}G_{\Sc_2^c,\Sc_1}G_{\Sc_2^c,\Sc_1}^T+I\Big\vert +\sum_{l\in \Sc_2}C_l-\frac{\vert \Sc_2\vert}{2}\log\left(1+\frac{1}{\sigma^2}\right) =: f_{\mathrm{in}}(\Sc_1,\Sc_2)\numberthis\label{upinngaussinformal}
\]
for all $\Sc_1\subseteq[K]$ and $\Sc_2\subseteq[L]$ for some $\sigma^2>0.$  Here, $G_{\Sc_2^c,\Sc_1}$ is the submatrix of $G$ formed by the rows with indices in $\Sc_2^c$ and the columns with indices in $\Sc_1.$ This follows by considering $X^K$ to be a vector of i.i.d.\@ $\N(0,P)$ random variables, and setting $\hat{Y}_l = Y_l+\hat{Z}_l$, $l\in[L]$, where $\hat{Z}_l\sim\N(0,\sigma^2)$. For convenience, for every $\sigma^2>0,$ we denote the set of tuples $(R_1, \ldots, R_K)$ satisfying \eqref{upinngaussinformal} and hence, achievable by network compress--forward (NCF), by $\Rr_\mathrm{up}^\text{NCF}(\sigma^2).$ We also denote the achievable sum-rate for each $\sigma^2>0$ by
\begin{align*}
R_\mathrm{sum}^\text{NCF}(\sigma^2) & :=\sup_{(R_1, \ldots, R_K)}\{R_1 + \cdots + R_K: (R_1, \ldots, R_K)\in\Rr_\mathrm{up}^\text{NCF}(\sigma^2)\}\numberthis\label{eq:uplinksumratedefinition}\\
& = \min_{\Sc_2\subseteq[L]}\Bigl(\frac{1}{2}\log\Big\vert\frac{P}{\sigma^2+1}G_{\Sc_2^c,[K]}G_{\Sc_2^c,[K]}^T+I\Big\vert +\sum_{l\in \Sc_2}C_l-\frac{\vert \Sc_2\vert}{2}\log\bigl(1+\frac{1}{\sigma^2}\bigr)\Bigr).\numberthis\label{upinngauss_sumrate_prev}
\end{align*} 

The cutset bound in Proposition~\ref{upgenouter} can also be specialized (see Lemma \ref{cutsetuplink} in Appendix~\ref{upproofs}) to the rate region characterized by
\[
\sum_{k\in \Sc_1}R_k \le \frac{1}{2}\log\left| PG_{\Sc_2^c,\Sc_1}G_{\Sc_2^c,\Sc_1}^T+I\right|+\sum_{l\in \Sc_2}C_l\numberthis\label{upoutgaussinformal}
 \]
for all $\Sc_1\subseteq[K]$ and $\Sc_2\subseteq[L].$ We denote the set of tuples $(R_1, \ldots, R_K)$ satisfying \eqref{upoutgaussinformal} by $\Rr_\mathrm{up}^\text{CS}.$ We also denote the sum-rate upper bound by
\[
R_\mathrm{sum}^\text{CS}:=\sup_{(R_1, \ldots, R_K)}\{R_1 + \cdots + R_K: (R_1, \ldots, R_K)\in\Rr_\mathrm{up}^\text{CS}.\}\numberthis\label{eq:uplinksumrateouterdefinition}
\] 
 
Our main goal of this section is to quantify how well network compress--forward performs for the Gaussian network, by comparing its achievable rates in \eqref{upinngaussinformal} with the cutset bound in \eqref{upoutgaussinformal}. In particular, we establish the following result, whose proof is deferred to Appendix~\ref{upproofs}.
\begin{theorem}\label{upgaptheo1}
For every $G\in\mathbbm{R}^{L\times K}$ and every $P\in\mathbbm{R}^+$, if a rate tuple $(R_1,\ldots,R_K)$ is in the cutset bound \eqref{upoutgaussinformal}, then the rate tuple $((R_1-\Delta)^+,\ldots,(R_K-\Delta)^+)$ is achievable, where 
\[
\Delta  \le \frac{1}{2}\log(eL) \approx\frac{1}{2}\log L+0.722.
\]
Moreover, the sum-rate gap between the cutset bound and the network compress--forward inner bound is upper-bounded as
\[
\Delta_\mathrm{sum}:=R_\mathrm{sum}^\text{CS} - \sup_{\sigma^2>0}R_\mathrm{sum}^\text{NCF}(\sigma^2)\le\begin{cases}\frac{L}{2}H(K/L)\le\frac{K}{2}\log(eL/K), & L\ge 2K,\\
				\frac{L}{2}, & L < 2K,\end{cases}
\]
irrespective of $P$ and $G,$ where $H(\cdot)$ is the binary entropy function.
\end{theorem}
\begin{remark}
Theorem~\ref{upgaptheo1} tightens the existing sum-rate gap of $L/2$ bits per real dimension \cite{zhouyujsa14}. 
\end{remark}

\subsection{Comparisons with Fronthaul-Unlimited Uplink}\label{updiscuss}      
In this section, we examine the effect of the capacities $C_l$ of the fronthaul links (in particular, their sum $C_{\sum}:=C_1 + \cdots + C_L$) on the capacity region of the uplink C-RAN. Recall from Remark~\ref{inffronthaullimit} that as the fronthaul link capacities approach infinity, the uplink C-RAN capacity region becomes the same as the fronthaul-unlimited uplink capacity region in the limit. However, as shown by the following example, this limit is in general unattainable when the link capacities are finite.
\begin{example}\label{example:toyexample}
Consider the single-sender, $2$-relay Gaussian uplink C-RAN with first hop given by
\begin{align*}
Y_1 & = gX+Z_1,\\
Y_2 & = gX+Z_2,
\end{align*}
where $g\in\Real\setminus\{0\}$ and $Z_1, Z_2$ are i.i.d.\@ $\N(0,1)$ noise components. Let us denote the fronthaul link capacities of this network by $C_1$ and $C_2,$ and let there be an average power constraint $P > 0$ on the sender. Then the first hop has conditionally i.i.d.\@ outputs $Y_1, Y_2$ given $X.$ If $C_2 = \infty,$ this network is equivalent to the relay channel model studied in the Gaussian version of Cover's problem \cite{covergopinathopenprobs87} and by the results of Wu, Barnes, and \"{O}zg\"{u}r \cite{wubarnesozgurarxiv18}, the capacity for any finite $C_1$ is strictly less than the capacity for $C_1 = \infty.$ Thus, even for this simple network, the fronthaul-unlimited uplink capacity $R_\mathrm{sum}^{\infty} := (1/2)\log(1+2g^2P)$ is unattainable unless both the fronthaul link capacities are infinite.  
\begin{figure}[h]
        	\centering
        	\begin{overpic}[width=0.72\textwidth,height=3.7cm]{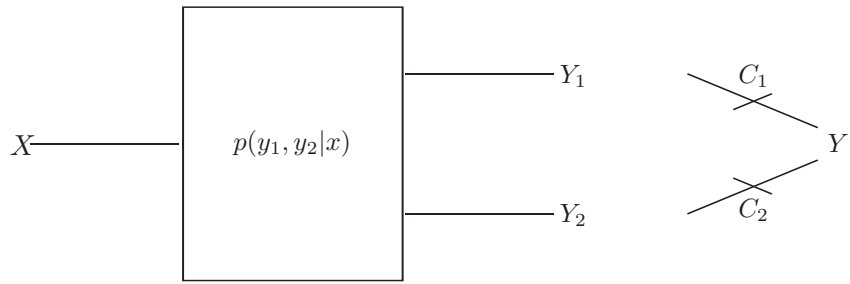}
	\put(9.5,14.5){$X$}
	\put(101,14.75){\small{$Y$}}
	\put(71,22.5){\small{$Y_1$}}
	\put(71,7){\small{$Y_2$}}
	\put(34.5,15){\small{$p(y_1, y_2|x)$}}
	\put(91,22.5){\small{$C_1$}}
	\put(91,7.5){\small{$C_2$}}
	\end{overpic}
        \caption{A single sender $2$-relay uplink C-RAN.}
	\label{fig:coverprob}
\end{figure}
On the positive side, it is possible to approximately achieve the fronthaul-unlimited uplink sum-rate for finite fronthaul capacities, provided we spend a sufficient amount of extra capacity on the fronthaul. Suppose that we have a certain amount of total capacity $C_{\sum}$ to spend on the fronthaul links, which we are free to allocate in any way among the two links. The cutset bound implies that we cannot hope to achieve the capacity $(1/2)\log(1+2g^2P)$ unless $C_{\sum}$ is at least equal to this amount. However, if we set
\[
C_1 = C_2 = \frac{1}{2} + \frac{1}{4}\log(1+2g^2P).
\]     
and thus spend the fronthaul sum-capacity of 
\[
C_{\sum} = \frac{1}{2}\log(1+2g^2P) + 1,
\]
then it can be shown, by taking $\sigma^2 = 1$ in \eqref{upinngauss_sumrate_prev}, that the uplink C-RAN sum-rate is 
\begin{align*}
& \min\left\{\frac{1}{2}\log(1 + g^2P), \frac{1}{2}\log\left(1+\frac{g^2P}{2}\right)+C_1 - \frac{1}{2}, \frac{1}{2}\log\left(1+\frac{g^2P}{2}\right)+C_2 - \frac{1}{2}, C_1 + C_2 - 1\right\}\\
& = \min\left\{\frac{1}{2}\log(1 + g^2P), \frac{1}{2}\log\left(1+\frac{g^2P}{2}\right)+\frac{1}{4}\log\left(1+2g^2P\right), \frac{1}{2}\log(1 + 2g^2P)\right\}\\
& =\frac{1}{2}\log(1+g^2P)\\
& \stackrel{(b)}{\ge} \frac{1}{2}\log(1+2g^2P) - \frac{1}{2},\numberthis\label{eq:toyexamplesumrate}
\end{align*}
where $(b)$ follows since $(1+2g^2P)\le 2(1+g^2P).$ Thus, using a total fronthaul link capacity only $1$ bit higher than the fronthaul-unlimited uplink capacity, we can achieve the fronthaul-unlimited uplink capacity within half a bit, irrespective of $P$ and $g.$ Thus we can achieve the fronthaul-unlimited uplink sum-capacity within a finite additive gap using a total fronthaul link capacity which is also finitely larger than the fronthaul-unlimited sum-capacity in the additive sense. This statement is formalized and generalized in Corollary~\ref{cor:additivegap} to Theorem~\ref{theorem:Cstaruplink}.

The result \eqref{eq:toyexamplesumrate} holds for every $P$ and therefore, with this fronthaul allocation strategy, we can achieve $R_\mathrm{sum}^\text{NCF}/R_\mathrm{sum}^\infty$ approaching $1$ as $P\to\infty,$ with a total fronthaul link capacity $C_{\sum}$ satisfying $C_{\sum}/R_\mathrm{sum}^\infty\to 1$ as $P\to\infty.$ In fact, we can go one step further and show, letting $P$ go to infinity in \eqref{eq:toyexamplesumrate}, that at high SNR, using a $C_{\sum}$ whose ratio to $R_\mathrm{sum}^\infty$ is $1$ within $O(1/\log P),$ network compress--forward can achieve a sum-rate whose ratio to $R_\mathrm{sum}^\infty$ is also $1$ within $O(1/\log P).$ Thus, in a multiplicative sense as well, only a slightly larger fronthaul capacity is sufficient to approximate the fronthaul-unlimited capacity for this network. This statement is formalized and explored in Corollary~\ref{cor:multiplicativegap}. 
\end{example}
%To see this, we write
%\begin{align*}
%1-\frac{R_\mathrm{sum}^\text{NCF}}{R_\mathrm{sum}^\mathrm{MIMO}} & = 1 - \frac{\log(1 + g^2P)}{\log(1+2g^2P)}\\
%& = \frac{\log(1 + 2g^2P)-\log(1 + g^2P)}{\log(1 + 2g^2P)}\\
%& \sim\frac{1}{\log P}
%\end{align*}
%and 
%\begin{align*}
%\frac{C_{\sum}}{R_\mathrm{sum}^\mathrm{MIMO}} - 1& = \frac{\log(1 + 2g^2P) + 2}{\log(1+2g^2P)} - 1\\
%& = \frac{2}{\log(1 + 2g^2P)}\\
%& \sim\frac{2}{\log P}.
%\end{align*}
%Thus, in an additive as well as a multiplicative sense, only a slightly larger fronthaul capacity is sufficient to approximate the centralized MIMO capacity for this network. These gap results are generalized in Theorem~\ref{theorem:Cstaruplink} and Corollaries~\ref{cor:additivegap} and \ref{cor:multiplicativegap}.  %To show this, we need the following.%To this end, we first rewrite the region \eqref{upinngauss}, for a fixed $\sigma^2 > 0,$ as the intersection of two \textit{polymatroids}.
%Using this, we can quantify the fronthaul requirement for noisy network coding to approximate the centralized MIMO uplink sum-capacity as follows.
We first quantify the fronthaul requirement for network compress--forward to approximate the fronthaul-unlimited uplink sum-capacity in Theorem~\ref{theorem:Cstaruplink}, from which the additive and multiplicative gap results follow as corollaries.
\begin{theorem}\label{theorem:Cstaruplink} 
If  
\[
C_{\sum} \ge \frac{1}{2}\log\left|\frac{P}{\sigma^2+1}GG^T + I\right| + \frac{L}{2}\log\left(1 + \frac{1}{\sigma^2}\right) =: C^*(\sigma^2)
\]
for some $\sigma^2>0,$ then there exist fronthaul link capacities $C_1, C_2, \ldots, C_L\ge 0$ with $\sum_{l\in[L]}C_l = C_{\sum}$ at which network compress--forward can achieve a sum-rate
\[
R_\mathrm{sum}^\text{NCF}(\sigma^2) = \frac{1}{2}\log\left|\frac{P}{\sigma^2+1}GG^T + I\right|.
\] 
Conversely, to achieve a sum-rate of $(1/2)\log|I+PGG^T|,$ we must have a total fronthaul capacity
\[
C_{\sum}\ge\frac{1}{2}\log|I+PGG^T|.
\] 
\end{theorem}
\begin{IEEEproof}
The achievable sum-rate can be written as
\begin{align*}
R_\mathrm{sum}^\text{NCF}(\sigma^2) & = \min_{\Sc_2\subseteq[L]}\Bigl(\frac{1}{2}\log\Big\vert\frac{P}{\sigma^2+1}G_{\Sc_2^c,[K]}G_{\Sc_2^c,[K]}^T+I\Big\vert +\sum_{l\in \Sc_2}C_l-\frac{\vert \Sc_2\vert}{2}\log\bigl(1+\frac{1}{\sigma^2}\bigr)\Bigr)\\
& = \min_{\Sc_2\subseteq[L]}\Bigl(\phi(\Sc_2^c) + \psi(\Sc_2)\Bigr),\numberthis\label{upinngauss_sumrate} 
\end{align*}
where 
\[
\phi(\Sc_2^c)  := \frac{1}{2}\log\left|\frac{P}{\sigma^2+1}G_{\Sc_2^c,[K]}G_{\Sc_2^c,[K]}^T+I\right|
\]
and
\[
\psi(\Sc_2)  := \sum_{l\in \Sc_2}\left(C_l - \frac{1}{2}\log\left(1+\frac{1}{\sigma^2}\right)\right).
\]
If $C_l\ge (1/2)\log(1 + 1/\sigma^2)$ for all $l\in[L],$ the set functions $\phi$ and $\psi$ are zero on the null set, monotonically increasing (with respect to the partial ordering defined by set inclusion), and are \emph{submodular}, i.e., we have $\phi(\emptyset) = 0,$ $\phi(\Sc)\le\phi(\Tc)$ if $\Sc\subseteq\Tc,$ and $\phi(\Sc\cup\Tc) + \phi(\Sc\cap\Tc) \le \phi(\Sc) + \phi(\Tc).$ The sets
\[
\Pr(\phi) := \Bigl\{(x_1, \ldots, x_L)\subseteq\Real_+^L: \sum_{l\in\Sc}x_l\le \phi(\Sc), \Sc\subseteq[L]\Bigr\}
\]
and $\Pr(\psi),$ defined in a similar manner, are referred to as \emph{polymatroids} \cite{edmonds1970}, which generalize two-dimensional pentagonal regions to $L$ dimensions. The following celebrated result can rewrite \eqref{upinngauss_sumrate} in an alternative form.
\begin{lemma}[Edmonds's polymatroid intersection theorem \cite{edmonds1970}]\label{lemma:edmonds}
If $\Pr(\phi)$ and $\Pr(\psi)$ are two polymatroids, then
\[
 \max\left\{\sum_{l\in[L]}x_l: (x_1, \ldots, x_L)\in\Pr(\phi)\cap\Pr(\psi)\right\} = \min_{\Sc\subseteq[L]}\left(\phi(\Sc) + \psi(\Sc^c)\right).
\]
\end{lemma} 
%The functions $\phi$ and $\psi$ in eq.~\eqref{upinngauss_sumrate} can be easily verified to be such that $\Pr(\phi)$ and $\Pr(\psi)$ are both polymatroids whenever $C_l\ge (1/2)\log(1 + 1/\sigma^2)$ for all $l\in[L].$ 
Using Lemma~\ref{lemma:edmonds}, we can rewrite \eqref{upinngauss_sumrate} as
\[
R_\mathrm{sum}^\text{NCF}(\sigma^2) = \max_{y^L}\left\{\sum_{l\in[L]}y_l: y_l\le \psi(\{l\}), l\in[L], \sum_{l\in \Sc_2}y_l\le\phi(\Sc_2), \Sc_2\subseteq[L]\right\}.
\] 
Now, let us fix 
\[
C_{\sum} \ge \phi([L]) + \frac{L}{2}\log\left(1+\frac{1}{\sigma^2}\right) = \frac{1}{2}\log\left|\frac{P}{\sigma^2+1}GG^T + I\right| + \frac{L}{2}\log\left(1 + \frac{1}{\sigma^2}\right)\numberthis\label{eq:Cstarinproof}
\]
 such that $C_1, \ldots, C_L$ are constrained to satisfy $C_1 + \ldots + C_L = C_{\sum}.$ Choose a point $\yb^*\equiv (y_1^*, \ldots, y_L^*)\in\Pr(\phi)$ such that $y_1^* + \ldots + y_L^* = \phi([L]).$ Such a point always exists since $\Pr(\phi)$ is a polymatroid. The point $\ybt\equiv(\yt_1, \ldots, \yt_L)$ defined by 
\[
\yt_l = \frac{C_{\sum} - \frac{L}{2}\log\left(1+\frac{1}{\sigma^2}\right)}{\phi([L])}y_l^*, \quad l\in[L],
\]
satisfies $\yt_1 + \cdots + \yt_L = C_{\sum} - (L/2)\log(1+1/\sigma^2).$ Therefore, choosing $C_l = \yt_l + (1/2)\log(1+1/\sigma^2)$ for each $l,$ $\Pr(\phi)$ becomes the cuboid 
\[
\{(y_1, \ldots, y_L): y_l\le \yt_l, l\in[L]\}
\]
with corner point $\ybt.$ Moreover, this cuboid includes the point $\yb^*,$ since $\yt_l\ge y_l^*$ for each $l$ by \eqref{eq:Cstarinproof}. Thus, the point $\yb^*$ lies in the intersection $\Pr(\phi)\cap\Pr(\psi)$ and therefore, network compress--forward, with this choice of $C_1, \ldots, C_L,$ achieves the sum-rate 
\[
y_1^* + \cdots + y_L^* = \phi([L]) = \frac{1}{2}\log\left|\frac{P}{\sigma^2+1}GG^T + I\right|,
\] 
establishing the result. The converse follows immediately from the cutset bound.
\end{IEEEproof}
\begin{remark}\label{rem:fronthaulallocLP}
Given $\sigma^2,$ $P,$ and $G,$ coming up with a specific allocation $(C_1, \ldots, C_L)$ satisfying the sum fronthaul constraint is equivalent to finding a point $\yb^*,$ as seen from the proof. Such a point can be found, moreover, by solving a linear feasibility problem
\begin{subequations}
\begin{alignat*}{2}
&\!\text{find}        &\qquad& (y_1^*, \ldots, y_L^*)\\
&\text{subject to} &      & \sum_{l\in \Sc}y_l^* \le\phi(\Sc),\quad \Sc\subsetneq[L],\\
&                  &      & \sum_{l\in[L]}y_l^* = \phi([L]).
\end{alignat*}
\end{subequations}
Thus, the fronthaul allocation problem is equivalent to checking the feasibility of a linear program with $2^L - 2$ inequalities and one equality.
\end{remark}
\begin{remark}\label{rem:uplinkCstartosumrate}
As an immediate consequence of the polymatroid representation, the best sum-rate achievable for a given total fronthaul capacity $C_{\sum}>0$ can be expressed as
\[
R_\mathrm{sum}^\mathrm{max}(C_{\sum}) = \sup_{\sigma^2 > 0}\min\left\{C_{\sum} - \frac{L}{2}\log\left(1+\frac{1}{\sigma^2}\right), \frac{1}{2}\log\left|\frac{P}{\sigma^2+1}GG^T+I\right|\right\}.
\]
The first term in the minimum increases monotonically from $0$ to $\infty$ as $\sigma^2$ increases from $0$ to $\infty,$ while for $G\ne 0,$ the second term decreases monotonically from $\infty$ to $0.$ Therefore, there is a unique $\sigma_*^2(C_{\sum})$ at which the supremum is attained and the two terms in the minimum are equal for $\sigma^2 = \sigma_*^2(C_{\sum}).$ This also shows that 
\[
\lim_{C_{\sum}\to\infty}\sigma_*^2(C_{\sum}) = 0
\] 
and hence, that
\[
\lim_{C_{\sum}\to\infty}R_\mathrm{sum}^\mathrm{max}(C_{\sum}) = \frac{1}{2}\log|PGG^T+I| = R_\mathrm{sum}^\infty.
\] 
Thus, our coding scheme is asymptotically optimal in the limit of large fronthaul sum-capacity. 
\end{remark}
Theorem~\ref{theorem:Cstaruplink} leads to a formalization of the achievable additive and multiplicative gaps from the fronthaul-unlimited uplink sum-capacity, that were briefly explored in Example~\ref{example:toyexample}. 
\begin{corollary}[Additive gap from fronthaul-unlimited uplink sum-capacity]\label{cor:additivegap}
Denote by $R_\mathrm{sum}^\infty$ the fronthaul-unlimited uplink sum-capacity, which is given by $(1/2)\log|I + PGG^T|.$ Then, for every $P$ and $G$ for some $\sigma^2>0,$ if $C_{\sum} = R_\mathrm{sum}^\infty + \Delta_1(\sigma^2),$ then there exist $C_1, \ldots, C_L$ with $\sum_{l\in[L]}C_l = C_{\sum},$ at which $R_\mathrm{sum}^\text{NCF}(\sigma^2)\ge R_\mathrm{sum}^\infty - \Delta_2(\sigma^2),$ where 
\[
\Delta_1(\sigma^2) = \frac{L}{2}\log\left(1+\frac{1}{\sigma^2}\right)
\]
and
\[
\Delta_2(\sigma^2) = \frac{\min\{K, L\}}{2}\log\left(1+\sigma^2\right).
\]
\end{corollary}
\begin{IEEEproof}
We have, from Theorem~\ref{theorem:Cstaruplink}, that if
\[
C_{\sum} = \frac{1}{2}\log\left|\frac{P}{\sigma^2+1}GG^T + I\right| + \frac{L}{2}\log\left(1 + \frac{1}{\sigma^2}\right),
\]
then
\[
C_{\sum} - R_\mathrm{sum}^\infty = \frac{1}{2}\log\frac{\left|\frac{P}{\sigma^2+1}GG^T + I\right|}{|PGG^T + I|} + \frac{L}{2}\log\left(1 + \frac{1}{\sigma^2}\right) \le \frac{L}{2}\log\left(1 + \frac{1}{\sigma^2}\right),\numberthis\label{eq:delta1}
\]
and 
\begin{align*}
R_\mathrm{sum}^\infty - R_\mathrm{sum}^\text{NCF}(\sigma^2) & = \frac{1}{2}\log\frac{|PGG^T + I|}{\left|\frac{P}{\sigma^2+1}GG^T + I\right|}\\
& \stackrel{(a)}{\le}\frac{\mathrm{rank}(G)}{2}\log(1+\sigma^2)\\
& \le\frac{\min\{K, L\}}{2}\log(1+\sigma^2),\numberthis\label{eq:delta2}
\end{align*}
where $(a)$ follows from the inequality $(1 + \alpha)/(1+\alpha/x)\le x$ for $x>1, \alpha > 0.$ Equations~\eqref{eq:delta1} and \eqref{eq:delta2} establish the result. 
\end{IEEEproof}
As a concrete illustration of the gaps established by Corollary~\ref{cor:additivegap}, taking $\sigma^2 = L$ in \eqref{eq:delta1} and \eqref{eq:delta2} yields
\begin{align*}
\Delta_1(\sigma^2) & = \frac{L}{2}\log\left(1+\frac{1}{L}\right)\\
& \stackrel{(a)}{\le} \frac{L}{2}\cdot\frac{\log e}{L}\\
& = \frac{\log e}{2}
\end{align*}
and 
\begin{align*}
\Delta_2(\sigma^2) & = \frac{\min{\{K, L\}}}{2}\log(1+L).
\end{align*}
Here, $(a)$ follows since $\log(1+x)\le x\log e$, $x > 0$. Similarly, setting $\sigma^2 = 1/L$ in \eqref{eq:delta1} and \eqref{eq:delta2} yields
\begin{align*}
\Delta_1(\sigma^2) & = \frac{L}{2}\log\left(1+L\right)
\end{align*}
and 
\begin{align*}
\Delta_2(\sigma^2) & = \frac{\min{\{K, L\}}}{2}\log\left(1+\frac{1}{L}\right)
 \le\frac{\min{\{K, L\}}}{2}\cdot\frac{\log e}{L}
 \le\frac{\log e}{2}.
\end{align*}
Various choices of $\sigma^2,$ as well as the corresponding tradeoffs between $\Delta_1$ and $\Delta_2,$ are summarized in Table~\ref{table:additivegap}.
 \begin{table}[h]
    \setlength{\tabcolsep}{20pt}
    \renewcommand{\arraystretch}{1.8}
    \centering
    \begin{tabular}{  c  c c} \hline
	$\sigma^2$ & $\Delta_1(\sigma^2)$ & $\Delta_2(\sigma^2)$ \\[1pt] \hline
       $L$ & $\displaystyle\frac{\log e}{2}$ & $\displaystyle\frac{\min\{K, L\}}{2}\log(1+L)$\\[1pt]
	$1$ & $\displaystyle\frac{L}{2}$ & $\displaystyle\frac{\min\{K, L\}}{2}$\\[1pt]
	$\displaystyle\frac{1}{L}$ & $\displaystyle\frac{L}{2}\log(1+L)$ & $\displaystyle\frac{\log e}{2}$\\[4pt]\hline
       \end{tabular}
\caption{Additive gap from fronthaul-unlimited uplink sum-capacity.}\label{table:additivegap}
\end{table} 
As noted before, Corollary~\ref{cor:additivegap}, being a channel- and SNR-independent result, implies that both $R_\mathrm{sum}^\text{NCF}/R_\mathrm{sum}^\infty$ and $C^*/R_\mathrm{sum}^\infty$ approach $1$ at high SNR. The next result is a further refinement of this statement.
\begin{corollary}[Multiplicative gap from fronthaul-unlimited uplink sum-capacity at high SNR]\label{cor:multiplicativegap} 
For a fixed channel gain matrix $G,$ let $P\to\infty$ and let $\sigma^2$ be chosen as $\sigma^2 = \sigma^2(P)$ such that
\[
\lim_{P\to\infty}P\sigma^2(P) = \infty
\]
and 
\[
\lim_{P\to\infty}\sigma^2(P)/P = 0.
\]  
Then, 
\[
1-\frac{R_\mathrm{sum}^\text{NCF}}{R_\mathrm{sum}^\infty}\sim\frac{\log\left(1+\sigma^2(P)\right)}{\log P}
\]
and 
\[
\frac{C^*}{R_\mathrm{sum}^\infty} - 1\sim\begin{cases}\displaystyle\frac{\frac{L\log e}{\sigma^2(P)} - \rank(G)\log(\sigma^2(P))}{\rank(G)\log P}, & \sigma^2(P)\xrightarrow[]{P\to\infty}\infty,\\
\displaystyle\frac{L\log(1/\sigma^2(P))}{\rank(G)\log P}, & \sigma^2(P)\xrightarrow[]{P\to\infty}0,\\
\displaystyle\frac{\rank(G)\log\left(\frac{1}{1+\sigma^2}\right) + L\log\left(1+\frac{1}{\sigma^2}\right)}{\rank(G)\log P}, & \sigma^2 > 0\text{ is fixed,}\end{cases}
\]
where $R_\mathrm{sum}^\text{NCF}, R_\mathrm{sum}^\infty,$ and $C^*$ depend on $P$ and $G$ (as well as $\sigma^2(\cdot)$ for $C^*$ and $R_\mathrm{sum}^\text{NCF}$).
\end{corollary} 
\begin{IEEEproof}
Let $\beta_1, \ldots, \beta_{\rank(G)}$ be the non-zero eigenvalues of $GG^T.$ Then, we have
\begin{align*}
1-\frac{R_\mathrm{sum}^\text{NCF}(P)}{R_\mathrm{sum}^\infty(P)} & = 1 - \frac{\sum_{l = 1}^{\rank(G)}\log\left(1+\frac{P\beta_l}{1+\sigma^2(P)}\right)}{\sum_{l = 1}^{\rank(G)}\log(1+P\beta_l)}\\
& = \frac{\sum_{l = 1}^{\rank(G)}\log\left(\frac{1+P\beta_l}{1+\frac{P\beta_l}{1+\sigma^2(P)}}\right)}{\sum_{l = 1}^{\rank(G)}\log(1+P\beta_l)}\\
& = \frac{\sum_{l = 1}^{\rank(G)}\log\left(1+\frac{P\sigma^2(P)\beta_l}{1+\frac{P\beta_l}{1+\sigma^2(P)}}\right)}{\sum_{l = 1}^{\rank(G)}\log(1+P\beta_l)}\\
& \sim\frac{\rank(G)\log(1+\sigma^2(P))}{\rank(G)\log P}\\
& = \frac{\log(1+\sigma^2(P))}{\log P},\numberthis\label{ncfmimomult}
\end{align*}
and
\begin{align*}
\frac{C^*(P)}{R_\mathrm{sum}^\infty(P)} - 1 & = \frac{\sum_{l = 1}^{\rank(G)}\log\left(\frac{1+\frac{P\beta_l}{1+\sigma^2(P)}}{1+P\beta_l}\right) + L\log\left(1+\frac{1}{\sigma^2(P)}\right)}{\sum_{l = 1}^{\rank(G)}\log(1+P\beta_l)}.\numberthis\label{Cstarmimomult1}
\end{align*}
If $\sigma^2(P)\to\infty$ as $P\to\infty,$ \eqref{Cstarmimomult1} leads to
\[
\frac{C^*(P)}{R_\mathrm{sum}^\infty(P)} - 1 \sim\frac{\rank(G)\log\left(1/\sigma^2(P)\right) + \frac{L\log e}{\sigma^2(P)}}{\rank(G)\log P}\sim\frac{\log\left(1/\sigma^2(P)\right)}{\log P},\numberthis\label{Cstarmimomult2}
\] 
and if $\sigma^2(P)\to 0$ as $P\to\infty,$ \eqref{Cstarmimomult1} leads to
\begin{align*}
\frac{C^*(P)}{R_\mathrm{sum}^\infty(P)} - 1 &  = \frac{\sum_{l = 1}^{\rank(G)}\log\left(1-\frac{P\beta_l\sigma^2(P)}{(1+\sigma^2(P))(1+P\beta_l)}\right) +L\log\left(1+\frac{1}{\sigma^2(P)}\right)}{\sum_{l = 1}^{\rank(G)}\log(1+P\beta_l)}\\
& \sim\frac{\rank(G)\sigma^2(P) + L\log(1/\sigma^2(P))}{\rank(G)\log P}\\
& \sim\frac{L\log(1/\sigma^2(P))}{\rank(G)\log P}.\numberthis\label{Cstarmimomult3}
\end{align*}
Similarly, if $\sigma^2 >0$ is fixed, \eqref{Cstarmimomult1} leads to 
\[
\frac{C^*(P)}{R_\mathrm{sum}^\infty(P)} - 1 \sim\frac{\rank(G)\log\left(\frac{1}{1+\sigma^2}\right) + L\log\left(1+\frac{1}{\sigma^2}\right)}{\rank(G)\log P}.\numberthis\label{Cstarmimomult4}
\]
\end{IEEEproof}
For various choices of $\sigma^2(P),$ \eqref{ncfmimomult}, \eqref{Cstarmimomult2}, \eqref{Cstarmimomult3}, and \eqref{Cstarmimomult4} enable us to make several statements about the behaviors of the ratios $R_\mathrm{sum}^\text{NCF}/R_\mathrm{sum}^\infty$ and $C^*/R_\mathrm{sum}^\infty$ at high SNR. These are summarized in Table~\ref{table:multgap}.
\begin{table}[t]
    \setlength{\tabcolsep}{20pt}
    \renewcommand{\arraystretch}{2.5}
    \centering
    \begin{tabular}{  c  c c} \hline
	$\sigma^2(P)$ & $\displaystyle\frac{C^*(P)}{R_\mathrm{sum}^\infty(P)} - 1$ & $1 - \displaystyle\frac{R_\mathrm{sum}^\text{NCF}(P)}{R_\mathrm{sum}^\infty(P)}$ \\[5pt] \hline
       $1$ & $O\left(\displaystyle\frac{1}{\log P}\right)$ & $O\left(\displaystyle\frac{1}{\log P}\right)$\\[5pt]
	$\log P$ & $O\left(\displaystyle\frac{\log\log P}{\log P}\right)$ & $O\left(\displaystyle\frac{\log\log P}{\log P}\right)$\\[5pt]
	$(\log P)^{-\epsilon},$ $\epsilon\in(0, 1)$ & $O\left(\displaystyle\frac{\log\log P}{\log P}\right)$ & $O\left(\displaystyle\frac{1}{(\log P)^{1+\epsilon}}\right)$\\[5pt]\hline
       \end{tabular}
\caption{Multiplicative gap from fronthaul-unlimited uplink sum-capacity.}\label{table:multgap}
\end{table} 
%In particular, choosing $\sigma^2(P) = 1$ in \eqref{ncfmimomult} and \eqref{Cstarmimomult4} shows that using 
%\[
%C^*(P) = R_\mathrm{sum}^\text{un--lc}(P)\left(1 + O(1/\log P)\right),
%\]
%one can obtain a C-RAN uplink sum-rate 
%\[
%R_\mathrm{sum}^\text{NCF}(P) = R_\mathrm{sum}^\text{un--lc}(P)\left(1-O(1/\log P)\right).
%\]
%Similarly, choosing $\sigma^2(P) = \log P$ leads to
%\[
%R_\mathrm{sum}^\text{NCF}(P) = R_\mathrm{sum}^\text{un--lc}(P)\left(1-O(\log\log P/\log P)\right)
%\]
%with
%\[
%C^*(P) = R_\mathrm{sum}^\text{un--lc}(P)\left(1 + O(\log\log P/\log P)\right),
%\]
%and choosing $\sigma^2(P) = (\log P)^{-\epsilon}$ for some $\epsilon\in(0, 1)$ yields
%\[
%R_\mathrm{sum}^\text{NCF}(P) = R_\mathrm{sum}^\text{un--lc}(P)\left(1-O\left(1/(\log P)^{1+\epsilon}\right)\right)
%\]
%with
%\[
%C^*(P) = R_\mathrm{sum}^\text{un--lc}(P)\left(1 + O(\log\log P/\log P)\right).
%\]
As another result that demonstrates the asymptotically optimal fronthaul link capacity, we examine how $R_\mathrm{sum}^\text{NCF},$ $R_\mathrm{sum}^\infty,$ and $C^*$ scale with network size for specific network models, in the next section.

\subsection{Capacity Scaling}\label{upscale}
In this section, as opposed to keeping the network size fixed and varying the SNR and the channel coefficients, we let the network size grow and examine how the sum-rates and the fronthaul capacity requirement behave under certain network models. In Section~\ref{upscaletheo}, we consider a channel model, referred to as the \emph{rich scattering model}, where the entries of the channel gain matrix $G$ are  i.i.d.\@ $\N(0,1)$ random variables. In Section~\ref{upstochgeom}, we study a \emph{stochastic geometry} model through simulations, where users and relays are physically distributed over a two-dimensional area at random, and the channel coefficient between a particular user--relay pair is determined by the Euclidean distance between the two.

In contrast to the current treatment, large network size asymptotics for achievable symmetric rates was considered in \cite{sanderovichsomekhpoorshitztit09} for $L = K$ and equal fronthaul link capacities, under various localized interference models such as the Wyner model \cite{wynertit94} and the soft-handoff model \cite{somekhzeidelshitztit07}. Specifically, under these models, the limit of the symmetric achievable rate was computed as the network size grows to infinity. The high- and low-SNR behaviors of this limit were then studied. Fading was incorporated into the same localized interference model and similar studies were made on the limit of the ergodic capacity.     

\subsubsection{Rich scattering}\label{upscaletheo}
We consider a \textit{rich scattering} network model with slow fading, where the entries of the channel gain matrix $G$ are i.i.d.\@ random variables with variance $1$ and are assumed fixed for the duration of transmission. Moreover, the average power constraint $P$ is kept fixed. We recall the following.
\begin{lemma}[Telatar \cite{telatarett99}, Silverstein \cite{silversteinjma95}]\label{lemma:limitdistribution}
Let $W$ be an $m\times n$ random matrix with i.i.d.\@ entries $W_{ij},$ each of which has unit variance. Then, as $n\to\infty$ such that $n/m\to\rho\in[1,\infty),$ the limiting density of the empirical distribution of the eigenvalues of $WW^T/m$ is given, almost surely, by
\[
f_\Lambda(\lambda) = \frac{\sqrt{(\lambda - \alpha(\rho))(\beta(\rho) - \lambda)}}{2\pi\lambda}\mathbf{1}_{[\alpha(\rho), \beta(\rho)]},
\] 
where $\alpha(\rho):= (\sqrt{\rho}-1)^2$ and $\beta(\rho):= (\sqrt{\rho}+1)^2.$ On the other hand, if $n/m\to\infty,$ all the eigenvalues of $WW^T/n$ approach $1$ a.s.
\end{lemma}   

Using Lemma~\ref{lemma:limitdistribution}, we can establish the following result on the large network size behavior of $R_\mathrm{sum}^\text{NCF}(\sigma^2),$ $C^*(\sigma^2),$ and $R_\mathrm{sum}^\infty.$  
\begin{theorem}\label{theorem:pggt}
Let the entries of the $L\times K$ channel gain matrix $G$ be distributed as i.i.d.\@ random variables with unit variance, and let $\sigma^2 = \sigma^2(L, K) > 0.$ 
If $L\to\infty$ such that $L/K\to\rho \in(1, \infty]$ and $L/\sigma^2\to \infty,$ then 
\[
R_\mathrm{sum}^\infty\sim \frac{K}{2}\log L
\]
and
\[
R_\mathrm{sum}^\text{NCF} = C^*- \frac{L}{2}\log\left(1+\frac{1}{\sigma^2}\right)\sim \frac{K}{2}\log(L/\sigma^2),
\]
a.s.\@ in $G.$ Similarly, if $K\to\infty$ such that $L/K\to\rho \in[0, 1)$ and $K/\sigma^2\to \infty,$ then
\[
R_\mathrm{sum}^\infty\sim \frac{L}{2}\log K
\]
and
\[
R_\mathrm{sum}^\text{NCF} = C^*- \frac{L}{2}\log\left(1+\frac{1}{\sigma^2}\right)\sim \frac{L}{2}\log(K/\sigma^2),
\]
a.s.\@ in $G.$
\end{theorem}
Theorem~\ref{theorem:pggt} acts as a powerful tool to examine the large network size asymptotics for $R_\mathrm{sum}^\text{NCF}, R_\mathrm{sum}^\infty,$ and the fronthaul link capacity requirement $C^*.$ We consider various scaling regimes of $L$ and $K,$ namely, $K$ fixed and $L$ growing, $L = \gamma K$ with $\gamma \notin\{0, 1\},$ and $L = K^\gamma$ with $\gamma \notin\{0,1\}.$ For each case, we choose $\sigma^2 = \sigma^2(L, K)$ appropriately and use Theorem~\ref{theorem:pggt} to establish the scaling laws for this rich scattering model. The results are summarized in Table~\ref{table:uplinkscaling}. The detailed derivation of the tabulated results are deferred to Appendix~\ref{upproofs}.
\begin{table*}[h]
    \setlength{\tabcolsep}{20pt}
    \renewcommand{\arraystretch}{2.5}
    \centering
    \begin{tabular}{ c  c | c  c  c c} \hline
	\multicolumn{2}{c|}{$L$ vs.\@ $K$} & $\sigma^2$ & $C^*$ & $R_\mathrm{sum}^\text{NCF}$ & $R_\mathrm{sum}^\infty$ \\[5pt] \hline
       \multirow{2}{*}{$L = \gamma K$} & $(\gamma > 1)$ & 1 & $\displaystyle\frac{K}{2}\log L$ & $\displaystyle\frac{K}{2}\log L$ & $\displaystyle\frac{K}{2}\log L$\\
							&   $(\gamma < 1)$    &  1  & $\displaystyle\frac{L}{2}\log K$ & $\displaystyle\frac{L}{2}\log K$ & $\displaystyle\frac{L}{2}\log K$\\[20pt]
	 \multirow{3}{*}{$L =  K^\gamma$} & \multirow{2}{*}{$( \gamma > 1 )$} & $K^{\gamma - 1}$ & $\displaystyle\frac{K}{2}\log K$ & $\displaystyle\frac{K}{2}\log K$ & \multirow{2}{*}{$\displaystyle\frac{K}{2}\log L$}  \\[-10pt]
           & & $K^{\gamma - 1 - \delta}$ & $\displaystyle\frac{K^{1+\delta}}{2}\log e$ &  $(1+\delta)\displaystyle\frac{K}{2}\log K$ & \\
  & $(\gamma < 1)$ & $1$ & $\displaystyle\frac{L}{2}\log K$ & $\displaystyle\frac{L}{2}\log K$ & $\displaystyle\frac{L}{2}\log K$\\[20pt] 
         $K$ fixed & & $L^\epsilon$ & $\displaystyle\frac{L^{1-\epsilon}}{2}\log e$ &$(1-\epsilon)\displaystyle\frac{K}{2}\log L$ & $\displaystyle\frac{K}{2}\log L$\\[5pt]
$L$ fixed & & $1$ & $\displaystyle\frac{L}{2}\log K$ & $\displaystyle\frac{L}{2}\log K$ & $\displaystyle\frac{L}{2}\log K$ \\[5pt] \hline
       \end{tabular}
\caption{Sum-rate scaling for fronthaul-limited and fronthaul-unlimited uplink C-RAN; $\gamma > 0,$ $\gamma\ne 1,$ $0<\delta<\gamma - 1,$ $0<\epsilon<1.$}\label{table:uplinkscaling}
\end{table*}
\begin{remark}\label{remark:LlessthanK}
When $L$ is fixed and $K$ is growing, or $L = K^\gamma$ with $\gamma<1,$ the sum-rates scale sublinearly in $K$ and therefore, the per-user rate is asymptotically zero if one attempts to serve all users fairly. 
\end{remark} 
\begin{remark}\label{remark:scalinghistory}
We note that most of the classical scaling results in the literature (see, for example, \cite{telatarett99, ozarowshitzwyner94, verdutit02} and the references therein), consider ergodic capacities and their limits. In contrast, our results focus on \emph{global} interference network models with certain known statistical properties of the channel and make high-probability predictions on achievable rate regions. 
\end{remark} 
\begin{remark}\label{remark:lequalsk}
The theory developed here does not lead to the same a.s.\@ statements for the case $L = K\to\infty.$ As a workaround, for every $\epsilon > 0,$ however small, one can choose to only serve $(1-\epsilon)K$ of the users, thereby leading to a sum-rate scaling of 
\[
(1-\epsilon)K\log K/2
\]
for this case, in accordance with Table~\ref{table:uplinkscaling}. 
\end{remark} 
\begin{IEEEproof}[Proof of Theorem~\ref{theorem:pggt}]
We will prove the first part, i.e., the case when $L\to\infty.$ The second part will follow from this by exchanging the roles of $K$ and $L.$ Note that 
\[
\log\left|I+\frac{P}{\sigma^2+1}GG^T\right| = \log\left|I+\frac{P}{\sigma^2+1}G^TG\right|.
\]
In the regime considered, $L\ge K$ eventually, therefore we can use $G^T$ in place of $W$ in Lemma~\ref{lemma:limitdistribution}. Assume first that $\rho<\infty.$ We can conclude from Lemma~\ref{lemma:limitdistribution} that w.p.\@ $1,$ for every $\delta>0,$ there exists $L^*(\delta)$ such that for all $L\ge L^*(\delta),$ all eigenvlaues $\Lambda_1, \ldots, \Lambda_K$ of $G^TG/K$ lie in
\[
\Bigl[(\sqrt{\rho}-1)^2 - \delta, (\sqrt{\rho}+1)^2 + \delta\Bigr].
\]
Therefore, w.p.\@ $1,$ for every $L\ge L^*(\delta),$ we have
\begin{align*}
&\log\left|I + \frac{P}{\sigma^2+1}GG^T\right| \\
& = \sum_{k=1}^K\log\left(1 + \frac{PK}{\sigma^2+1}\Lambda_k\right)\\
& \in\left[K\log\left(1+\frac{PK}{\sigma^2+1}((\sqrt{\rho}-1)^2 - \delta)\right), K\log\left(1+\frac{PK}{\sigma^2+1}((\sqrt{\rho}+1)^2 + \delta)\right)\right].\numberthis\label{eq:asinterval}
\end{align*}
Further, we have
\begin{align*}
& \lim_{L\to\infty}\frac{K\log\left(1+\frac{PK}{\sigma^2+1}((\sqrt{\rho}\pm 1)^2 \pm \delta)\right)}{K\log\left(\frac{PL}{\sigma^2}\right)}\\
& = \lim_{L\to\infty}\frac{\log\left(1+\frac{PL}{\sigma^2+1}\cdot\frac{((\sqrt{\rho}\pm 1)^2 \pm \delta)}{L/K}\right)}{\log\left(\frac{PL}{\sigma^2+1}\right)}\\
& = \lim_{L\to\infty}\Bigg[\frac{\log\left(1+\frac{PL}{\sigma^2+1}\cdot\frac{((\sqrt{\rho}\pm 1)^2 \pm \delta)}{L/K}\right)}{\log\left(1 + \frac{PL}{\sigma^2+1}\cdot\frac{((\sqrt{\rho}\pm 1)^2 \pm \delta)}{\rho}\right)}\times\frac{\log\left(1 + \frac{PL}{\sigma^2+1}\cdot\frac{((\sqrt{\rho}\pm 1)^2 \pm \delta)}{\rho}\right)}{\log\left(\frac{PL}{\sigma^2+1}\cdot\frac{((\sqrt{\rho}\pm 1)^2 \pm \delta)}{\rho}\right)}\\
& \quad\quad\quad\quad\quad\quad\times\frac{\log\left(\frac{PL}{\sigma^2+1}\cdot\frac{((\sqrt{\rho}\pm 1)^2 \pm \delta)}{\rho}\right)}{\log\left(\frac{PL}{\sigma^2}\right)}\Bigg]\\
& = 1,
\end{align*}
since each factor approaches $1.$ Therefore, from \eqref{eq:asinterval}, we obtain the limiting behavior
\[
\log\left|I+\frac{P}{\sigma^2+1}GG^T\right|\sim K\log\left(\frac{PL}{\sigma^2}\right)\sim K\log(L/\sigma^2)\quad \text{w.p. }1.
\]
For the case $L/K\to\infty,$ since all $K$ eigenvalues of $G^TG/L$ approach $1,$ we have
\[
\log\left|I + \frac{P}{\sigma^2+1}GG^T\right|\sim K\log\left(\frac{PL}{\sigma^2}\right)\sim K\log(L/\sigma^2)\quad \text{w.p. }1.
\]
A similar line of reasoning with $P/(\sigma^2+1)$ replaced by $P$ yields the result for $R_\mathrm{sum}^\infty.$
\end{IEEEproof}

\subsubsection{Stochastic geometry}\label{upstochgeom}
We now consider an alternative network model based on stochastic geometry \cite{baccelligamaltsetit11}. In this model, users and relays are distributed over a $100\mathrm{m}\times 100\mathrm{m}$ area according to independent Poisson point processes with intensities $\lambda_u$ and $\lambda_r$ (per $10^4$ square meters) respectively. All channel gains are assumed to be real and unchanged for the duration of transmission. 

As an initial simple model, the gain $G_{lk}$ from sender $k$ to relay $l,$ separated by Euclidean distance $r_{lk},$ is modeled by $\max\{r_0, r_{lk}\}^{-\beta},$ where $\beta$ is the \emph{path loss exponent} and $r_0$ (set to 1 meter) is a minimum link distance to prohibit the singularity of the path loss for $r_{lk}\rightarrow0$. In contrast to the rich scattering model, the channel randomness now comes exclusively from the placement of user and relay nodes, and once the nodes are fixed, the channel coefficients become deterministic. Therefore, this simple model approximates line-of-sight (LOS) propagation with no multipath component. 

Following \cite{lu:hal-01269564}, we also study a more practical channel model based on stochastic geometry, where multipath effects such as blockage, shadowing, and fast fading are considered. More specifically, the multipath channel gain is given by
\[
G_{lk} = \begin{cases} G_{lk}^{(\mathrm{LOS})} & \text{w.p. }p_\mathrm{LOS}\left(r_{lk}\right),\\
 G_{lk}^{(\mathrm{NLOS})} & \text{w.p. }1-p_\mathrm{LOS}\left(r_{lk}\right),\end{cases}
\] 
where
\[
G_{lk}^{(\mathrm{LOS})} = \frac{A_{lk}^{(\mathrm{LOS})}\Theta_{lk}^{(\mathrm{LOS})}}{\kappa\left(\max\{r_0, r_{lk}\}\right)^{\beta^{(\mathrm{LOS})}}}
\]
and
\[
G_{lk}^{(\mathrm{NLOS})} = \frac{A_{lk}^{(\mathrm{NLOS})}\Theta_{lk}^{(\mathrm{NLOS})}}{\kappa\left(\max\{r_0, r_{lk}\}\right)^{\beta^{(\mathrm{NLOS})}}}.
\]
Here, NLOS stands for non-LOS. The random variable $A_{lk}$ represents the fast fading component for modeling small-scale fluctuations in the envelope of the links in LOS and in NLOS. $A_{lk}^{(\mathrm{LOS})}$ and $A_{lk}^{(\mathrm{NLOS})}$ follow a \emph{Nakagami-}$m$ distribution with $m=2$ and scale parameter $\Omega=1,$ and a \emph{Rayleigh} distribution with scale parameter $\Omega=1$, respectively. The factor $\Theta_{lk}$ models the shadowing effect due to changes in the surrounding environment. We consider a typical log-normal shadowing and set $\Theta_{lk}^{(\mathrm{LOS})}$ and $\Theta_{lk}^{(\mathrm{NLOS})}$ as log-normal random variables with means and standard deviations as specified in \cite{lu:hal-01269564}. We also assume that $A_{lk}$ and $\Theta_{lk}$ are independently distributed. The parameter $\kappa$ is the free-space path loss at a distance of 1 meter from the sender at the center frequency $f_c$ (which is set to 2.1 GHz here), and $\beta^{(\mathrm{LOS})}$ and $\beta^{(\mathrm{NLOS})}$ denote the path loss exponent for LOS and NLOS scenarios, respectively. We take $\beta^{(\mathrm{LOS})}=2.5$ and $\beta^{(\mathrm{NLOS})}=3.5.$ Finally, $p_\mathrm{LOS}\left(r_{lk}\right)$ represents the probability that the link is in LOS and is modeled according to the 3GPP urban micro (UMi) channel model \cite{3gpp.36.873} as
\[
p_\mathrm{LOS}\left(r_{lk}\right) = \min\{18/r_{lk},1\}\left(1 - e^{-r_{lk}/36}\right)+e^{-r_{lk}/36},
\]
where $r_{lk}$ is measured in meters.

\begin{figure}[h]
\begin{subfigure}[t]{0.33\textwidth}
\center
\includegraphics[scale=0.33]{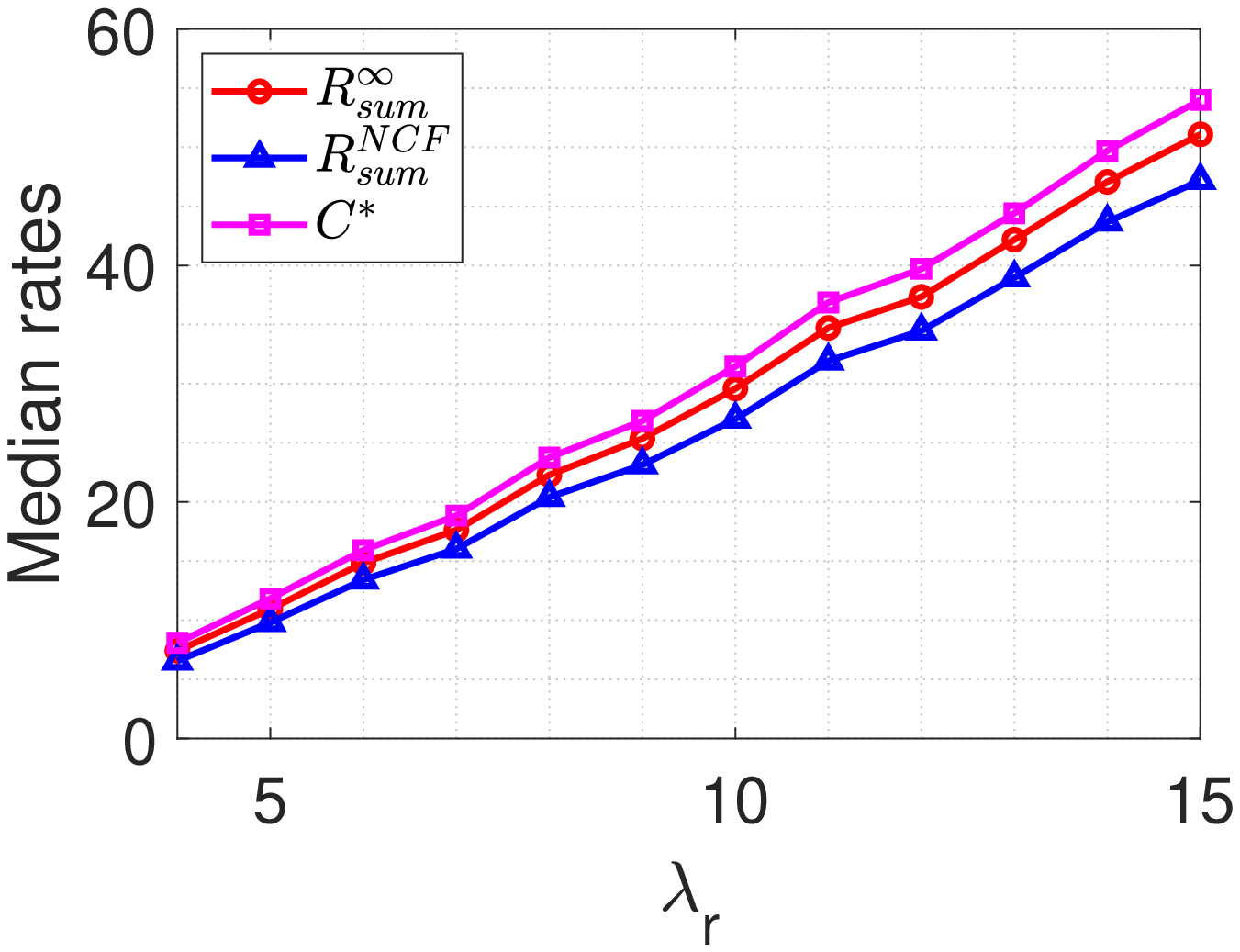}
\caption{$\beta = 2.5,$ $\lambda_r = 2\lambda_u.$}
\label{fig:plot2a}
\end{subfigure}%
\begin{subfigure}[t]{0.33\textwidth}
\center
\includegraphics[scale=0.33]{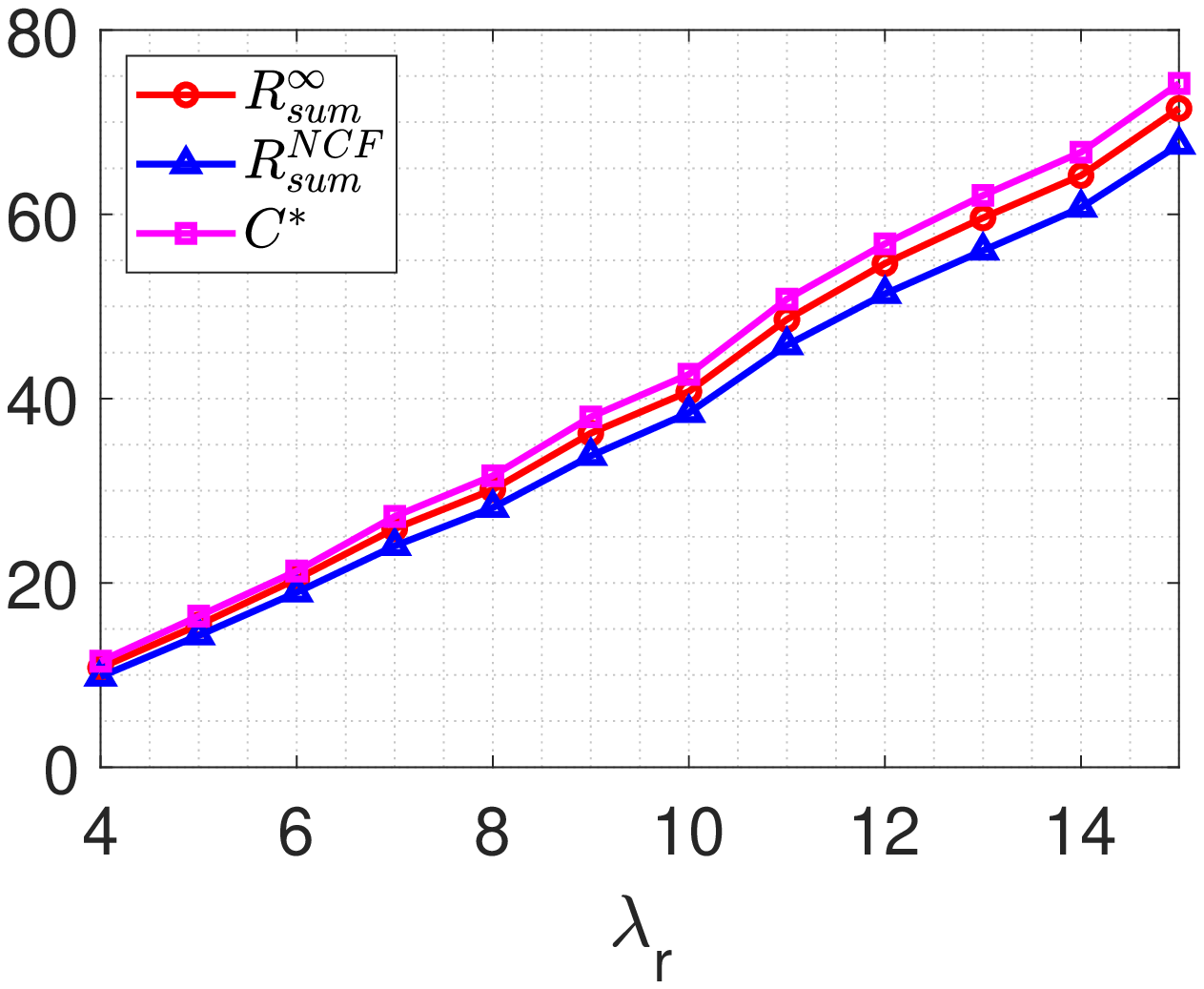}
\caption{$\beta = 3.5,$ $\lambda_r = 2\lambda_u.$}
\label{fig:plot2b}
\end{subfigure}
\begin{subfigure}[t]{0.33\textwidth}
\center
\includegraphics[scale=0.33]{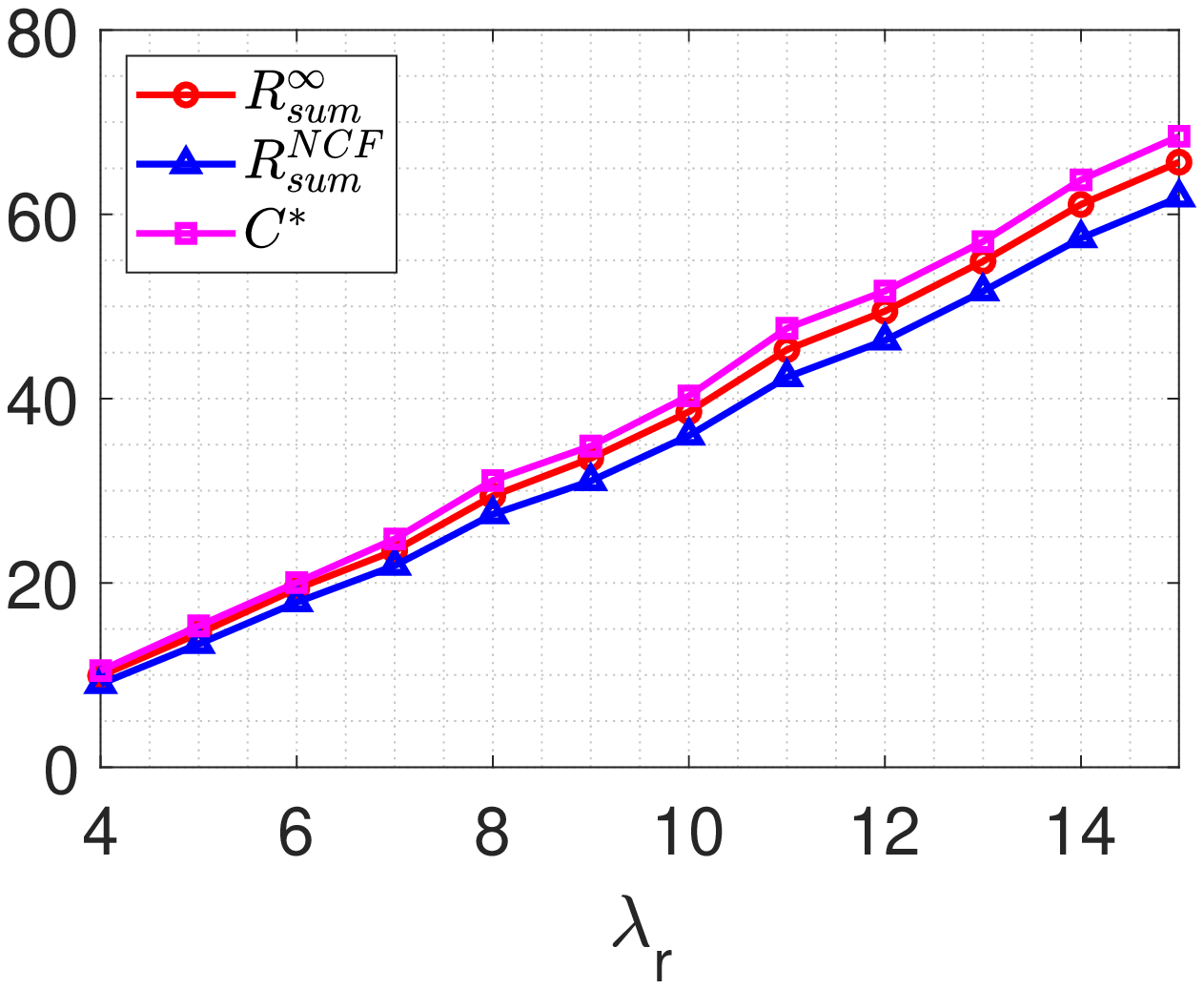}
\caption{Multipath, $\lambda_r = 2\lambda_u.$}
\label{fig:plot2c}
\end{subfigure}
\vskip\baselineskip
\begin{subfigure}[t]{0.33\textwidth}
\center
\includegraphics[scale=0.33]{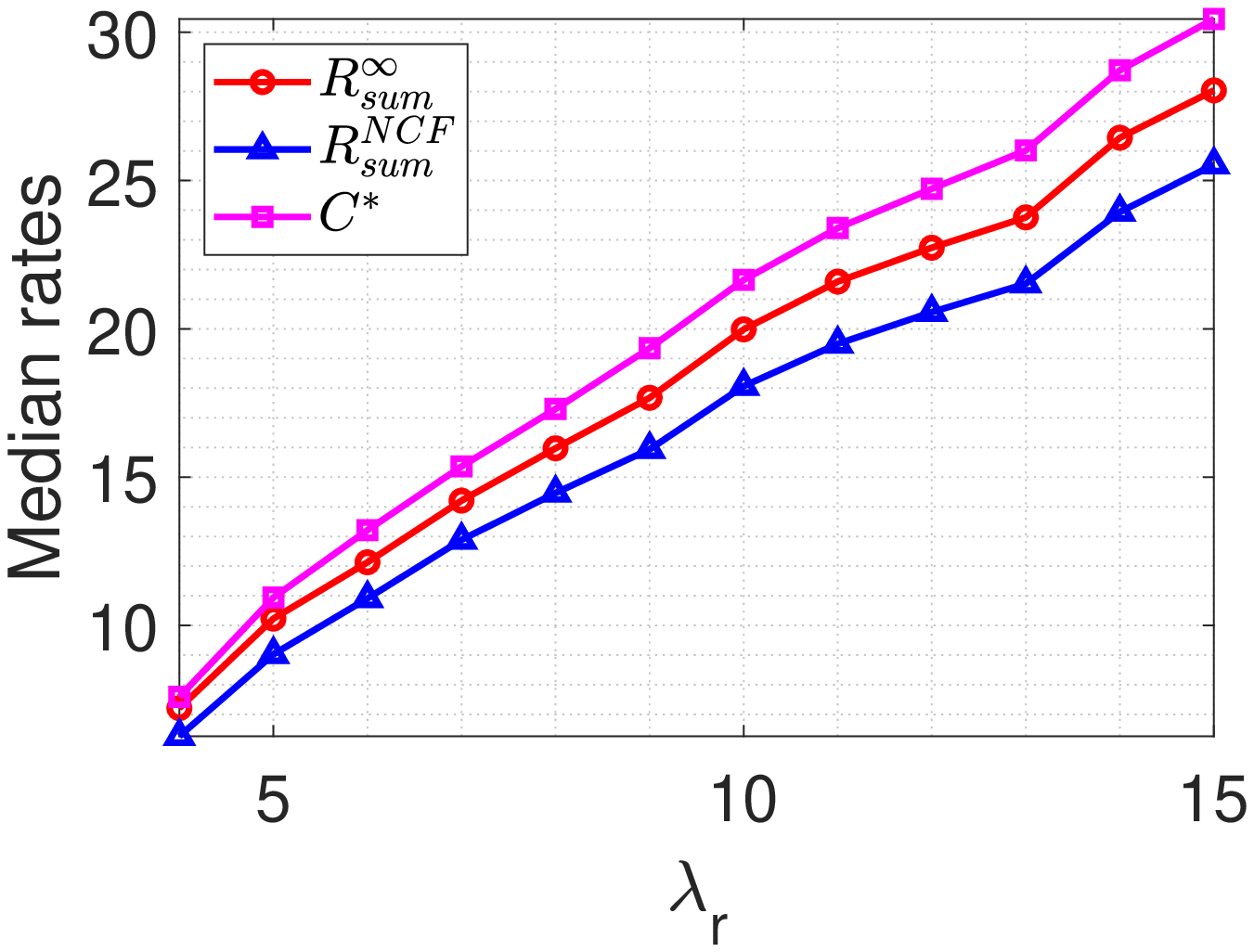}
\caption{$\beta = 2.5,$ $\lambda_r = \lambda_u^2.$}
\label{fig:plot2d}
\end{subfigure}%
\begin{subfigure}[t]{0.33\textwidth}
\center
\includegraphics[scale=0.33]{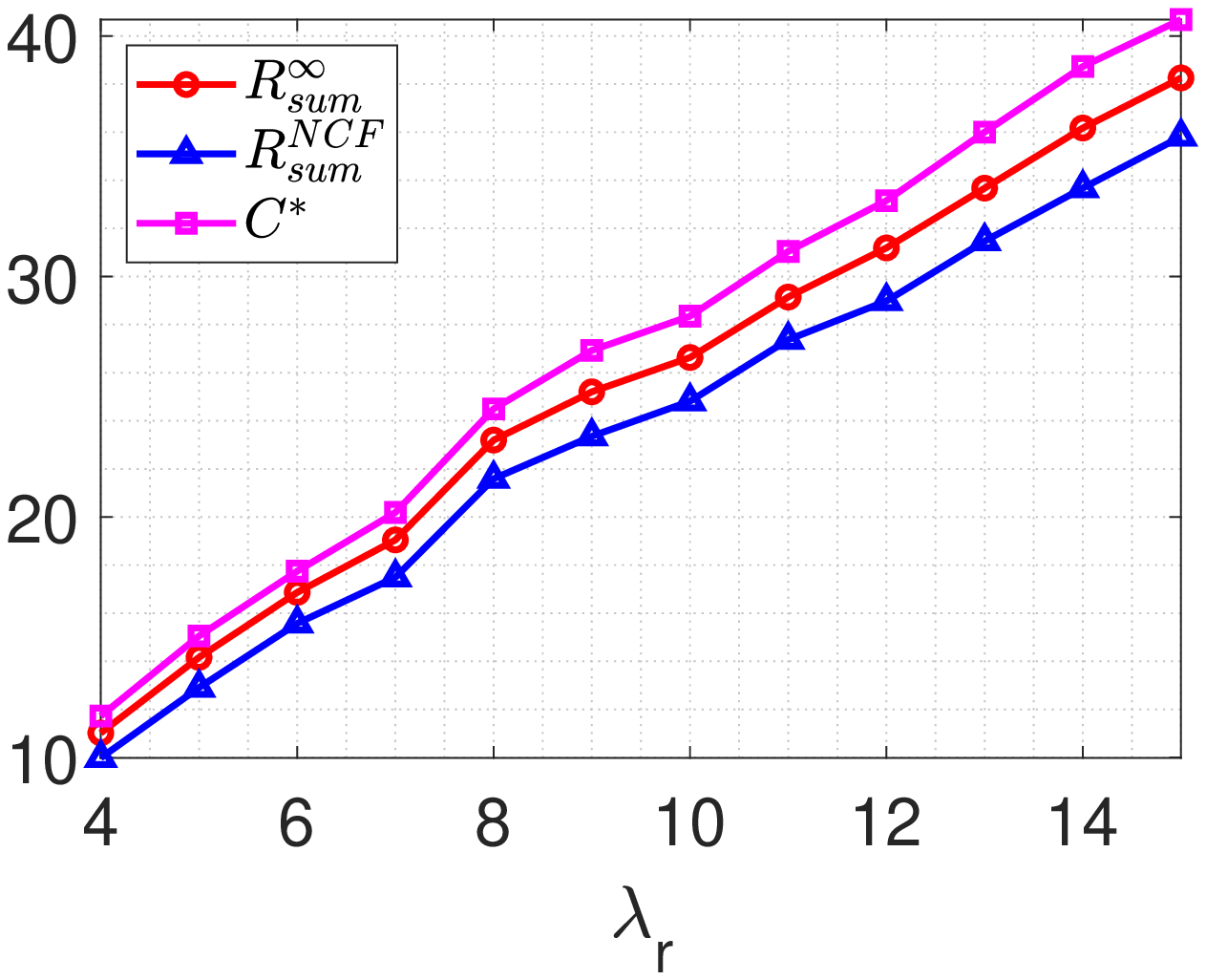}
\caption{$\beta = 3.5,$ $\lambda_r = \lambda_u^2.$}
\label{fig:plot2e}
\end{subfigure}
\begin{subfigure}[t]{0.33\textwidth}
\center
\includegraphics[scale=0.33]{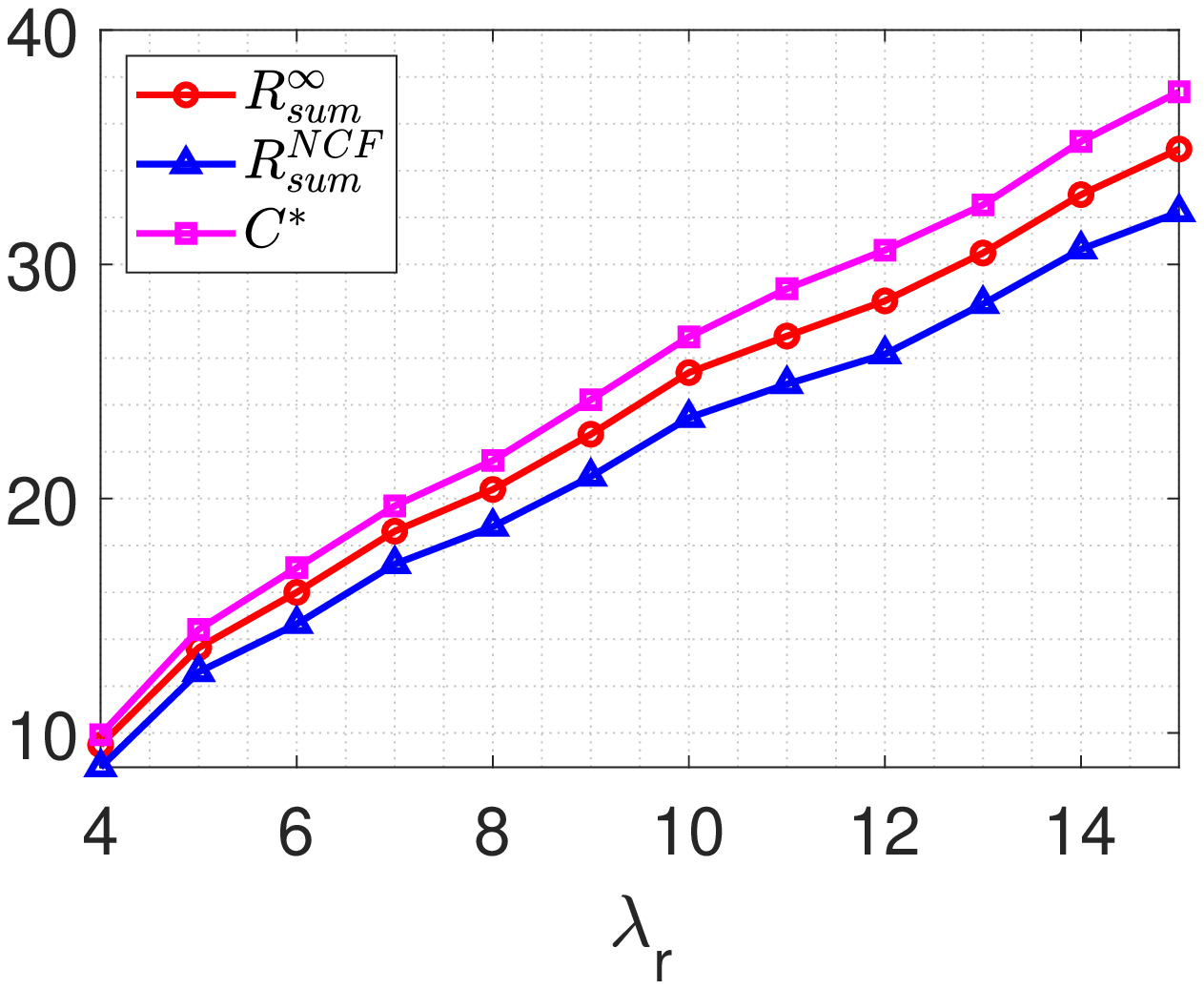}
\caption{Multipath, $\lambda_r = \lambda_u^2.$}
\label{fig:plot2f}
\end{subfigure}
\vskip\baselineskip
\begin{subfigure}[t]{0.33\textwidth}
\center
\includegraphics[scale=0.33]{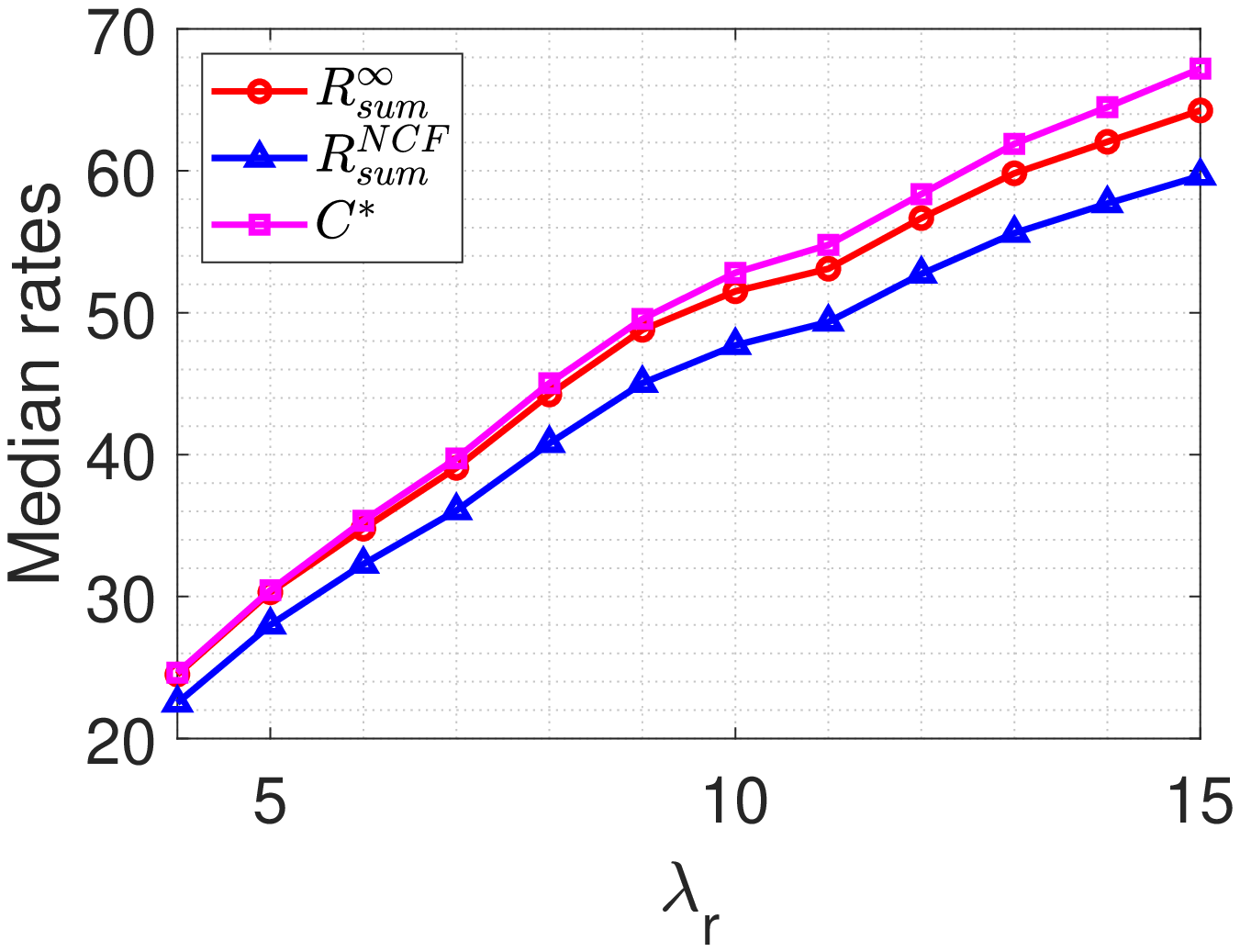}
\caption{$\beta = 2.5,$ $\lambda_u = 10.$}
\label{fig:plot2g}
\end{subfigure}%
\begin{subfigure}[t]{0.33\textwidth}
\center
\includegraphics[scale=0.33]{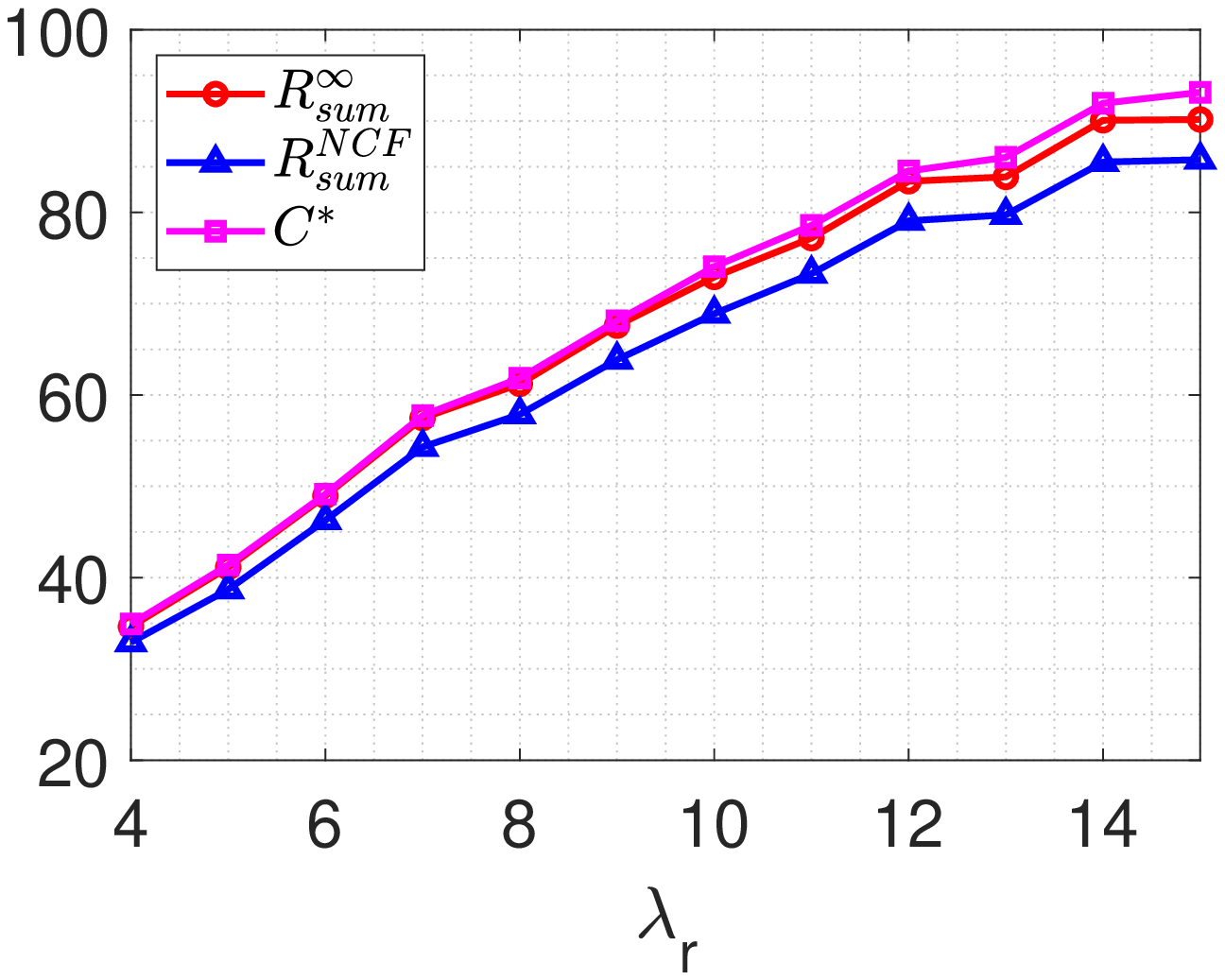}
\caption{$\beta = 3.5,$ $\lambda_u = 10.$}
\label{fig:plot2h}
\end{subfigure}
\begin{subfigure}[t]{0.33\textwidth}
\center
\includegraphics[scale=0.33]{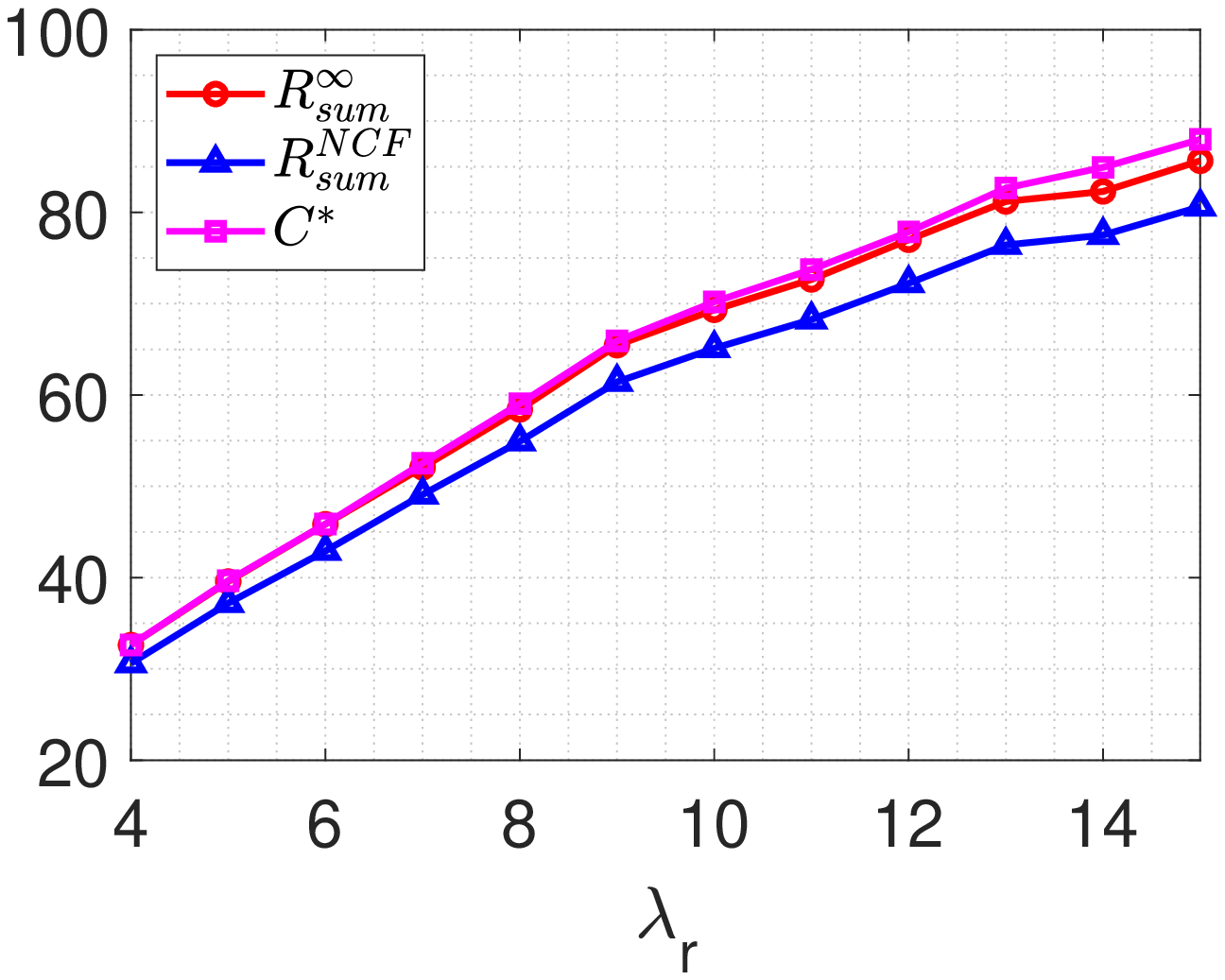}
\caption{Multipath, $\lambda_u = 10$.}
\label{fig:plot2i}
\end{subfigure}
\caption{Uplink capacity scaling under stochastic geometry.}\label{fig:plot2}
\end{figure}

For simulating the large network asymptotics for all the aforementioned channel models, we examine the cases $\lambda_r = 2\lambda_u,$ $\lambda_r = \lambda_u^2,$ and $\lambda_u$ fixed. The corresponding median sum-rates for fronthaul-limited and fronthaul-unlimited C-RAN uplink, as well as the corresponding $C^*$ required, are plotted as functions of $\lambda_u$ in Fig.~\ref{fig:plot2} for different values of $\beta.$ The median values are taken over $1000$ runs of the simulations. For each simulation run, $\sigma^2$ is chosen so as to (numerically) minimize 
\[
\max\left\{C^*(\sigma^2) - R_\mathrm{sum}^\infty, R_\mathrm{sum}^\infty - R_\mathrm{sum}^\text{NCF}(\sigma^2)\right\}.
\] 
From the plots, we observe that $R_\mathrm{sum}^\text{NCF}$ scales in a similar fashion as $R_\mathrm{sum}^\infty$, remaining only slightly lower, provided we have a slightly larger amount to spend on the fronthaul. Moreover, for $\lambda_r = 2\lambda_u$ as well as for $\lambda_r = \lambda_u^2,$ the sum-rates show an approximately linear scaling with $\lambda_r$ (and hence with $L$), unlike the $K\log L$ scaling observed for the rich scattering case. This loss seems to be caused by the dependence among the channel coefficients, which are still identically distributed but not independent of each other.

%\begin{figure}[h]
%\begin{subfigure}[t]{0.5\textwidth}
%\center
%\includegraphics[scale=0.32]{Figures/beta25_Llinear_uplink.eps}
%\caption{$\beta = 2.5,$ $\lambda_r = 2\lambda_u.$}
%\label{fig:plot1a}
%\end{subfigure}%
%\begin{subfigure}[t]{0.5\textwidth}
%\center
%\includegraphics[scale=0.32]{Figures/beta3_Llinear_uplink.eps}
%\caption{$\beta = 3,$ $\lambda_r = 2\lambda_u.$}
%\label{fig:plot1b}
%\end{subfigure}
%\vskip\baselineskip
%\begin{subfigure}[t]{0.5\textwidth}
%\center
%\includegraphics[scale=0.32]{Figures/beta25_Lexp_uplink.eps}
%\caption{$\beta = 2.5,$ $\lambda_r = \lambda_u^2.$}
%\label{fig:plot1c}
%\end{subfigure}%
%\begin{subfigure}[t]{0.5\textwidth}
%\center
%\includegraphics[scale=0.32]{Figures/beta3_Lexp_uplink.eps}
%\caption{$\beta = 3,$ $\lambda_r = \lambda_u^2.$}
%\label{fig:plot1d}
%\end{subfigure}
%\vskip\baselineskip
%\begin{subfigure}[t]{0.5\textwidth}
%\center
%\includegraphics[scale=0.32]{Figures/beta25_Kfixed_uplink.eps}
%\caption{$\beta = 2.5,$ $\lambda_u$ fixed.}
%\label{fig:plot1e}
%\end{subfigure}%
%\begin{subfigure}[t]{0.5\textwidth}
%\center
%\includegraphics[scale=0.32]{Figures/beta3_Kfixed_uplink.eps}
%\caption{$\beta = 3,$ $\lambda_u$ fixed.}
%\label{fig:plot1f}
%\end{subfigure}
%\caption{Uplink capacity scaling under stochastic geometry.}\label{fig:plot1}
%\end{figure}

%-----------------------------------------------------------------

\section{Downlink Communication}\label{download}
\subsection{General Model}\label{downgen}
\begin{figure}[h]
\center
\includegraphics[scale=1.0]{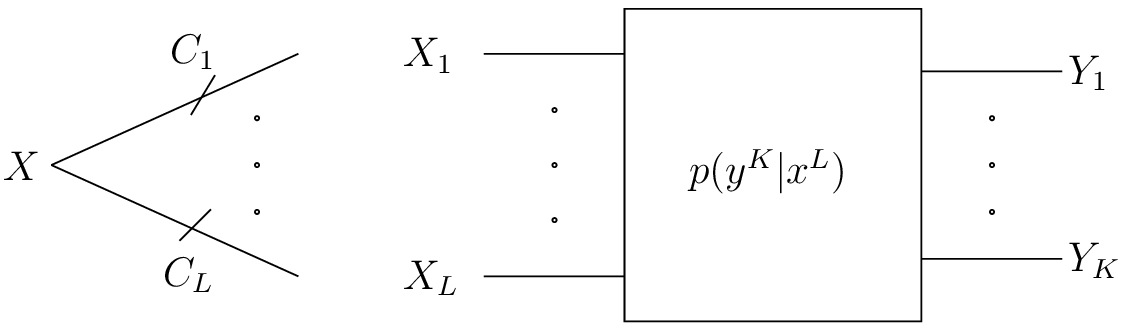}
\caption{Downlink network model.}\label{downfig}
\end{figure}

Similar to the uplink case, we model the downlink C-RAN as a two-hop relay network in Fig.~\ref{downfig}, where the first hop (central processor to radio heads) consists of orthogonal noiseless links of capacities $C_1,\ldots,C_L$ bits per real dimension and the second hop (radio heads to user devices or receivers) is modeled as a discrete memoryless network $p(y^K\cond x^L)$.

A $(2^{nR_1},\ldots,2^{nR_K},n)$ code for this network consists of $K$ message sets $[2^{nR_1}],\ldots,[2^{nR_k}]$; an encoder $w^L(m_1,\ldots,m_K)\in\prod_{l=1}^L[2^{nC_l}]$; relay encoders $x_l^n(w_l)$, $l\in[L]$; and decoders $\hat{m}_k(y_k^n)\in[2^{nR_k}]$, $k\in[K]$. The average probability of error, achievability of a rate tuple, and the capacity region are defined similar to Section~\ref{upgen}. 

One can specialize the distributed decode--forward coding scheme to this network model. To transmit messages $(M_1, \ldots, M_K),$ the sender comes up with precoding sequences $u_1^n(M_1, T_1), \ldots, u_K^n(M_K, T_K)$ and \textit{intended} transmit codewords $x_1^n(W_1), \ldots, x_L^n(W_L),$ and sends $W_1, \ldots, W_L$ to the relays. %Here, $T_k\in[2^{n\Rt_k}]$ are auxiliary indices. 
For randomly generated codebooks, the encoding succeeds w.h.p.\@ if
\[
\sum_{l\in \Sc_1^c}C_l+\sum_{k\in \Sc_2^c}\Rt_k  > I^*(X(\Sc_1^c),U(\Sc_2^c))\numberthis\label{dlddfinformal1}
\]
for every $\Sc_1\subseteq[L]$ and $\Sc_2\subseteq[K].$ Upon receiving index $W_l,$ relay $l$ transmits the codeword $x_l^n(W_l).$ Receiver $k$ recovers $M_k$ based on the received sequence $Y_k^n.$ The decoding at receiver $k$ succeeds w.h.p.\@ if 
\[
R_k + \Rt_k < I(U_k; Y_k).\numberthis\label{dlddfinformal2}
\] 
Combining \eqref{dlddfinformal1} and \eqref{dlddfinformal2} to eliminate the auxiliary rates $\Rt_1, \ldots, \Rt_K$ leads to the following inner bound on the capacity region of this network. (See Appendix~\ref{downproofs} for a complete proof.)

\begin{prop}[Distributed decode--forward inner bound for the downlink C-RAN]\label{propo3}  
A rate tuple $(R_1,\ldots,R_K)$ is achievable for the downlink C-RAN if
\[ 
\sum_{k\in \Sc_2^c}R_k <  I(X(\Sc_1);U(\Sc_2^c)|X(\Sc_1^c))+\sum_{l\in \Sc_1^c}C_l-\sum_{k\in \Sc_2^c}I(U_k;X^L|Y_k) - I^*(U(\Sc_2^c)\cond X^L)-I^*(X(\Sc_1^c))\numberthis\label{eq:downlinkgeneral}
\]
for all $\Sc_1\subseteq[L]$ and $\Sc_2\subseteq[K]$ for some pmf $p(x^L, u^K),$ such that
\[
I^*(X(\Sc_1))\le\sum_{l\in \Sc_1}C_l
\] 
for all $\Sc_1\subseteq[L].$
\end{prop}    
\begin{remark}\label{inffronthaullimitdownlink}
As the fronthaul capacities $C_1, \ldots, C_L$ tend to infinity, this downlink C-RAN channel model becomes identical to the ``fronthaul-unlimited'' downlink channel from a single sender with $L$ transmit antennas to $K$ receivers, i.e., the broadcast channel $p(y^K\cond x^L)$ with one sender $X^L$ and $K$ receivers $Y_1, \ldots, Y_K.$ In this regime, the distributed decode--forward inner bound converges to the Marton coding inner bound with no common messages \cite{nit}, characterized by rate tuples $(R_1, \ldots, R_K)$ satisfying
\[
\sum_{k\in \Sc_2}R_k\le \sum_{k\in \Sc_2}I(U_k; Y_k)-I^*(U(\Sc_2))
\]
for every $\Sc_2\subseteq[K]$ for some pmf $p(u^K)$ and some function $x^L(u^K).$ 
\end{remark} 

Specializing the cutset bound \cite{gamalcutset} to the downlink C-RAN model leads to the following.
\begin{prop}[Cutset outer bound for the downlink C-RAN]\label{downgenouter}
If a rate tuple $(R_1,\ldots,R_K)$ is achievable for the downlink C-RAN, then
\begin{align*}
\sum_{k\in \Sc_2^c}R_k & \le I(X(\Sc_1);Y(\Sc_2^c)|X(\Sc_1^c))+\sum_{l\in \Sc_1^c}C_l\numberthis\label{cutsetdown}
\end{align*}
for all $\Sc_1\subseteq[L]$ and $\Sc_2\subseteq[K]$ for some pmf $p(x^L)$. 
\end{prop}
We provide a proof of Proposition~\ref{downgenouter} in Appendix~\ref{downproofs}.

\subsection{Gaussian Model}\label{downgauss}      
We now assume that the second hop of the network is Gaussian, i.e., $Y^K=HX^L+Z^K$, where $H\in\mathbbm{R}^{K\times L}$ is a channel gain matrix and $Z^K$ is a vector of i.i.d.\@ $\N(0,1)$ noise components. We also impose the average power constraint $P$ on each relay. For this Gaussian network model, the distributed decode--forward inner bound in Proposition~\ref{propo3} can be specialized to establish the achievability of all rate tuples $(R_1,\ldots,R_K)$ such that   
\[
\sum_{k\in \Sc_2^c}R_k \le \frac{1}{2}\log\Big|\frac{P}{\sigma^2}H_{\Sc_2^c,\Sc_1}H_{\Sc_2^c,\Sc_1}^T+I\Big|+\sum_{l\in \Sc_1^c}C_l -\frac{|\Sc_2^c|}{2}\log\left(1+\frac{1}{\sigma^2}\right) =: F_{\mathrm{in}}(\Sc_1,\Sc_2)\numberthis\label{downinngaussinformal}
\]
for all $\Sc_1\subseteq[L]$ and $\Sc_2\subseteq[K]$ for some $\sigma^2 > 0.$  This follows by setting $X^L$ to be a vector of i.i.d.\@ $\N(0,P)$ random variables and defining $U^K = GX^L+\hat{Z}^K$, where $\hat{Z}^K\sim\N(0,\sigma^2I)$ is independent of $Z^K$. For every $\sigma^2 > 0,$ we denote the set of rate tuples $(R_1, \ldots, R_K)$ satisfying \eqref{downinngaussinformal} by $\Rr_{\mathrm{down}}^{\mathrm{DDF}}(\sigma^2).$ We also denote the achievable sum-rate for each $\sigma^2>0$ by
\begin{align*}
R_\mathrm{sum}^\text{DDF}(\sigma^2) & :=\sup_{(R_1, \ldots, R_K)}\{R_1 + \cdots + R_K: (R_1, \ldots, R_K)\in\Rr_\mathrm{down}^\text{DDF}(\sigma^2)\}\numberthis\label{eq:downlinksumratedefinition}\\
& = \min_{\Sc_1\subseteq[L]}\Bigl(\frac{1}{2}\log\Big\vert\frac{P}{\sigma^2}H_{[K],\Sc_1}H_{[K],\Sc_1}^T+I\Big\vert +\sum_{l\in \Sc_1^c}C_l\Bigr)-\frac{K}{2}\log\bigl(1+\frac{1}{\sigma^2}\bigr).\numberthis\label{downinngauss_sumrate_prev}
\end{align*} 

Similar to Section~\ref{upgauss}, the cutset bound in Proposition~\ref{downgenouter} can be specialized to the rate region characterized by
\begin{align*}
\sum_{k\in \Sc_2^c}R_k & \le \frac{1}{2}\log\Big\vert H_{\Sc_2^c,\Sc_1}\Gamma_{\Sc_1|\Sc_1^c}H_{\Sc_2^c,\Sc_1}^T+I\Big\vert+\sum_{l\in \Sc_1^c}C_l\\
& =: F_\mathrm{out}(\Sc_1,\Sc_2)\numberthis\label{cutsetdowngauss}
\end{align*}
for all $\Sc_1\subseteq[L]$ and $\Sc_2\subseteq[K]$ for some covariance matrix $\Gamma\succeq 0$ satisfying $\Gamma_{ll}\le P$ for all $l\in[L].$ Here, $\Gamma_{\Sc_1|\Sc_1^c}$ is the conditional covariance matrix given by 
\[
\Gamma_{\Sc_1|\Sc_1^c} = \Gamma_{\Sc_1, \Sc_1} - \Gamma_{\Sc_1, \Sc_1^c}\Gamma_{\Sc_1^c, \Sc_1^c}^{-1}\Gamma_{\Sc_1^c, \Sc_1}.
\]
For each covariance matrix $\Gamma,$ we denote the set of rate tuples $(R_1, \ldots, R_K)$ satisfying \eqref{cutsetdowngauss} by $\Rr_{\mathrm{down}}^\text{CS}(\Gamma).$ 
We denote the sum-rate upper bound by
\[
R_\mathrm{sum}^\text{CS} := \sup_{(R_1, \ldots, R_K), \Gamma}\{R_1 + \cdots + R_K: (R_1, \ldots, R_K)\in\Rr_\mathrm{down}^\text{CS}(\Gamma)\text{ for some }\Gamma\}.\numberthis\label{eq:uplinksumrateouterdefinition}
\] 
The achievable per-user rate gap $\Delta,$ as well as the sum-rate gap $\Delta_\mathrm{sum}$ between the cutset bound and the distributed decode--forward inner bound \eqref{downinngaussinformal}, can be bounded as in the following result, whose proof is deferred to Appendix~\ref{downproofs}. 
\begin{theorem}\label{downgaptheo1}
For every $H\in\Real^{K\times L}$ and $P\in\Real^+,$ if a rate tuple $(R_1,\ldots,R_K)$ is in the cutset bound \eqref{cutsetdowngauss}, then the rate tuple $((R_1-\Delta)^+,\ldots,(R_K-\Delta)^+)$ is achievable, where 
\[
\Delta\le \frac{1}{2}\log(eKL) \approx \frac{1}{2}\log(KL) + 0.722.
\]
Moreover, the sum-rate gap between the cutset bound and the distributed decode--forward inner bound is upper-bounded as 
\[
\Delta_\mathrm{sum}:= R_\mathrm{sum}^\text{CS} - \sup_{\sigma^2>0}R_\mathrm{sum}^\text{DDF}(\sigma^2) \le\frac{K}{2} + \frac{\min\{L, K\}}{2}\log L
\]
irrespective of $P$ and $H.$
\end{theorem}
%The proof of Theorem~\ref{downgaptheo1} is deferred to Appendix~\ref{downproofs}.

\subsection{Comparisons with Fronthaul-Unlimited Downlink}\label{downdiscuss}      

Similar to Section~\ref{updiscuss}, we can use Edmonds's polymatroid intersection theorem to quantify the total fronthaul $C_{\sum}:=C_1+\cdots+C_L$ required to approximate the fronthaul-unlimited downlink sum-capacity.
\begin{theorem}\label{theorem:Cstardownlink} 
If 
\[
C_{\sum} \ge \frac{1}{2}\log\left|\frac{P}{\sigma^2}HH^T + I\right| =: C^*(\sigma^2) 
\]
for some $\sigma^2>0,$ then there exist $C_1, C_2, \ldots, C_L\ge 0$ with $\sum_{l\in[L]}C_l = C_{\sum}$ at which distributed decode--forward can achieve a sum-rate
\[
R_\mathrm{sum}^\text{DDF}(\sigma^2) = \frac{1}{2}\log\left|\frac{P}{\sigma^2}HH^T + I\right|-\frac{K}{2}\log\left(1 + \frac{1}{\sigma^2}\right).
\] 
Conversely, to achieve a sum-rate of $(1/2)\log|I+PHH^T|,$ we must have a total fronthaul capacity 
\[
C_{\sum}\ge\frac{1}{2}\log|I+PHH^T|.
\] 
\end{theorem}
\begin{IEEEproof}
We assume that 
\[
\frac{1}{2}\log\left|\frac{P}{\sigma^2}HH^T + I\right| \ge \frac{K}{2}\log\left(1 + \frac{1}{\sigma^2}\right),\numberthis\label{firstconddownlinkpolymatroid}
\]
since a negative sum-rate has no physical meaning. Define $r_k:= R_k + (1/2)\log(1 + 1/\sigma^2),$ $k\in[K].$ We will work with the tuple $(r_1, \ldots, r_K)$ instead of $(R_1, \ldots, R_K).$ The maximum sum $r_1 + \cdots+r_K$ corresponding to $\Rr_{\mathrm{down}}^{\mathrm{DDF}}(\sigma^2)$ can be written as
\begin{align*}
r_\mathrm{max} & = \min_{\Sc_1\subseteq[L]}\Bigl(\frac{1}{2}\log\Big\vert\frac{P}{\sigma^2}H_{[K],\Sc_1}H_{[K],\Sc_1}^T+I\Big\vert +\sum_{l\in \Sc_1^c}C_l\Big)\\
& = \min_{\Sc_1\subseteq[L]}\Bigl(\phi(\Sc_1) + \psi(\Sc_1^c)\Bigr),\numberthis\label{downinngauss_sumrate} 
\end{align*}
where 
\[
\phi(\Sc_1)  := \frac{1}{2}\log\Big\vert\frac{P}{\sigma^2}H_{[K],\Sc_1}H_{[K],\Sc_1}^T+I\Big\vert
\]
and
\[
\psi(\Sc_1^c)  := \sum_{l\in \Sc_1^c}C_l
\]
are such that $\Pr(\phi)$ and $\Pr(\psi)$ are both polymatroids. Therefore, by Edmonds's polymatroid intersection theorem, 
\[
r_\text{max} = \max_{y^L}\left\{\sum_{l\in[L]}y_l: y_l\le \psi(\{l\}), l\in[L], \sum_{l\in \Sc_1}y_l\le\phi(\Sc_1), \Sc_1\subseteq[L]\right\}.
\] 
Now, let us fix 
\[
C_{\sum} \ge \phi([L]) = \frac{1}{2}\log\left|\frac{P}{\sigma^2}HH^T + I\right|\numberthis\label{eq:Cstarinproofdown}
\]
 such that $C_1, \ldots, C_L$ are constrained to satisfy $C_1 + \cdots + C_L = C_{\sum}.$ Choose a point $\yb^*\equiv (y_1^*, \cdots, y_L^*)\in\Pr(\phi)$ such that $y_1^* + \ldots + y_L^* = \phi([L])$ and $y_l^* \ge (1/2)\log(1 + 1/\sigma^2)$ for each $l.$ Such a point always exists since $\Pr(\phi)$ is a polymatroid and since \eqref{firstconddownlinkpolymatroid} holds. The point $\ybt\equiv(\yt_1, \ldots, \yt_L)$ defined by 
\[
\yt_l = \frac{C_{\sum}}{\phi([L])}y_l^*, \quad l\in[L],
\]
satisfies $\yt_1 + \cdots + \yt_L = C_{\sum}.$ Therefore, choosing $C_l = \yt_l$ for each $l,$ $\Pr(\phi)$ becomes the cuboid with corner point $\ybt.$ Moreover, this cuboid includes the point $\yb^*,$ since $\yt_l\ge y_l^*$ for each $l$ by \eqref{eq:Cstarinproofdown}. Thus, the point $\yb^*$ lies in the intersection $\Pr(\phi)\cap\Pr(\psi)$ and therefore, 
\[
r_\text{max} \ge y_1^* + \cdots + y_L^* = \phi([L]) = \frac{1}{2}\log\left|\frac{P}{\sigma^2}HH^T + I\right|,
\] 
which implies that distributed decode--forward with the same fronthaul link capacities $(C_1, \ldots, C_L)$ can achieve
\[
R_\mathrm{sum}^\text{DDF} = \phi([L]) - \frac{K}{2}\log\left(1 + \frac{1}{\sigma^2}\right) = \frac{1}{2}\log\left|\frac{P}{\sigma^2}HH^T + I\right|-\frac{K}{2}\log\left(1 + \frac{1}{\sigma^2}\right),
\]
establishing the result. The converse follows immediately from the cutset bound.
\end{IEEEproof}
%A remark analogous to Remark~\ref{rem:fronthaulallocLP} in Section~\ref{updiscuss} holds for this case as well.
\begin{remark}\label{rem:downlinkCstartosumrate}
The best sum-rate achievable by our coding scheme for a given total fronthaul capacity $C_{\sum}>0$ can be expressed as
\[
R_\mathrm{sum}^\mathrm{max}(C_{\sum}) = \sup_{\sigma^2 > 0}\left(\min\left\{C_{\sum}, \frac{1}{2}\log\left|\frac{P}{\sigma^2}HH^T+I\right|\right\} - \frac{K}{2}\log\left(1+\frac{1}{\sigma^2}\right)\right).
\]
\end{remark}
\begin{remark}\label{rem:downlinksumrateduality}
As demonstrated in \cite{yutit06}, one can write the sum-capacity of the fronthaul-unlimited downlink with channel gain matrix $H\in\Real^{K\times L}$ as the solution of the optimization problem
\begin{subequations}
\begin{alignat*}{2}
&     &\qquad & \min_Q\max_{\Sigma}\frac{1}{2}\log\frac{|H^T\Sigma H + Q|}{|Q|}\\
&\text{subject to} &    & \Sigma\succeq 0\quad\text{diagonal},\quad \tr(\Sigma)\le 1,\\
& & &                            Q\succeq 0 \quad\text{diagonal},\quad \tr(Q)\le 1/P.
\end{alignat*}
Taking $Q = (1/PL)I$ and $\Sigma = I$ therefore yields an upper bound 
\[
R_\mathrm{sum}^\infty \le \frac{1}{2}\log|PLHH^T + I|. 
\]
\end{subequations}
\end{remark}

\subsection{Capacity Scaling}\label{downscale}

\begin{figure}[h]
\begin{subfigure}[t]{0.33\textwidth}
\center
\includegraphics[scale=0.33]{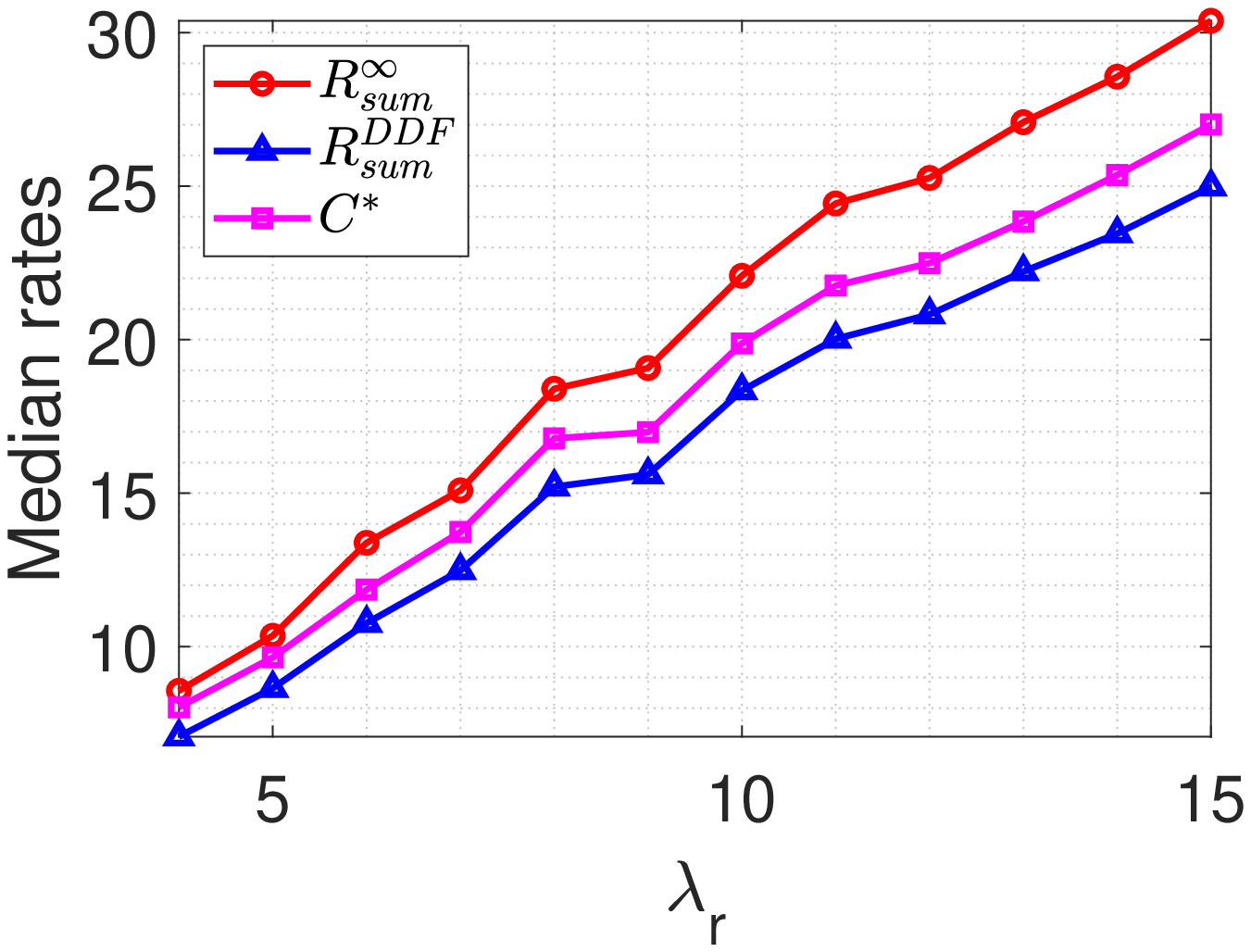}
\caption{$\beta = 2.5,$ $\lambda_r = 2\lambda_u.$}
\label{fig:plot2a}
\end{subfigure}%
\begin{subfigure}[t]{0.33\textwidth}
\center
\includegraphics[scale=0.33]{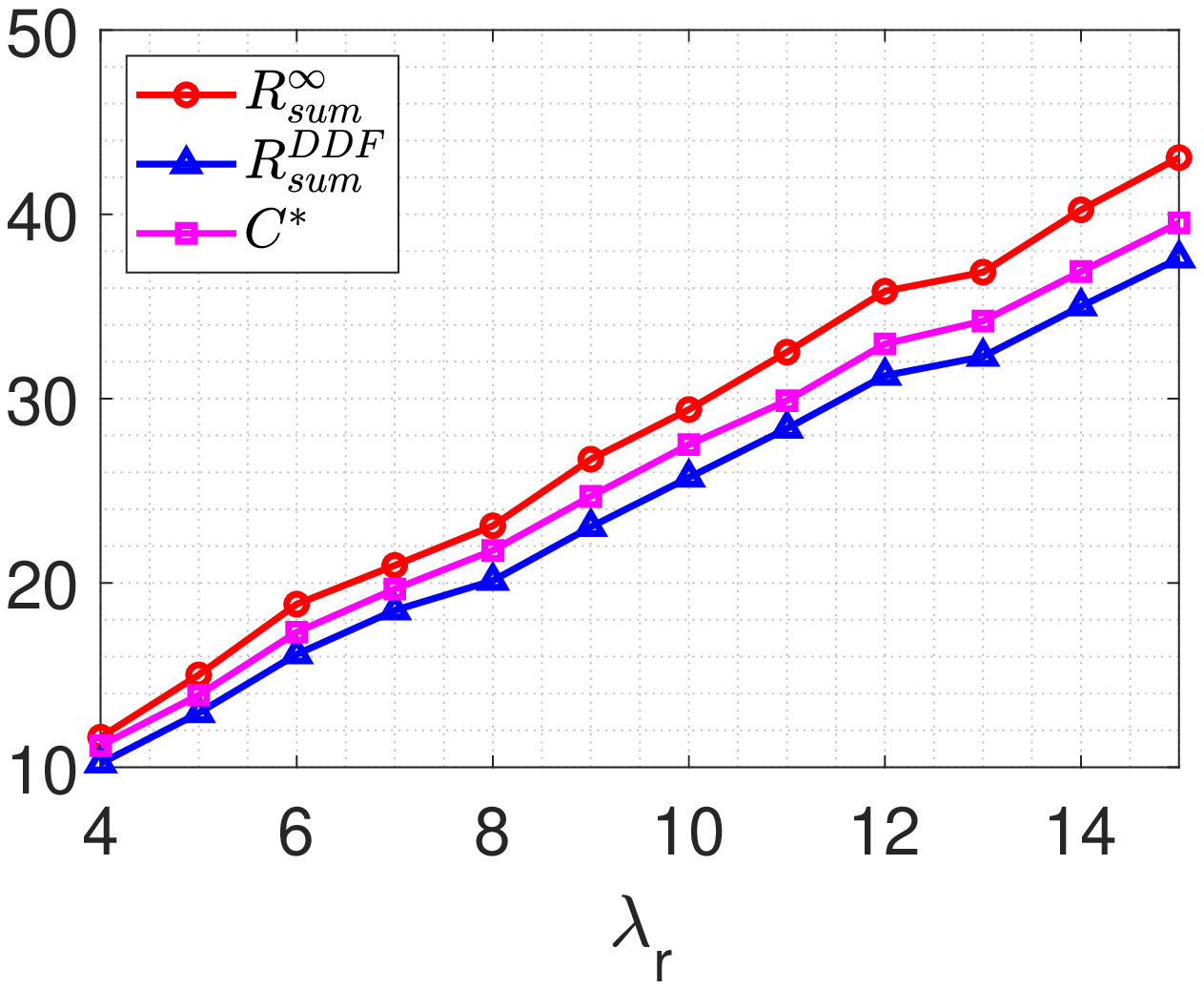}
\caption{$\beta = 3.5,$ $\lambda_r = 2\lambda_u.$}
\label{fig:plot2b}
\end{subfigure}
\begin{subfigure}[t]{0.33\textwidth}
\center
\includegraphics[scale=0.33]{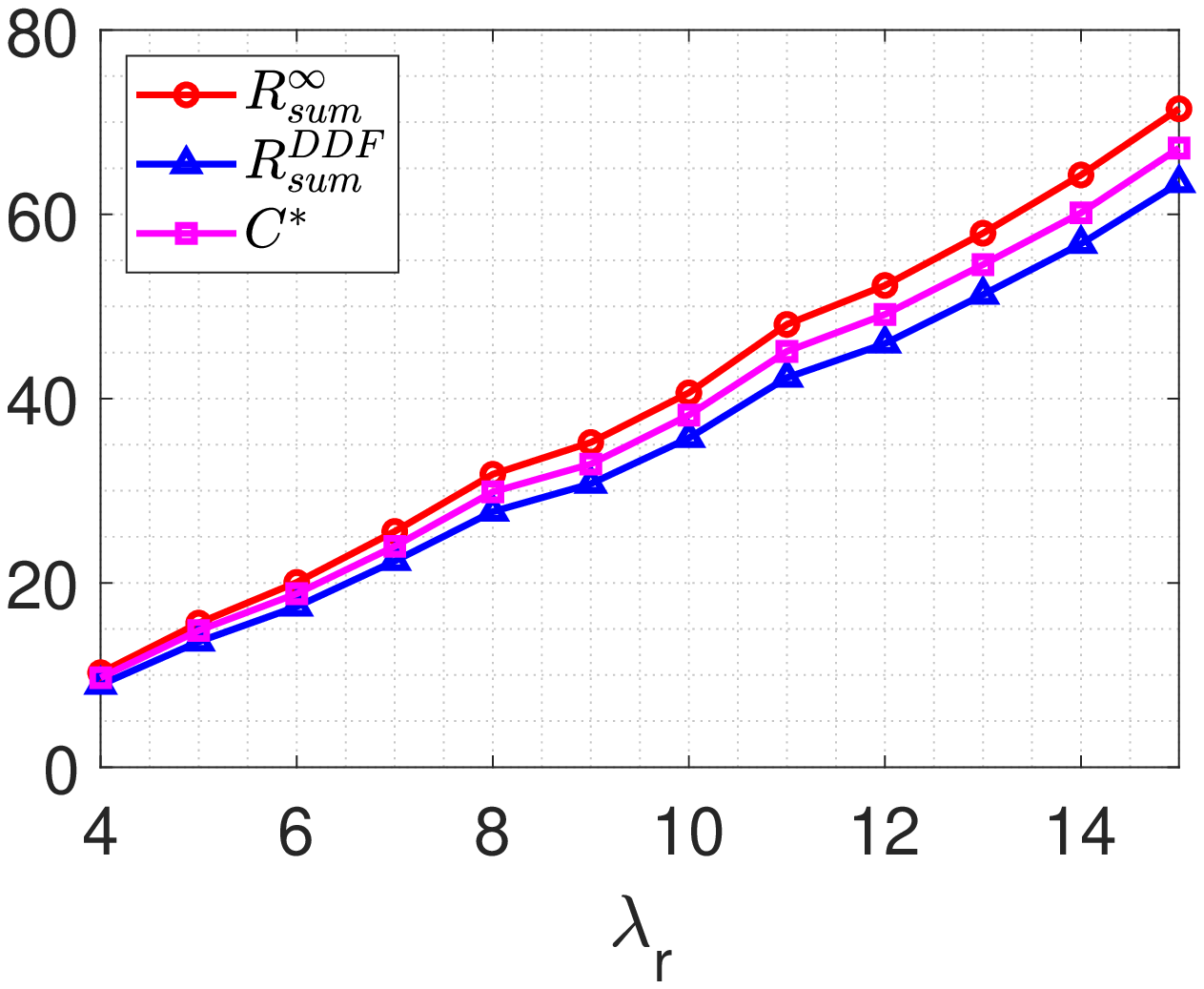}
\caption{Multipath, $\lambda_r = 2\lambda_u.$}
\label{fig:plot2c}
\end{subfigure}
\vskip\baselineskip
\begin{subfigure}[t]{0.33\textwidth}
\center
\includegraphics[scale=0.33]{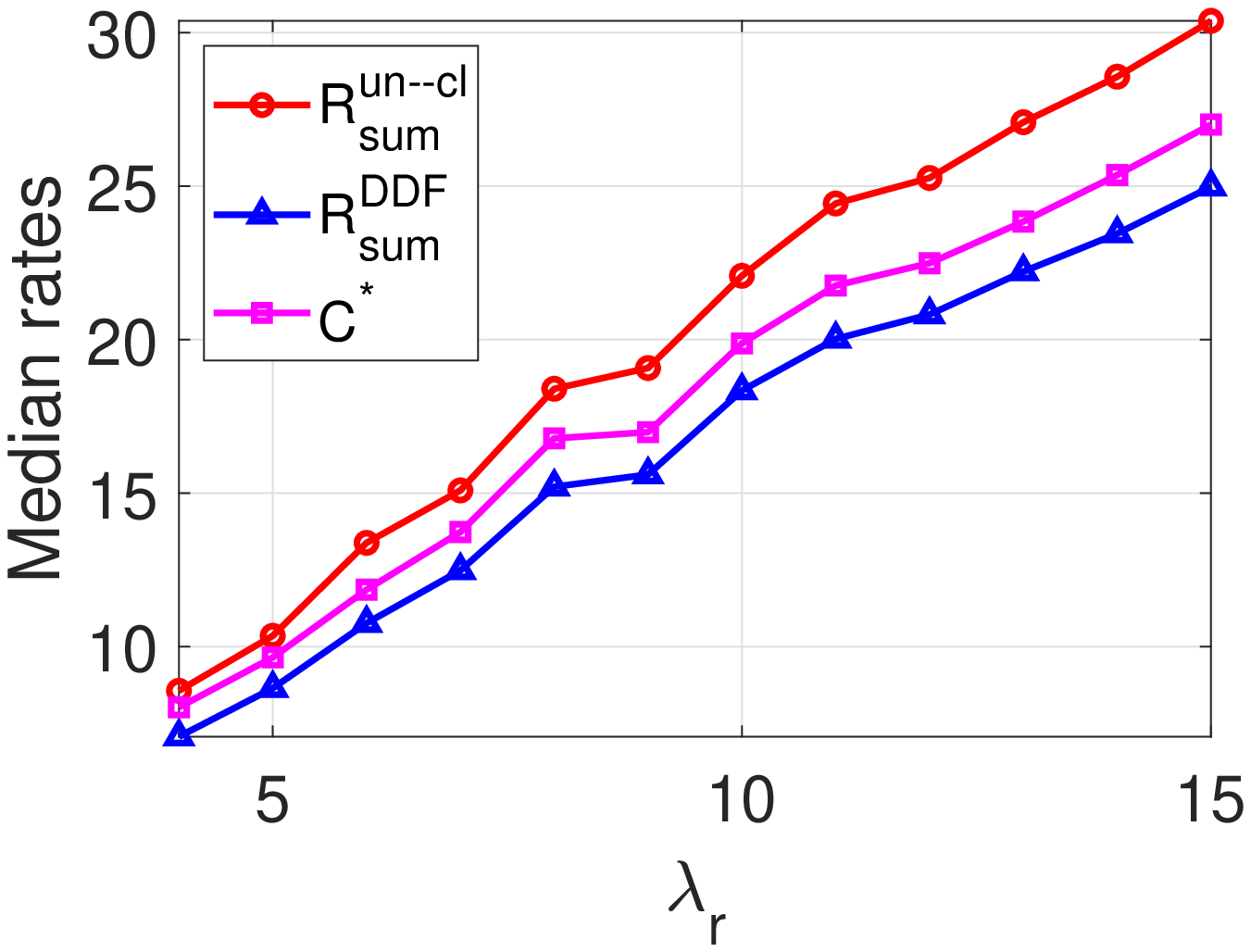}
\caption{$\beta = 2.5,$ $\lambda_r = \lambda_u^2.$}
\label{fig:plot2d}
\end{subfigure}%
\begin{subfigure}[t]{0.33\textwidth}
\center
\includegraphics[scale=0.33]{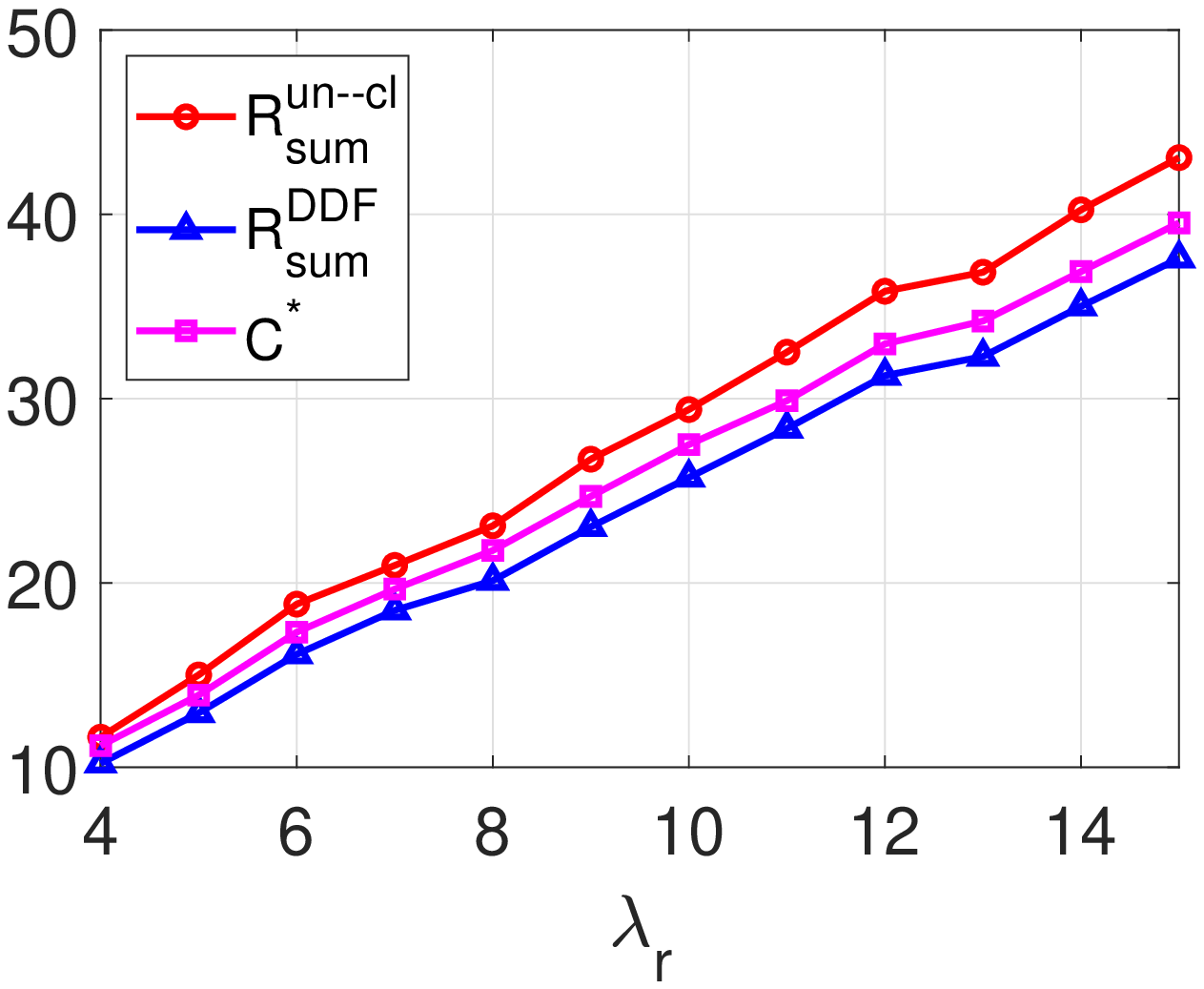}
\caption{$\beta = 3.5,$ $\lambda_r = \lambda_u^2.$}
\label{fig:plot2e}
\end{subfigure}
\begin{subfigure}[t]{0.33\textwidth}
\center
\includegraphics[scale=0.33]{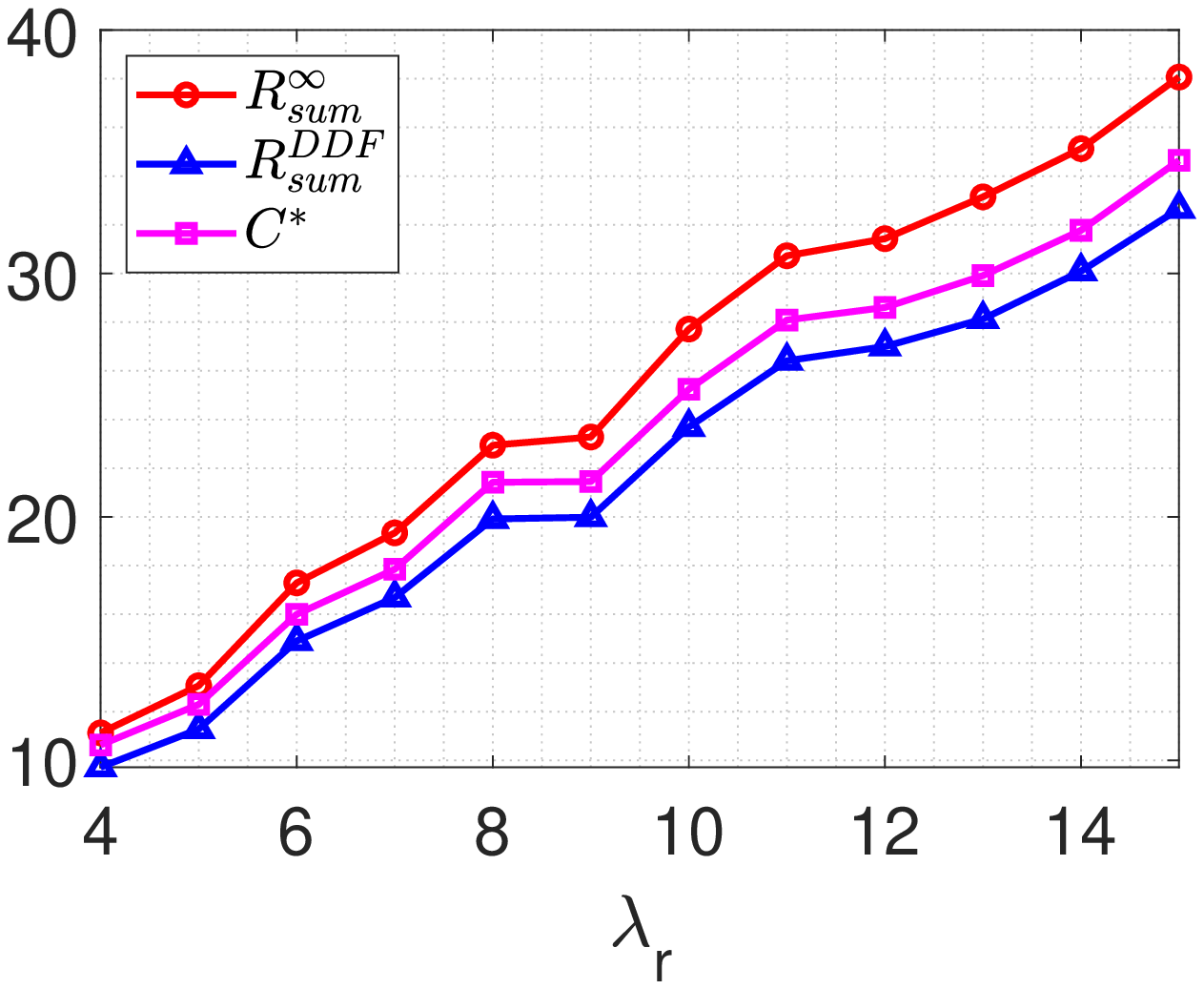}
\caption{Multipath, $\lambda_r = \lambda_u^2.$}
\label{fig:plot2f}
\end{subfigure}
\vskip\baselineskip
\begin{subfigure}[t]{0.33\textwidth}
\center
\includegraphics[scale=0.33]{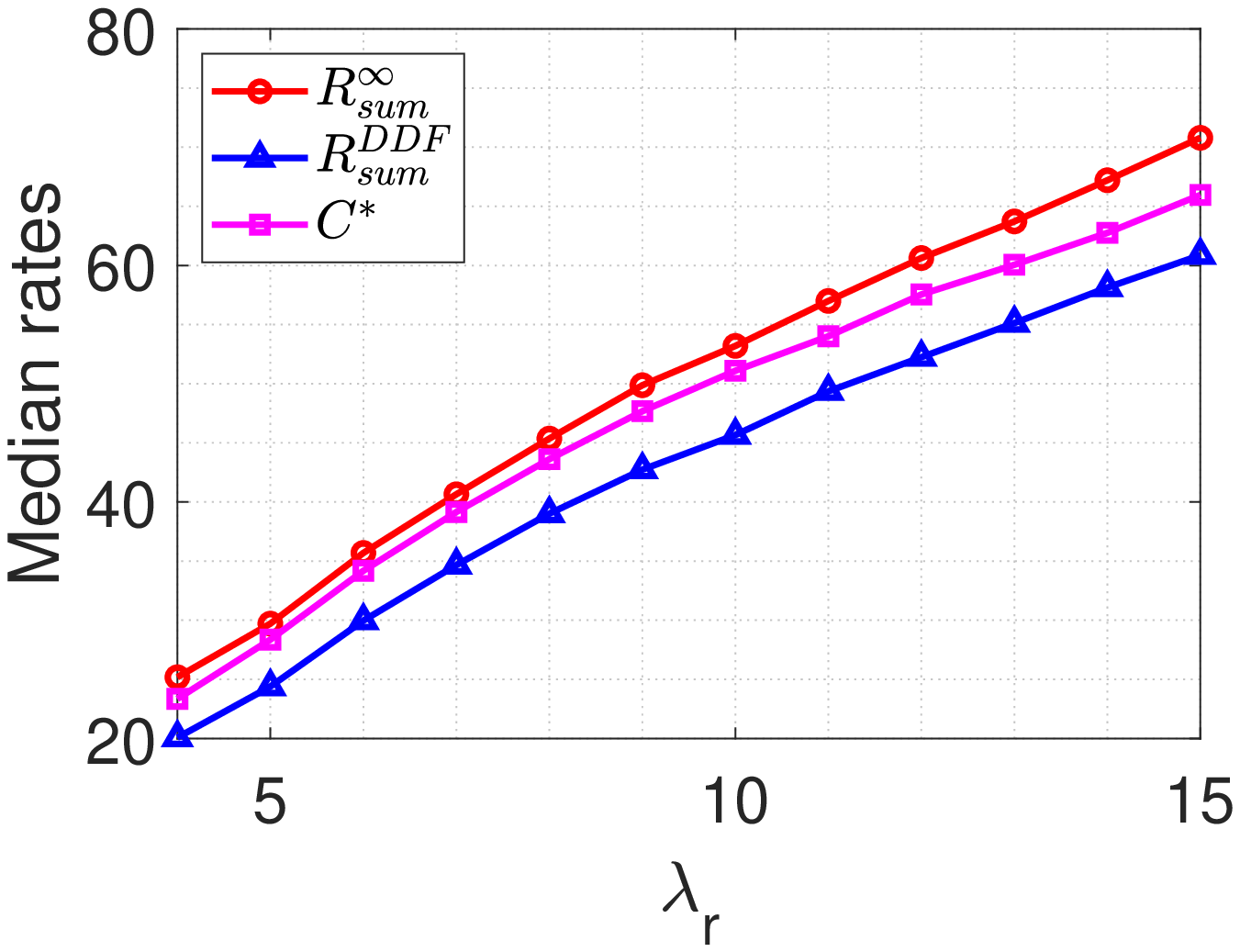}
\caption{$\beta = 2.5,$ $\lambda_u = 10.$}
\label{fig:plot2g}
\end{subfigure}%
\begin{subfigure}[t]{0.33\textwidth}
\center
\includegraphics[scale=0.33]{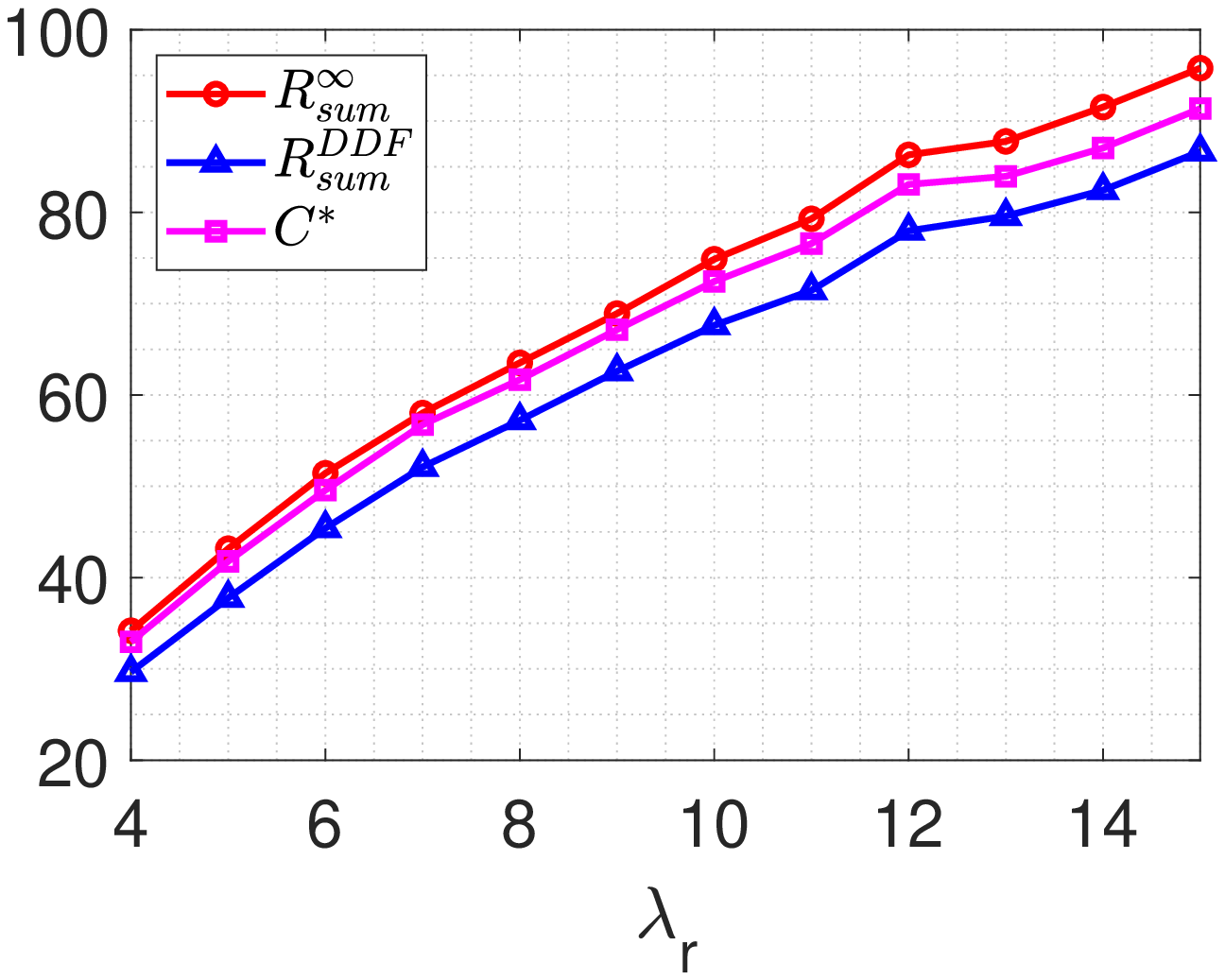}
\caption{$\beta = 3.5,$ $\lambda_u = 10.$}
\label{fig:plot2h}
\end{subfigure}
\begin{subfigure}[t]{0.33\textwidth}
\center
\includegraphics[scale=0.33]{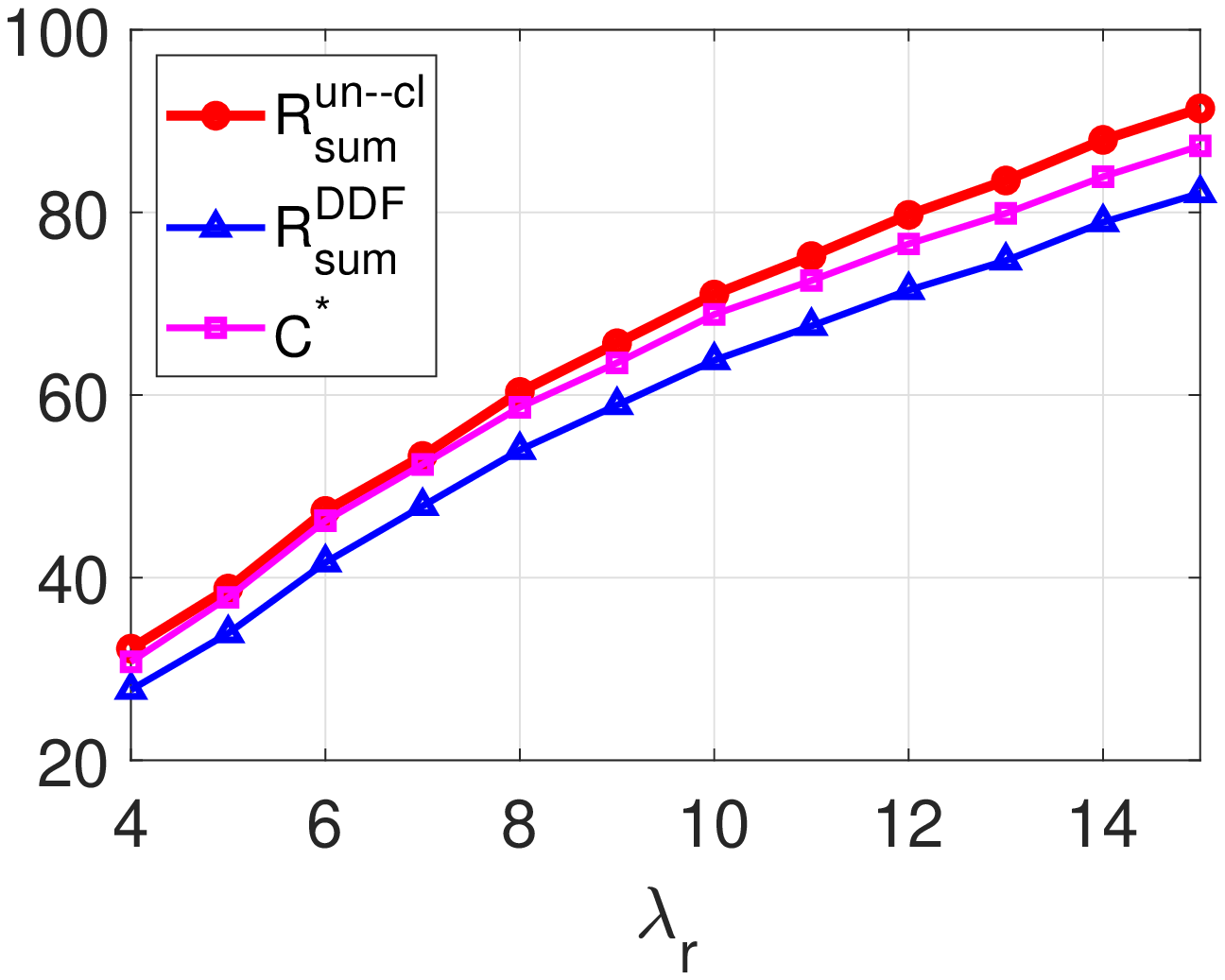}
\caption{Multipath, $\lambda_u = 10$.}
\label{fig:plot2i}
\end{subfigure}
\caption{Downlink capacity scaling under stochastic geometry.}\label{fig:plot3}
\end{figure}

Similar to Section~\ref{upscaletheo}, we first consider a rich scattering model. We can use Lemma~\ref{lemma:limitdistribution} to establish the following theorem on the large network size behavior of $R_\mathrm{sum}^\text{DDF}(\sigma^2),$ $C^*(\sigma^2),$ and $R_\mathrm{sum}^\infty.$ The proof is similar to that of Theorem~\ref{theorem:pggt} and is omitted.  
\begin{theorem}\label{theorem:phht}
Let the entries of the $K\times L$ channel gain matrix $H$ be distributed as i.i.d.\@ random variables with variance $1$, and let $\sigma^2 = \sigma^2(L, K) > 0.$ 
If $L\to\infty$ such that $L/K\to\rho \in(1, \infty]$ and $L/\sigma^2\to \infty,$ then 
\[
\frac{1}{2}\le\liminf\frac{R_\mathrm{sum}^\infty}{K\log L}\le\limsup\frac{R_\mathrm{sum}^\infty}{K\log L}\le 1
\]
and
\[
R_\mathrm{sum}^\text{DDF} = C^*- \frac{K}{2}\log\left(1+\frac{1}{\sigma^2}\right)\sim \frac{K}{2}\log(L/\sigma^2) - \frac{K}{2}\log\left(1+\frac{1}{\sigma^2}\right),
\]
a.s.\@ in $H.$ Similarly, if $K\to\infty$ such that $L/K\to\rho \in[0, 1)$ and $K/\sigma^2\to \infty,$ then
\[
\frac{1}{2}\le\liminf\frac{R_\mathrm{sum}^\infty}{L\log K}\le\limsup\frac{R_\mathrm{sum}^\infty}{L\log K}\le 1
\]
and
\[
R_\mathrm{sum}^\text{DDF} = C^*- \frac{K}{2}\log\left(1+\frac{1}{\sigma^2}\right)\sim \frac{L}{2}\log(K/\sigma^2)- \frac{K}{2}\log\left(1+\frac{1}{\sigma^2}\right),
\]
a.s.\@ in $H.$
\end{theorem}
Using Theorem~\ref{theorem:phht} and choosing $\sigma^2 = \sigma^2(L, K)$ appropriately, we summarize the scaling laws for $R_\mathrm{sum}^\text{DDF}, R_\mathrm{sum}^\infty,$ and the fronthaul link capacity requirement $C^*$ in Table~\ref{table:downlinkscaling}. %The detailed derivation of the tabulated results are deferred to Appendix~\ref{upproofs}.
\begin{table*}[t]
    \setlength{\tabcolsep}{20pt}
    \renewcommand{\arraystretch}{2.5}
    \centering
    \begin{tabular}{ c  c | c  c  c c} \hline
	\multicolumn{2}{c|}{$L$ vs.\@ $K$} & $\sigma^2$ & $C^*$ & $R_\mathrm{sum}^\text{DDF}$ & $R_\mathrm{sum}^\infty$ \\[5pt] \hline
       \multirow{2}{*}{$L = \gamma K$} & $(\gamma > 1)$ & 1 & $\displaystyle\frac{K}{2}\log L$ & $\displaystyle\frac{K}{2}\log L$ & $\alpha K\log L$\\
							&   $(\gamma < 1)$    &  1  & $\displaystyle\frac{L}{2}\log K$ & $\displaystyle\frac{L}{2}\log K$ & $\alpha L\log K$\\[20pt]
	 \multirow{3}{*}{$L =  K^\gamma$} & \multirow{2}{*}{$( \gamma < 1 )$} & $L^{1/\gamma - 1}$ & $\displaystyle\frac{L}{2}\log L$ & $\displaystyle\frac{L}{2}\log L$ & \multirow{2}{*}{$\alpha L\log K$}  \\[-10pt]
           & & $L^{1/\gamma - 1 + \delta}$ & $\displaystyle\frac{(1-\delta)L\log L}{2}$ &  $\displaystyle\frac{(1-\delta)L\log L}{2}$ & \\
  & $(\gamma > 1)$ & $1$ & $\displaystyle\frac{K}{2}\log L$ & $\displaystyle\frac{K}{2}\log L$ & $\alpha K\log L$\\[20pt] 
         %$L$ fixed & & $K^\epsilon$ & $\displaystyle\frac{L^{1-\epsilon}}{2}\log e$ &$(1-\epsilon)\displaystyle\frac{L}{2}\log K$ & $\displaystyle\frac{L}{2}\log K$\\[5pt]
$K$ fixed & & $1$ & $\displaystyle\frac{K}{2}\log L$ & $\displaystyle\frac{K}{2}\log L$ & $\alpha K\log L$ \\[5pt] \hline
       \end{tabular}
\caption{Sum-rate scaling for fronthaul-limited and fronthaul-unlimited downlink C-RAN; $\alpha\in[1/2, 1],$ $\gamma > 0,$ $\gamma\ne 1,$ $0<\delta<\gamma - 1,$ $0<\epsilon<1.$}\label{table:downlinkscaling}
\end{table*}
\begin{remark}
Unlike Table~\ref{table:uplinkscaling}, Table~\ref{table:downlinkscaling} does not have an exact coefficient in the scaling law for $R_\mathrm{sum}^\infty$ for downlink. The upper bound in Remark~\ref{rem:downlinksumrateduality} scales as $L\log K$ or $K\log L,$ while $R_\mathrm{sum}^\text{DDF}$ serves as a lower bound on $R_\mathrm{sum}^\infty.$ 
\end{remark}
For a stochastic geometry model similar to that in Section~\ref{upstochgeom}, Fig.~\ref{fig:plot3} plots the median sum-rates obtained experimentally over $1000$ simulation runs each, for different scaling regimes and different path loss exponents. The power constraint $P$ at each relay is kept fixed. For each realization of the channel gain matrix $H,$ $\sigma^2$ is chosen to (numerically) maximize $R_\mathrm{sum}^\text{DDF},$ and then $C^*$ is calculated using this value of $\sigma^2.$ As before, the C-RAN downlink sum-rate closely tracks the fronthaul-unlimited downlink sum-capacity using a similar amount of fronthaul capacity. We note here that the plots show an \emph{upper} bound on $R_\mathrm{sum}^\infty,$ corresponding to choosing randomized values of the entries of the matrix $Q$ in the dual characterization mentioned in Remark~\ref{rem:downlinksumrateduality} and maximizing over the input covariance matrix $\Sigma$ using the singular value decomposition of the channel gain matrix and water-filling power allocation (see, for example, \cite[Section~9.1]{nit}). 
\section{MIMO C-RANs}\label{mimocran}
In this section, we extend the results of Sections~\ref{upload} and \ref{download} %, in particular, Theorems~\ref{upgaptheo1}--\ref{theorem:phht}, as well as Tables~\ref{table:uplinkscaling} and \ref{table:downlinkscaling}, 
to the situation in which each user has $N_u$ local antennas and each relay has $N_r$ local antennas. The apparently more general situation in which users and/or relays have different numbers of antennas can be handled by setting the channel gain matrix appropriately. We assume a total average transmit power constraint $P$ at each node. The objects $\Rr_\mathrm{up}^\text{NCF},$ $R_\mathrm{sum}^\text{NCF},$ $\Rr_\mathrm{up}^\text{CS},$ $R_\mathrm{sum}^\text{CS},$ $\Rr_\mathrm{down}^\text{DDF},$ $R_\mathrm{sum}^\text{DDF},$ and $\Rr_\mathrm{down}^\text{CS}$ are defined as before. We consider channel matrices $G\in\Real^{N_rL\times N_uK}$ and $H\in\Real^{N_uK\times N_rL}$ for uplink and downlink C-RANs, respectively. Similar to \eqref{upinngaussinformal} in Section~\ref{upgauss}, the network compress--forward  inner bound $\Rr_\mathrm{up}^\text{NCF}$ for the uplink Gaussian MIMO C-RAN is characterized by the rate tuples $(R_1, \ldots, R_K)$ satisfying
\[
\sum_{k\in \Sc_1}R_k \le \frac{1}{2}\log\Big\vert\frac{\sum_{k\in \Sc_1}G_{\Sc_2^c,k}\Gamma_kG_{\Sc_2^c,k}^T}{\sigma^2+1}+I\Big\vert +\sum_{l\in \Sc_2}C_l-\frac{N_r\vert \Sc_2\vert}{2}\log\left(1+\frac{1}{\sigma^2}\right)\numberthis\label{upinngaussinformalmimo}
\]
for all $\Sc_1\subseteq[K]$ and $\Sc_2\subseteq[L]$ for some $\sigma^2>0$ and for some covariance matrices $\Gamma_1, \ldots, \Gamma_K\succeq 0$ such that $\tr(\Gamma_k) = P,$ $k\in[K].$ Here, for $\Sc_2\subseteq[L]$ and $k\in[K],$ $G_{\Sc_2^c,k}$ denotes the $N_r|\Sc_2^c|\times N_u$ channel gain matrix between user $k$ and the relays in $\Sc_2^c$. For each fixed $\sigma^2, \Gamma_1, \ldots, \Gamma_K,$ we denote this region by $\Rr_\mathrm{up}^\text{NCF}(\sigma^2, \Gamma_1,\ldots,\Gamma_K).$ The cutset bound $\Rr_\mathrm{up}^\text{CS}$ is characterized by rate tuples $(R_1, \ldots, R_K)$ satisfying
\[
\sum_{k\in \Sc_1}R_k \le \frac{1}{2}\log\Big\vert\sum_{k\in \Sc_1}G_{\Sc_2^c,k}\Gamma_kG_{\Sc_2^c,k}^T+I\Big\vert +\sum_{l\in \Sc_2}C_l.\numberthis\label{upoutgaussinformalmimo}
\]
Similar to \eqref{downinngaussinformal} in Section~\ref{downgauss}, the distributed decode--forward inner bound $\Rr_\mathrm{down}^\text{DDF}$ for the downlink Gaussian MIMO C-RAN is characterized by rate tuples $(R_1, \ldots, R_K)$ satisfying
\[
\sum_{k\in \Sc_2^c}R_k \le \frac{1}{2}\log\Big\vert\frac{\sum_{l\in \Sc_1}H_{\Sc_2^c,l}\Gamma_lH_{\Sc_2^c,l}^T}{\sigma^2}+I\Big\vert +\sum_{l\in \Sc_1^c}C_l-\frac{N_u\vert \Sc_2^c\vert}{2}\log\left(1+\frac{1}{\sigma^2}\right)\numberthis\label{downinngaussinformalmimo}
\]
for all $\Sc_1\subseteq[L]$ and $\Sc_2\subseteq[K]$ for some $\sigma^2>0$ and for some covariance matrices $\Gamma_1, \ldots, \Gamma_L\succeq 0$ satisfying $\tr(\Gamma_l) = P,$ $l\in[L].$ For each fixed $\sigma^2, \Gamma_1, \ldots, \Gamma_L,$ we denote this region by$\Rr_\mathrm{down}^\text{DDF}(\sigma^2, \Gamma_1,\ldots,\Gamma_L).$ The cutset bound $\Rr_\mathrm{down}^\text{CS}$ is characterized by rate tuples $(R_1, \ldots, R_K)$ satisfying
\[
\sum_{k\in \Sc_2^c}R_k \le \frac{1}{2}\log\Big\vert H_{\Sc_2^c,\Sc_1}\tilde{\Gamma}_{\Sc_1\cond \Sc_1^c}H_{\Sc_2^c,\Sc_1}^T+I\Big\vert +\sum_{l\in \Sc_1^c}C_l,\numberthis\label{downoutgaussinformalmimo}
\] 
where $\tilde{\Gamma}$ is a general $N_rL\times N_rL$ input covariance matrix satisfying the block trace constraints. Here, $H_{\Sc_2^c,l}$ denotes the $N_u|\Sc_2^c|\times N_r$ channel gain matrix between relay $l$ and the users in $\Sc_2^c,$ and $H_{\Sc_2^c,\Sc_1}$ denotes the $N_u|\Sc_2^c|\times N_r|\Sc_1|$ channel gain matrix between the relays in $\Sc_1$ and the users in $\Sc_2^c.$ 

We have the following result on the achievable per-user gaps from the capacity of the MIMO C-RAN, the proof of which is sketched in Appendix~\ref{mimoproofs}.       
\begin{prop}\label{mimogap}
For every $G\in\mathbbm{R}^{N_rL\times N_uK}$ and every $P\in\mathbbm{R}^+$, if a rate tuple $(R_1,\ldots,R_K)$ is in $\Rr_\mathrm{up}^\text{CS}$, then the rate tuple $((R_1-\Delta^\mathrm{up})^+,\ldots,(R_K-\Delta^\mathrm{up})^+)$ is achievable, where 
\[
\Delta^\mathrm{up}\le \frac{N_u}{2}\log\left(\frac{eN_rL}{N_u}\right).
\]
Moreover, 
\[
\Delta_\mathrm{sum}^\mathrm{up} := R_\mathrm{sum}^\text{CS} - \sup_{\sigma^2,\Gamma_1,\ldots,\Gamma_K}R_\mathrm{sum}^\text{NCF}(\sigma^2,\Gamma_1,\ldots,\Gamma_K)\le\begin{cases}\frac{N_rL}{2}H\left(\frac{N_uK}{N_rL}\right), & N_rL\ge 2N_uK,\\
				\frac{N_rL}{2}, & N_rL < 2N_uK.\end{cases}
\]
Similarly, for every $H\in\Real^{N_uK\times N_rL}$ and $P\in\Real^+,$ if a rate tuple $(R_1,\ldots,R_K)$ is in $\Rr_\mathrm{down}^\text{CS}$, then the rate tuple $((R_1-\Delta^\mathrm{down})^+,\ldots,(R_K-\Delta^\mathrm{down})^+)$ is achievable, where 
\[
\Delta^\mathrm{down}\le \begin{cases}\frac{N_u}{2}\log(eN_rLK), & N_u < N_rL,\\ \frac{N_rL}{2}\log(eN_uK), & N_u\ge N_rL, \quad N_uK\ge 2N_rL,\\ \frac{N_u}{2}+\frac{N_rL}{2}\log(N_rL), & K = 1, \quad N_rL \le N_u < 2N_rL.\end{cases}
\]
Moreover, 
\[
\Delta_\mathrm{sum}^\mathrm{down} := R_\mathrm{sum}^\text{CS} - \sup_{\sigma^2,\Gamma_1,\ldots,\Gamma_L}R_\mathrm{sum}^\text{DDF}(\sigma^2,\Gamma_1,\ldots,\Gamma_L)\le\frac{N_uK}{2} + \frac{\min\{N_rL, N_uK\}}{2}\log(N_rL).
\]
\end{prop}
\begin{remark}
Proposition~\ref{mimogap} recovers the results of Theorems~\ref{upgaptheo1} and \ref{downgaptheo1} when we set $N_u = N_r = 1.$ From the expressions for $\Delta^\mathrm{up}$ and $\Delta^\mathrm{down},$ we observe that the capacity gaps are the same as if there were $N_r\times L$ single-antenna relays. The sum-rate gaps are the same as if there were $N_u\times K$ single-antenna users, while $\Delta^\mathrm{up}$ and $\Delta^\mathrm{down}$ are in general larger than the gaps obtained with $N_u\times K$ single-antenna users.  
\end{remark}
Similar to Sections~\ref{updiscuss} and \ref{downdiscuss}, we can characterize the achievable sum-rates $R_\mathrm{sum}^\text{NCF}(\sigma^2, \Gamma_1, \ldots, \Gamma_K)$ and $R_\mathrm{sum}^\text{DDF}(\sigma^2, \Gamma_1, \ldots, \Gamma_L)$ as
\begin{align*}
& R_\mathrm{sum}^\text{NCF}(\sigma^2, \Gamma_1, \ldots, \Gamma_K)\\
& = \min_{\Sc_2\subseteq[L]}\Biggl(\frac{1}{2}\log\Big\vert\frac{1}{\sigma^2+1}\sum_{k\in[K]}G_{\Sc_2^c,k}\Gamma_kG_{\Sc_2^c,k}^T+I\Big\vert +\sum_{l\in \Sc_2}C_l-\frac{N_r\vert \Sc_2\vert}{2}\log\left(1+\frac{1}{\sigma^2}\right)\Biggr)\\
& = \max_{y^L}\Biggl\{\sum_{l\in[L]}y_l: y_l\le C_l - \frac{N_r}{2}\log\left(1+\frac{1}{\sigma^2}\right), l\in[L], \\
& \quad\quad\quad\quad\quad\quad\quad\quad\quad \sum_{l\in \Sc_2}y_l\le\log\Big\vert\frac{1}{\sigma^2+1}\sum_{k\in[K]}G_{\Sc_2^c,k}\Gamma_kG_{\Sc_2^c,k}^T+I\Big\vert, \Sc_2\subseteq[L]\Biggr\}
\end{align*}
and 
\begin{align*}
& R_\mathrm{sum}^\text{DDF}(\sigma^2, \Gamma_1, \ldots, \Gamma_L) + \frac{N_uK}{2}\log\left(1+\frac{1}{\sigma^2}\right)\\ 
& = \min_{\Sc_1\subseteq[L]}\left(\frac{1}{2}\log\Big\vert\frac{1}{\sigma^2}\sum_{l\in \Sc_1}H_{[K],l}\Gamma_lH_{[K],l}^T+I\Big\vert +\sum_{l\in \Sc_1^c}C_l\right)\\ 
& = \max_{y^L}\left\{\sum_{l\in[L]}y_l: y_l\le C_l, l\in[L], \sum_{l\in \Sc_1}y_l\le\frac{1}{2}\log\Big\vert\frac{1}{\sigma^2}\sum_{l\in \Sc_1}H_{[K],l}\Gamma_lH_{[K],l}^T+I\Big\vert, \Sc_1\subseteq[L]\right\},
\end{align*}
leading to the following extension of Theorems~\ref{theorem:Cstaruplink} and \ref{theorem:Cstardownlink}.
\begin{prop}\label{prop:Cstarmimo} 
If  
\[
C_{\sum} \ge \frac{1}{2}\log\left|\frac{1}{\sigma^2+1}\sum_{k\in[K]}G_{[L], k}\Gamma_kG_{[L], k}^T + I\right| + \frac{N_rL}{2}\log\left(1 + \frac{1}{\sigma^2}\right) =: C^*(\sigma^2,\Gamma_1,\ldots,\Gamma_K)
\]
for some $\sigma^2, \Gamma_1, \ldots, \Gamma_K,$ then there exist fronthaul link capacities $C_1, C_2, \ldots, C_L\ge 0$ with $\sum_{l\in[L]}C_l = C_{\sum}$ at which network compress--forward can achieve a sum-rate
\[
R_\mathrm{sum}^\text{NCF}(\sigma^2, \Gamma_1, \ldots, \Gamma_K) = \frac{1}{2}\log\left|\frac{1}{\sigma^2+1}\sum_{k\in[K]}G_{[L], k}\Gamma_kG_{[L], k}^T + I\right|.
\] 
If 
\[
C_{\sum} \ge \frac{1}{2}\log\left|\frac{1}{\sigma^2}\sum_{l\in[L]}H_{[K],l}\Gamma_lH_{[K],l}^T+I\right| =: C^*(\sigma^2, \Gamma_1, \ldots, \Gamma_L) 
\]
for some $\sigma^2, \Gamma_1, \ldots, \Gamma_L,$ then there exist $C_1, C_2, \ldots, C_L\ge 0$ with $\sum_{l\in[L]}C_l = C_{\sum}$ at which distributed decode--forward can achieve a sum-rate
\[
R_\mathrm{sum}^\text{DDF}(\sigma^2, \Gamma_1, \ldots, \Gamma_L) = \frac{1}{2}\log\left|\frac{1}{\sigma^2}\sum_{l\in[L]}H_{[K],l}\Gamma_lH_{[K],l}^T+I\right|-\frac{N_uK}{2}\log\left(1 + \frac{1}{\sigma^2}\right).
\] 
\end{prop}
\begin{remark}\label{rem:mimodownlinksumrateduality}
Following a similar line of reasoning as in \cite[Section IV-B]{yutit06}, one can write the sum-capacity of the fronthaul-unlimited MIMO downlink with channel gain matrix $H\in\Real^{N_uK\times N_rL}$ as the solution of the optimization problem
\begin{subequations}
\begin{alignat*}{2}
&     &\qquad & \min_Q\max_{\Sigma}\frac{1}{2}\log\frac{|H^T\Sigma H + Q|}{|Q|}\\
&\text{subject to} &    & Q \quad\text{non-negative diagonal},\\
& & & \Sigma = \begin{bmatrix}\Sigma_1 & & & \\ & \Sigma_2 & & \\ & & \ddots & \\ & & & \Sigma_K\end{bmatrix},\\
& & & \text{submatrix}\quad Q([(l-1)N_r+1:lN_r], [(l-1)N_r+1:lN_r])\quad\text{having equal diagonals}\\
& & & \quad\quad\quad\quad\quad\quad\quad\quad\quad\quad\quad\quad\quad\quad\quad\quad\quad\quad\quad\quad\quad\quad\quad\quad\quad\quad\text{for} \quad l = 1,\ldots, L,\\ 
&                  &      & \tr(Q)\le N_r/P,\\
& & &                       \tr(\Sigma)\le 1.
\end{alignat*}
\end{subequations}
Here, $\Sigma_k\succeq 0$ is of size $N_u\times N_u$ for every $k\in[K]$. 
\end{remark}
\begin{remark}\label{rem:mimoCstartosumrate}
Generalizing Remarks~\ref{rem:uplinkCstartosumrate} and \ref{rem:downlinkCstartosumrate}, the best sum-rates achievable for a total fronthaul capacity $C_{\sum}>0$ can be expressed as
\[
R_\mathrm{sum}^\mathrm{max}(C_{\sum}) = \sup_{\sigma^2 > 0}\min\left\{C_{\sum} - \frac{N_rL}{2}\log\left(1+\frac{1}{\sigma^2}\right), \max_{\Gamma_1, \ldots,\Gamma_K}\frac{1}{2}\log\left|\frac{1}{\sigma^2+1}\sum_{k\in[K]}G_{[L], k}\Gamma_kG_{[L], k}^T + I\right|\right\}
\]
for the uplink, and
\[
R_\mathrm{sum}^\mathrm{max}(C_{\sum}) = \sup_{\sigma^2 > 0}\left(\min\left\{C_{\sum}, \max_{\Gamma_1, \ldots,\Gamma_L}\frac{1}{2}\log\left|\frac{1}{\sigma^2}\sum_{l\in[l]}H_{[K], l}\Gamma_lH_{[K], l}^T+I\right|\right\} - \frac{N_uK}{2}\log\left(1+\frac{1}{\sigma^2}\right)\right)
\]
for the downlink.
\end{remark}

\begin{figure}[h]
\begin{subfigure}[t]{0.33\textwidth}
\center
\includegraphics[scale=0.33]{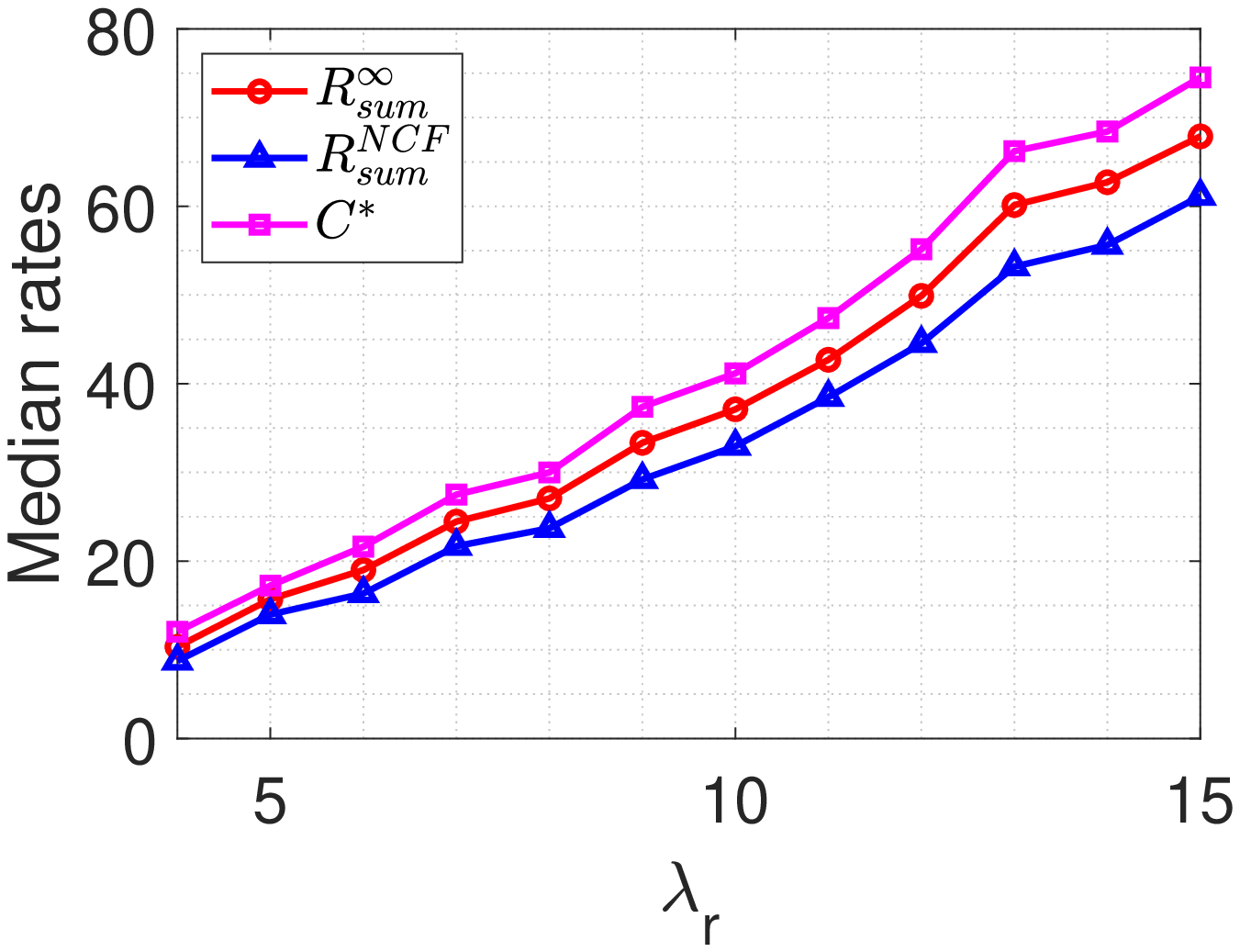}
\caption{$\beta = 2.5,$ $\lambda_r = 2\lambda_u.$}
\label{fig:plot2a}
\end{subfigure}%
\begin{subfigure}[t]{0.33\textwidth}
\center
\includegraphics[scale=0.33]{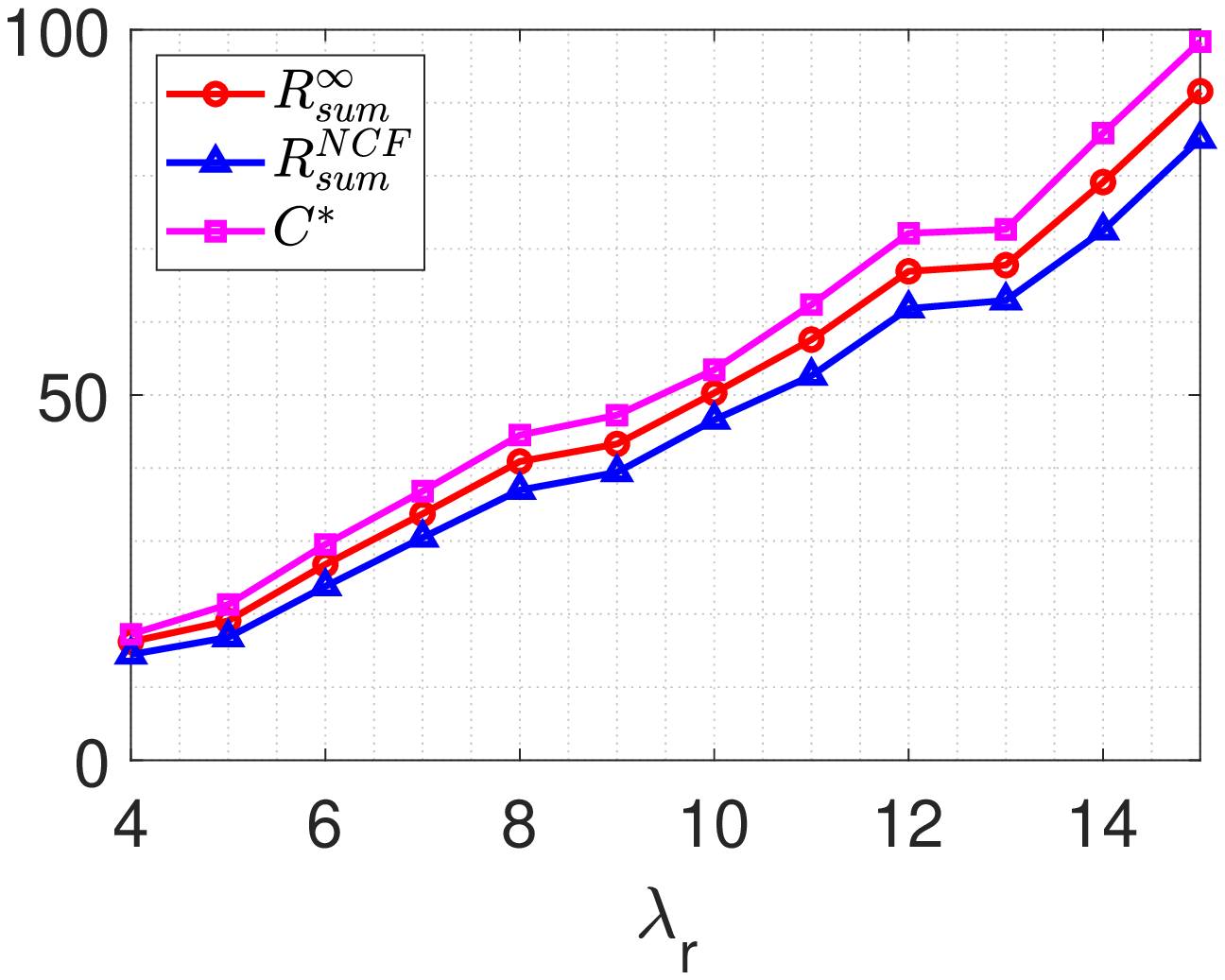}
\caption{$\beta = 3.5,$ $\lambda_r = 2\lambda_u.$}
\label{fig:plot2b}
\end{subfigure}
\begin{subfigure}[t]{0.33\textwidth}
\center
\includegraphics[scale=0.33]{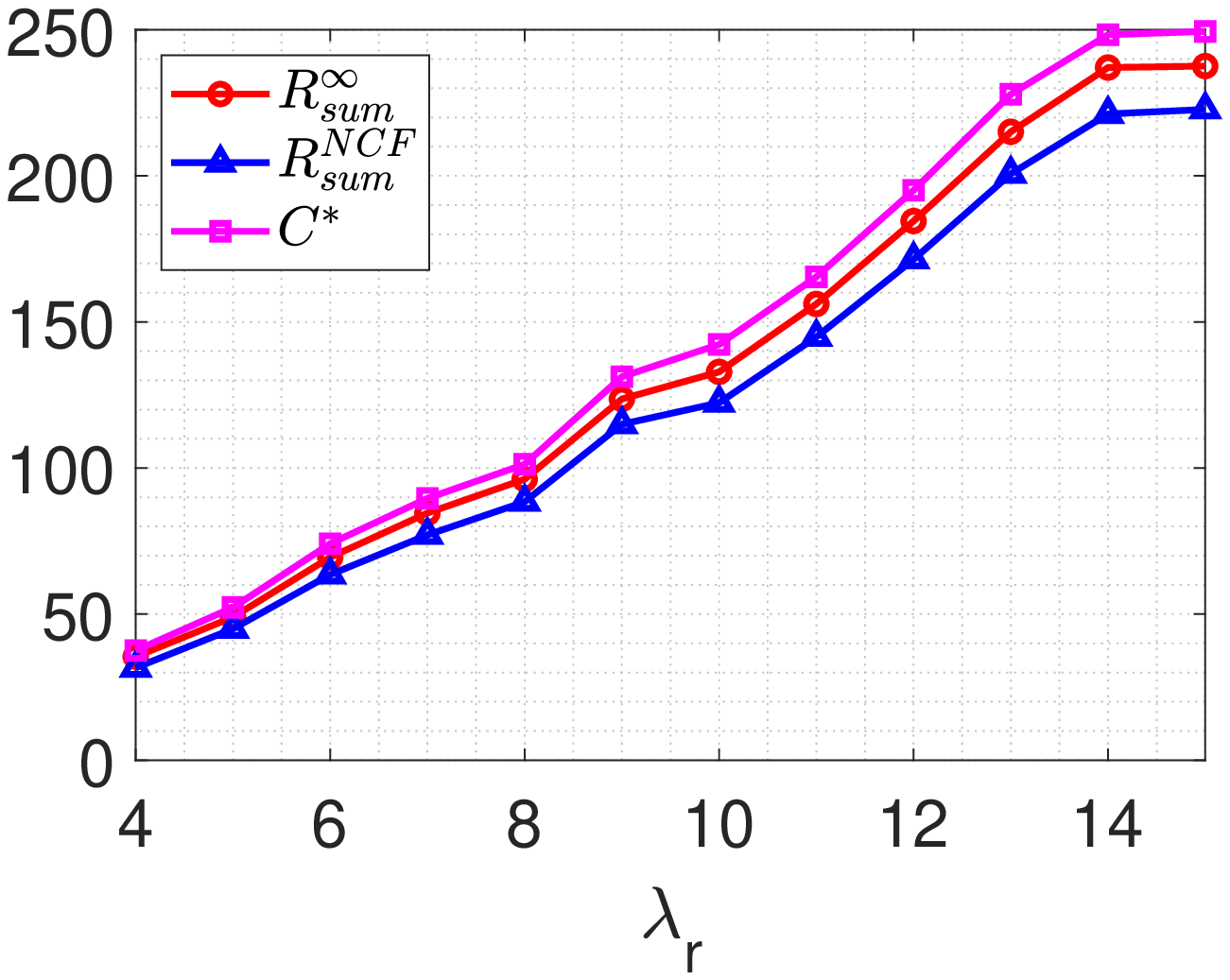}
\caption{Multipath, $\lambda_r = 2\lambda_u.$}
\label{fig:plot2c}
\end{subfigure}
\vskip\baselineskip
\begin{subfigure}[t]{0.33\textwidth}
\center
\includegraphics[scale=0.33]{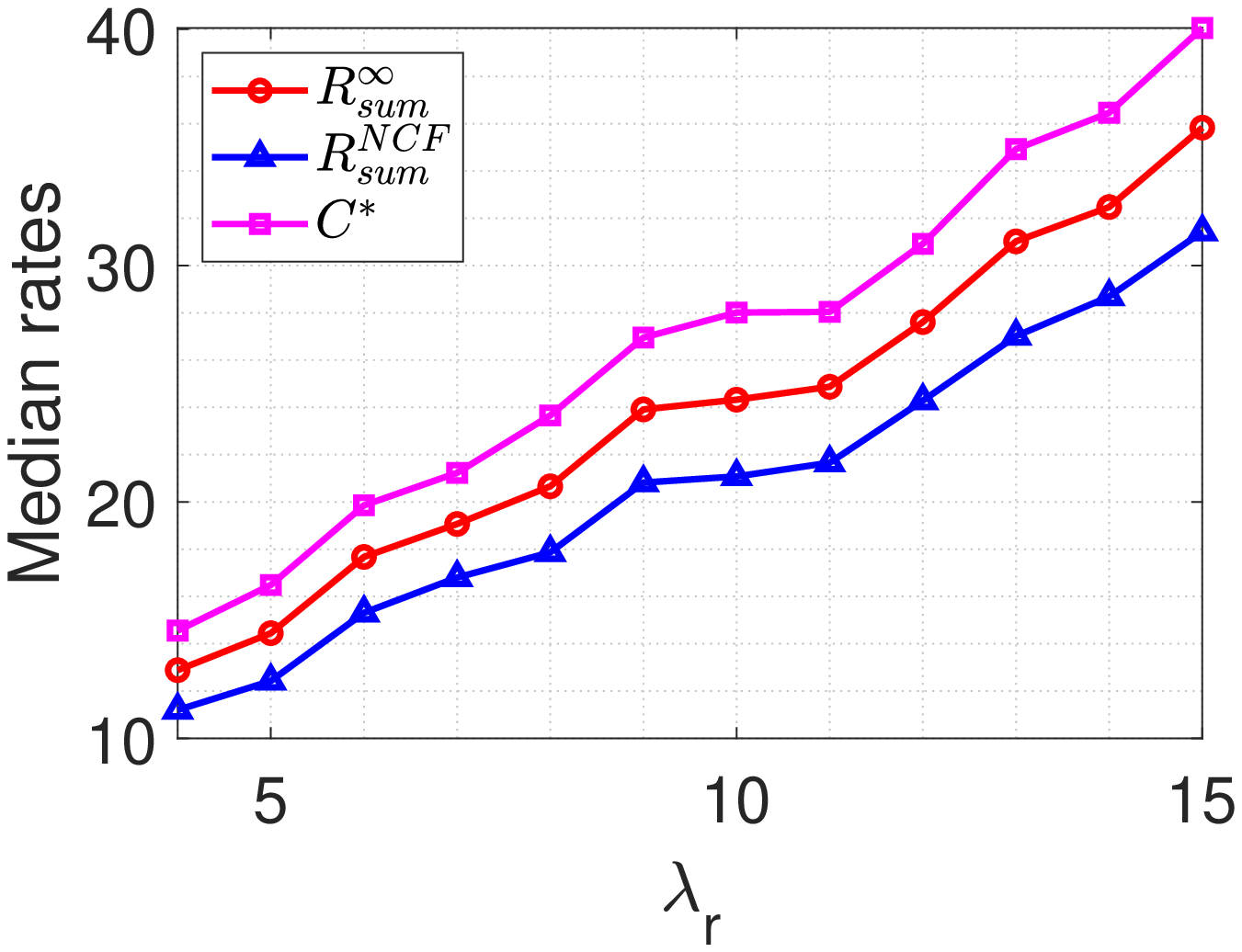}
\caption{$\beta = 2.5,$ $\lambda_r = \lambda_u^2.$}
\label{fig:plot2d}
\end{subfigure}%
\begin{subfigure}[t]{0.33\textwidth}
\center
\includegraphics[scale=0.33]{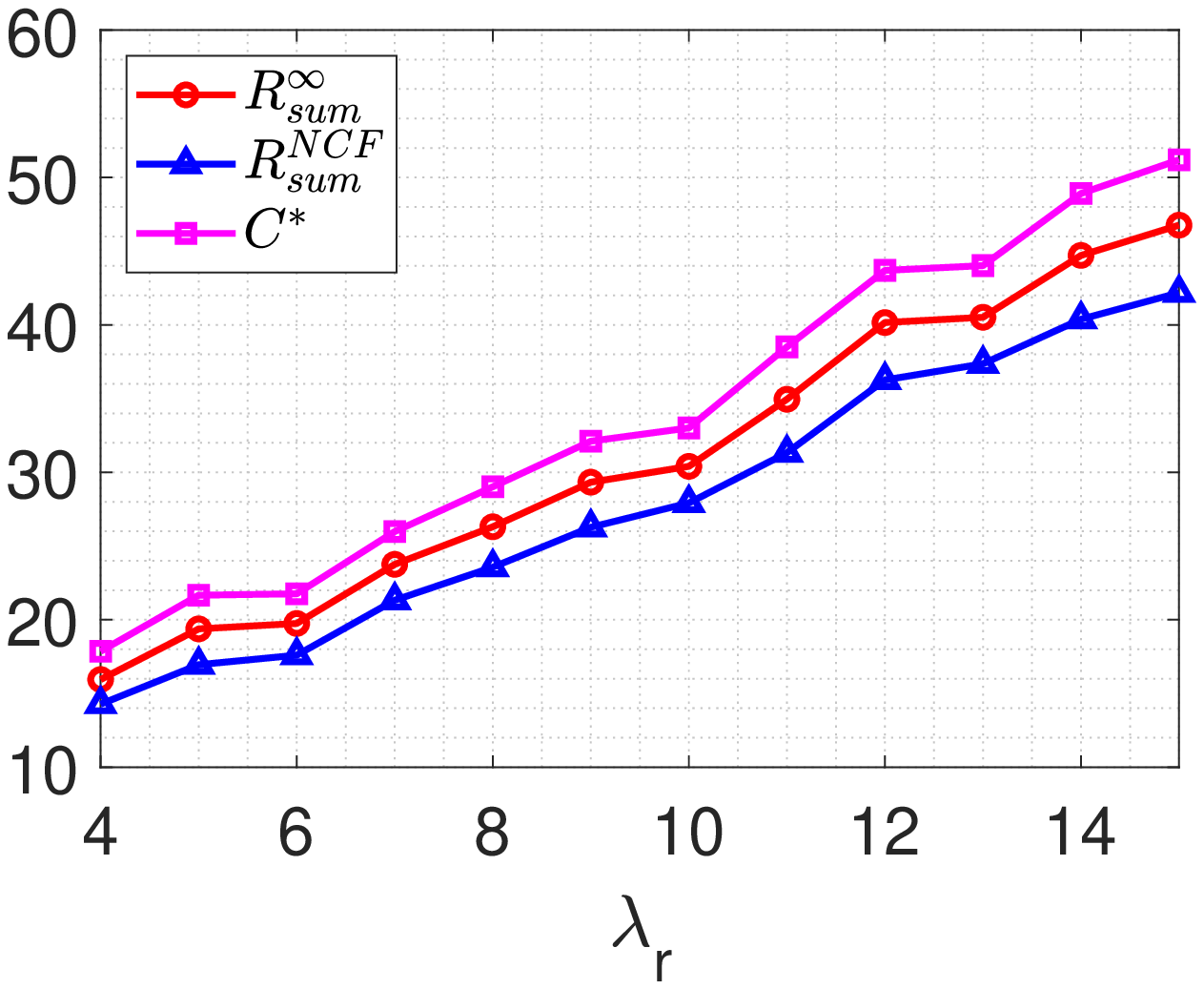}
\caption{$\beta = 3.5,$ $\lambda_r = \lambda_u^2.$}
\label{fig:plot2e}
\end{subfigure}
\begin{subfigure}[t]{0.33\textwidth}
\center
\includegraphics[scale=0.33]{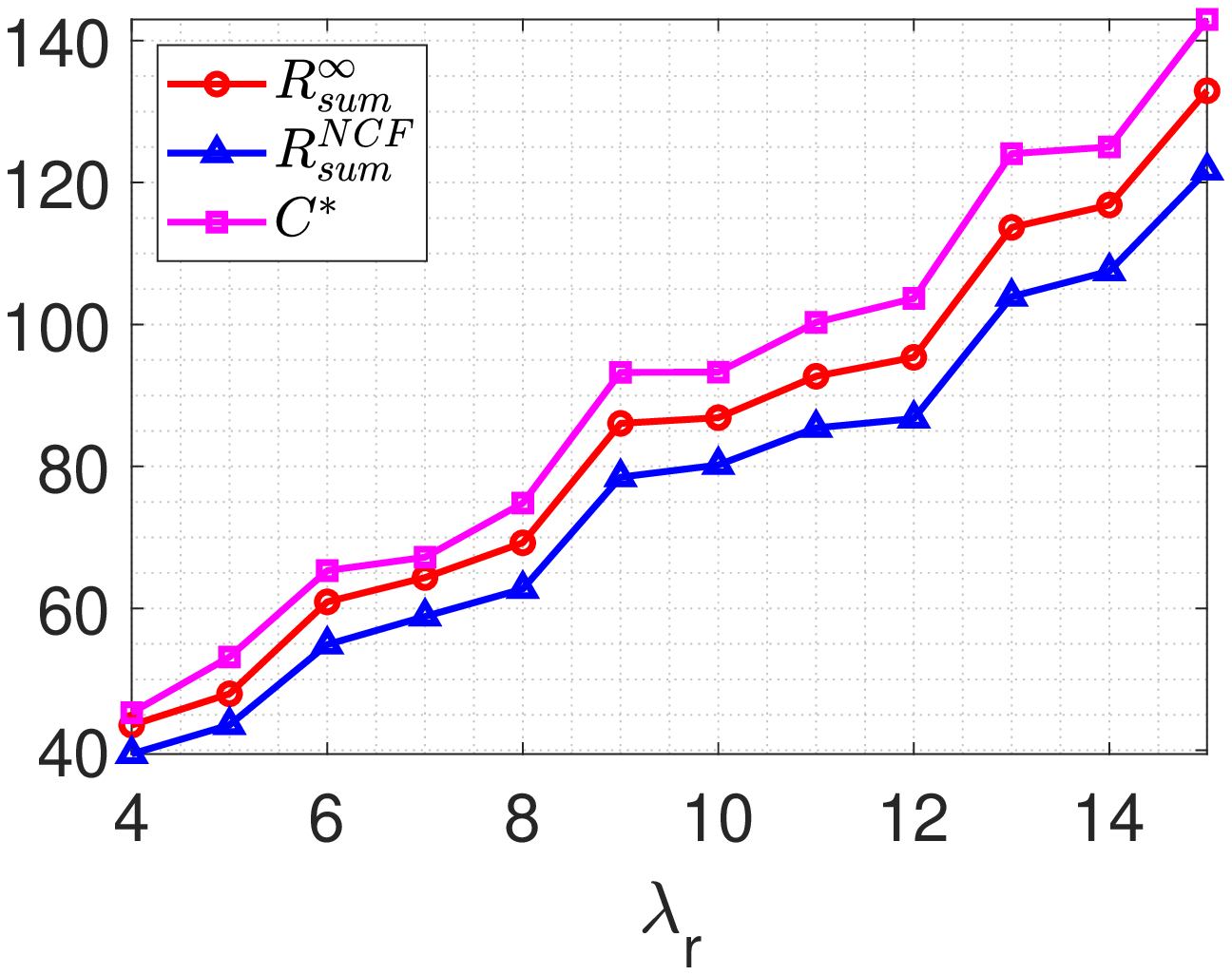}
\caption{Multipath, $\lambda_r = \lambda_u^2.$}
\label{fig:plot2f}
\end{subfigure}
\vskip\baselineskip
\begin{subfigure}[t]{0.33\textwidth}
\center
\includegraphics[scale=0.33]{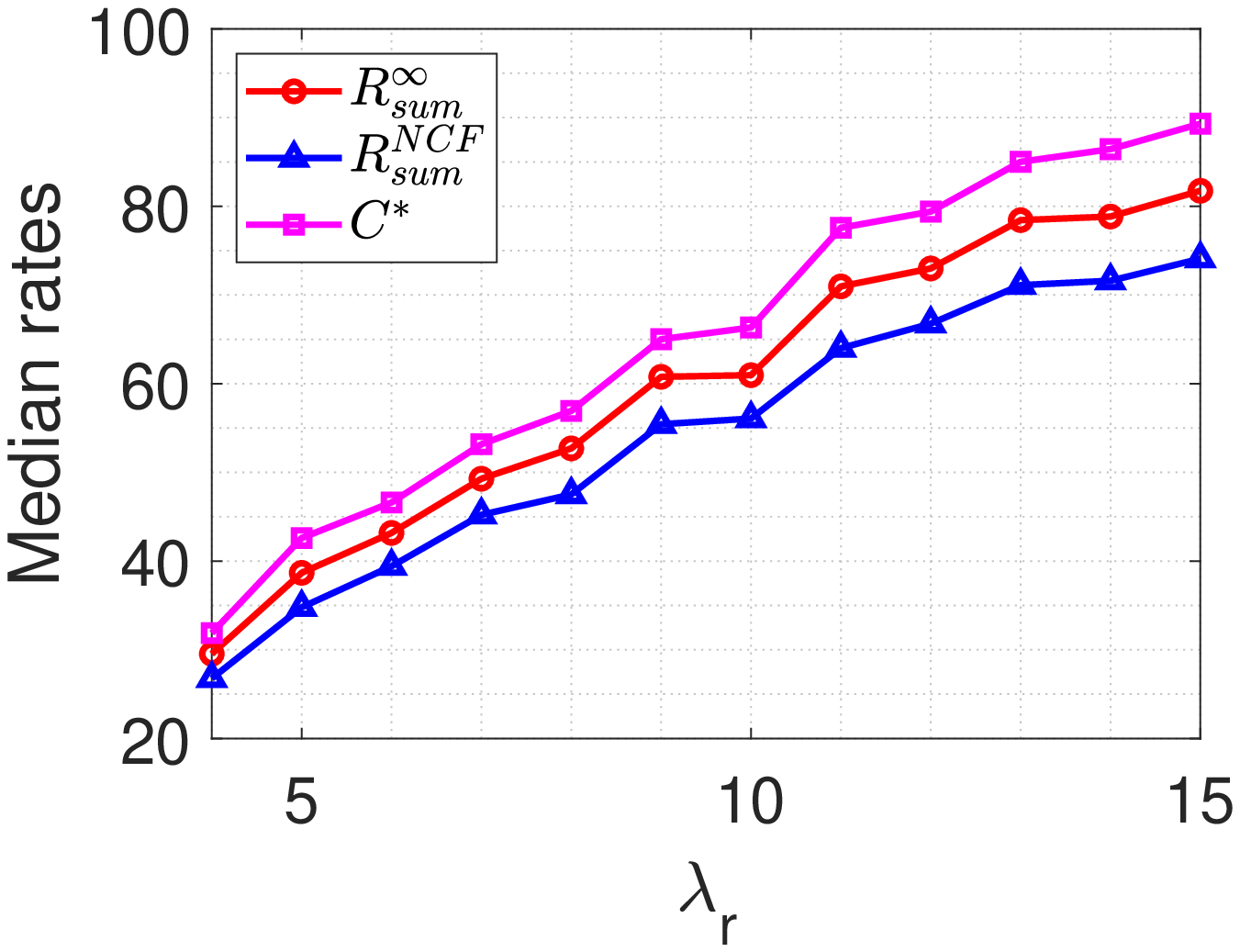}
\caption{$\beta = 2.5,$ $\lambda_u = 10.$}
\label{fig:plot2g}
\end{subfigure}%
\begin{subfigure}[t]{0.33\textwidth}
\center
\includegraphics[scale=0.33]{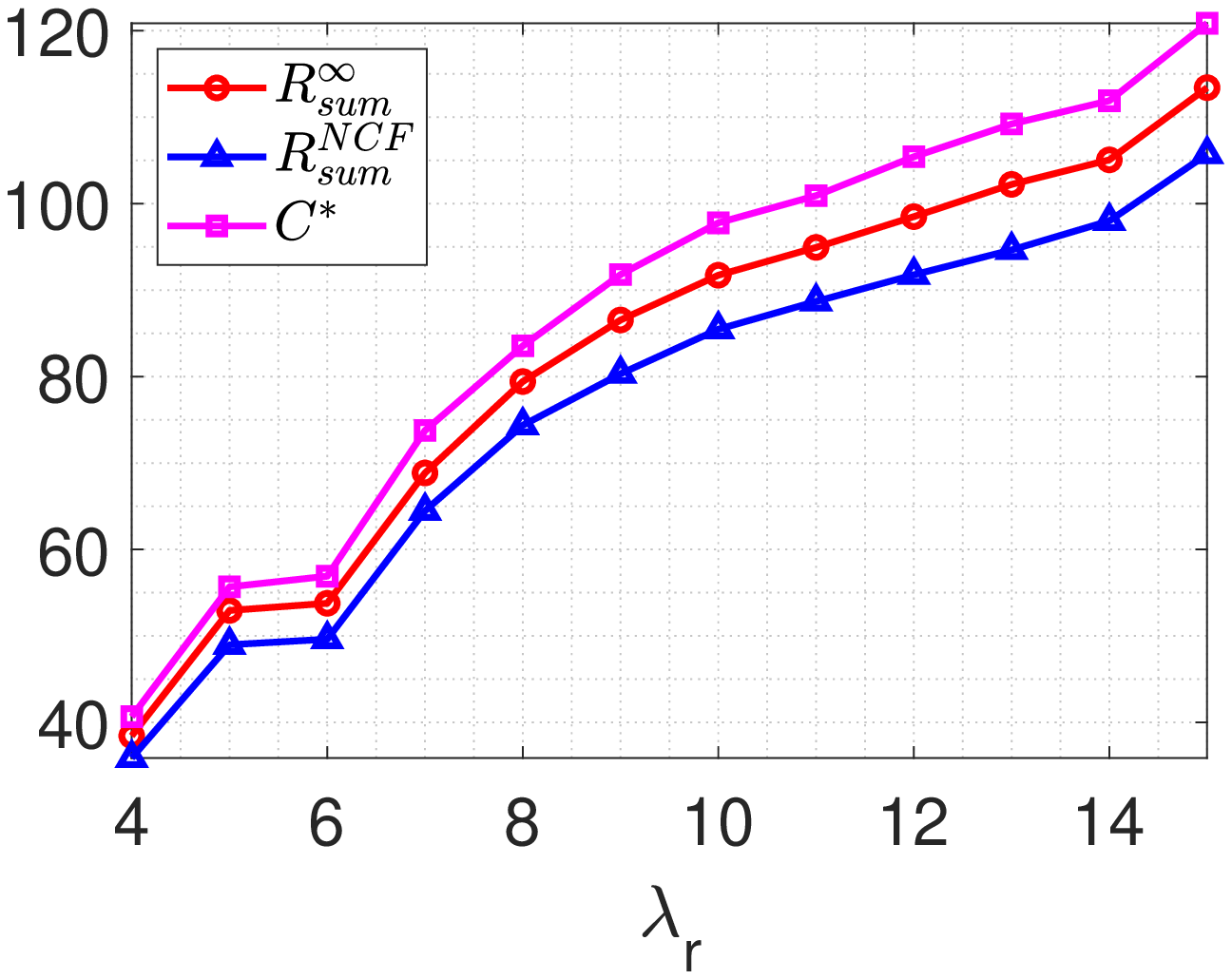}
\caption{$\beta = 3.5,$ $\lambda_u = 10.$}
\label{fig:plot2h}
\end{subfigure}
\begin{subfigure}[t]{0.33\textwidth}
\center
\includegraphics[scale=0.33]{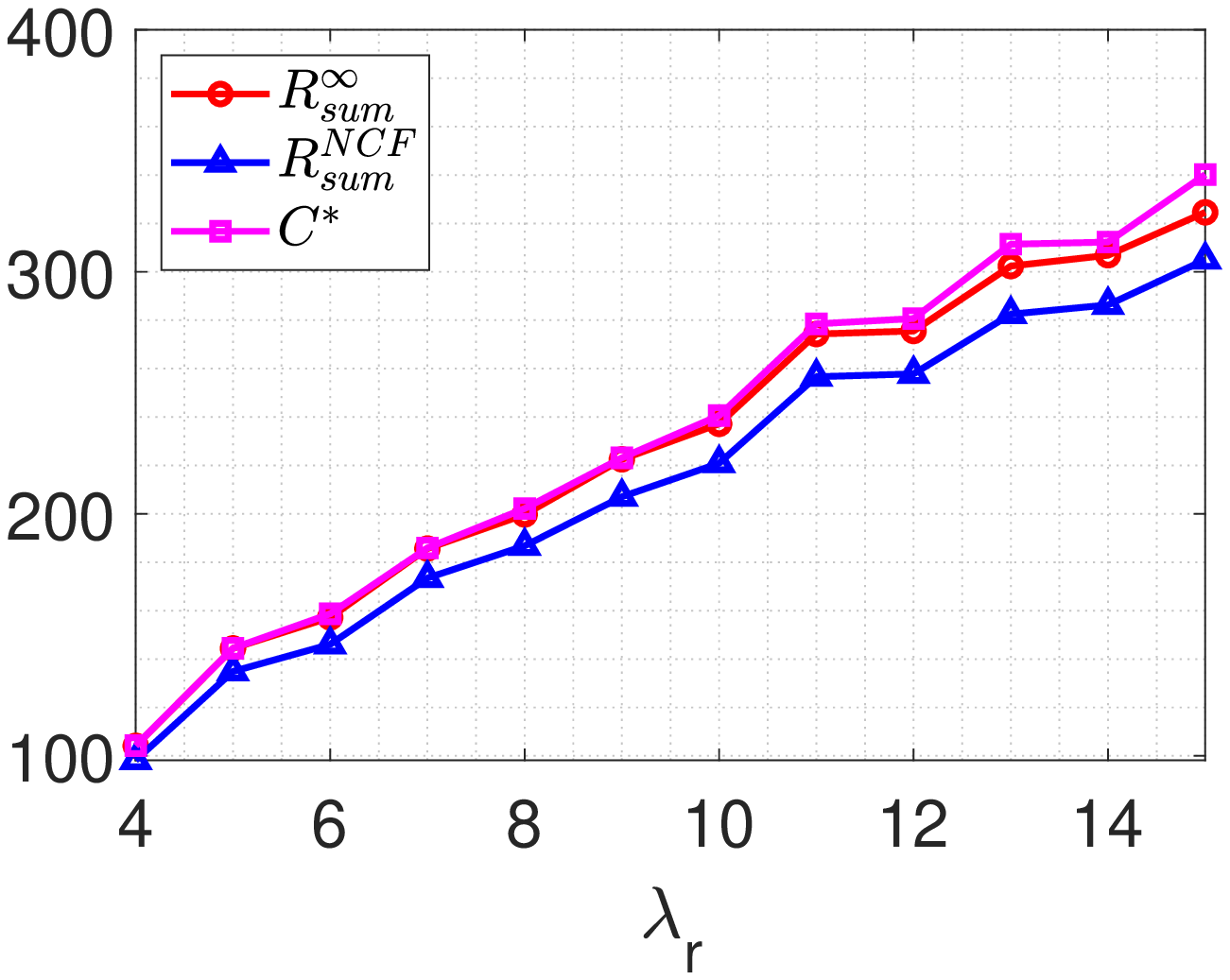}
\caption{Multipath, $\lambda_u = 10$.}
\label{fig:plot2i}
\end{subfigure}
\caption{MIMO uplink capacity scaling under stochastic geometry.}\label{fig:plotulmimo}
\end{figure}

\begin{figure}[h]
\begin{subfigure}[t]{0.33\textwidth}
\center
\includegraphics[scale=0.33]{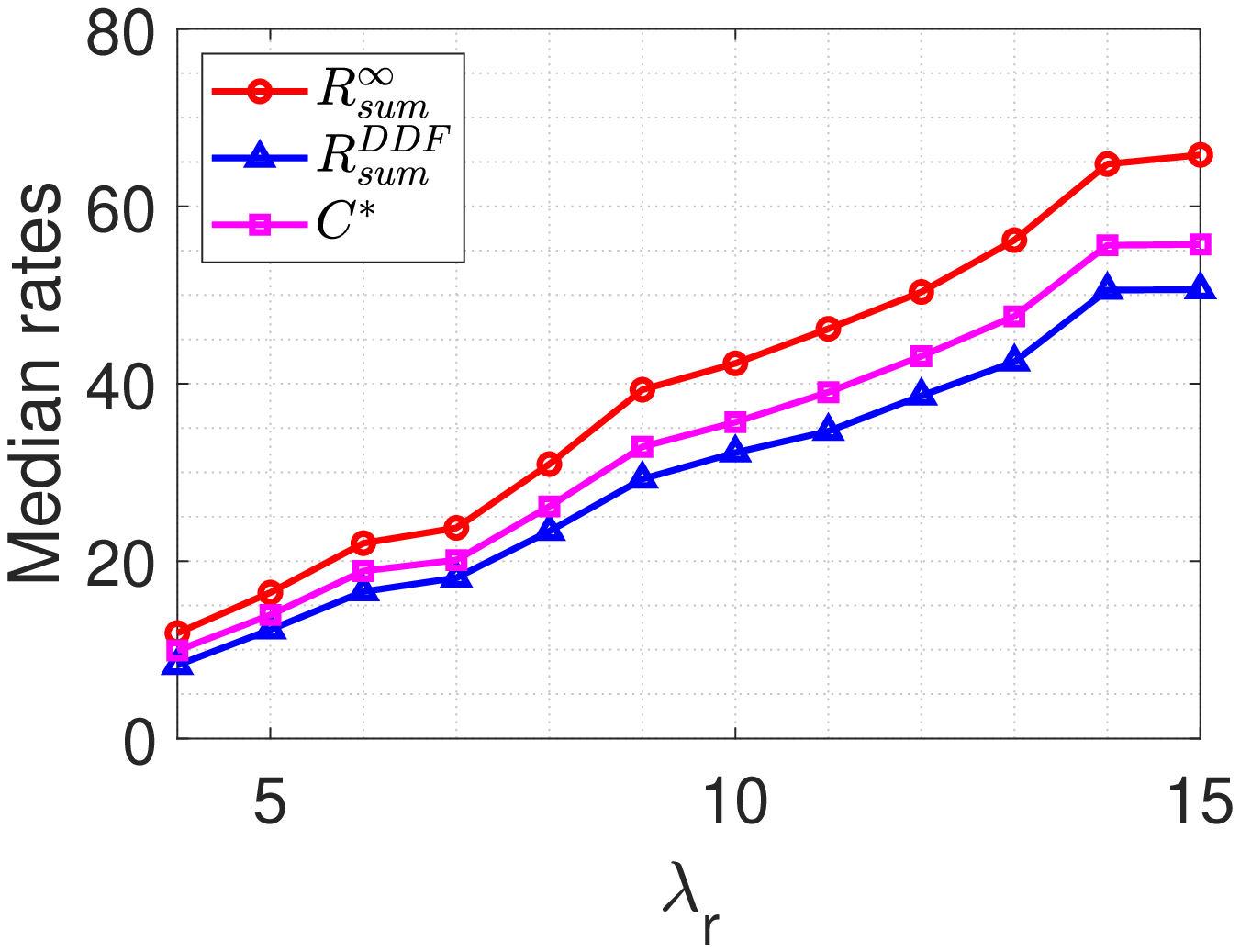}
\caption{$\beta = 2.5,$ $\lambda_r = 2\lambda_u.$}
\label{fig:plot2a}
\end{subfigure}%
\begin{subfigure}[t]{0.33\textwidth}
\center
\includegraphics[scale=0.33]{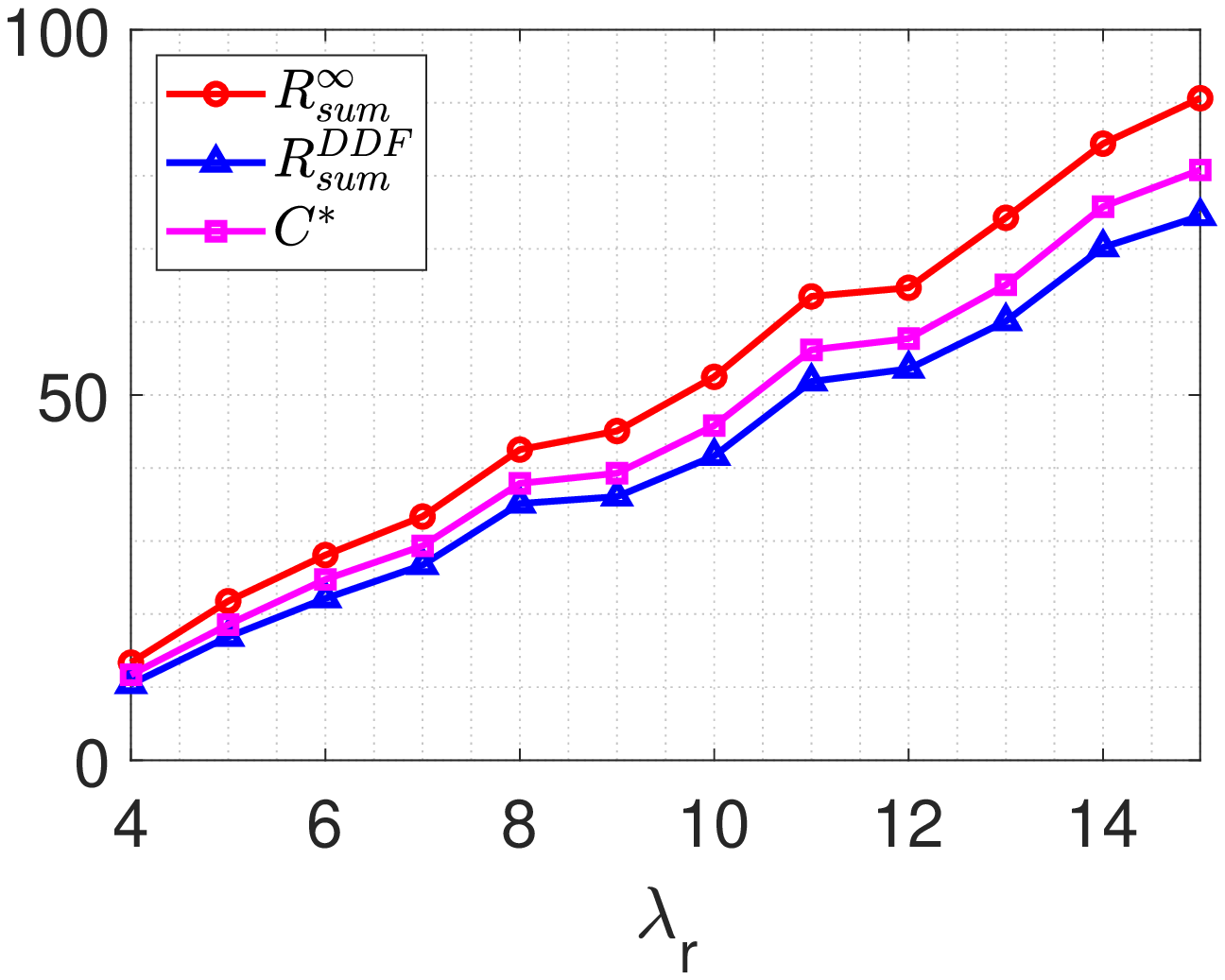}
\caption{$\beta = 3.5,$ $\lambda_r = 2\lambda_u.$}
\label{fig:plot2b}
\end{subfigure}
\begin{subfigure}[t]{0.33\textwidth}
\center
\includegraphics[scale=0.33]{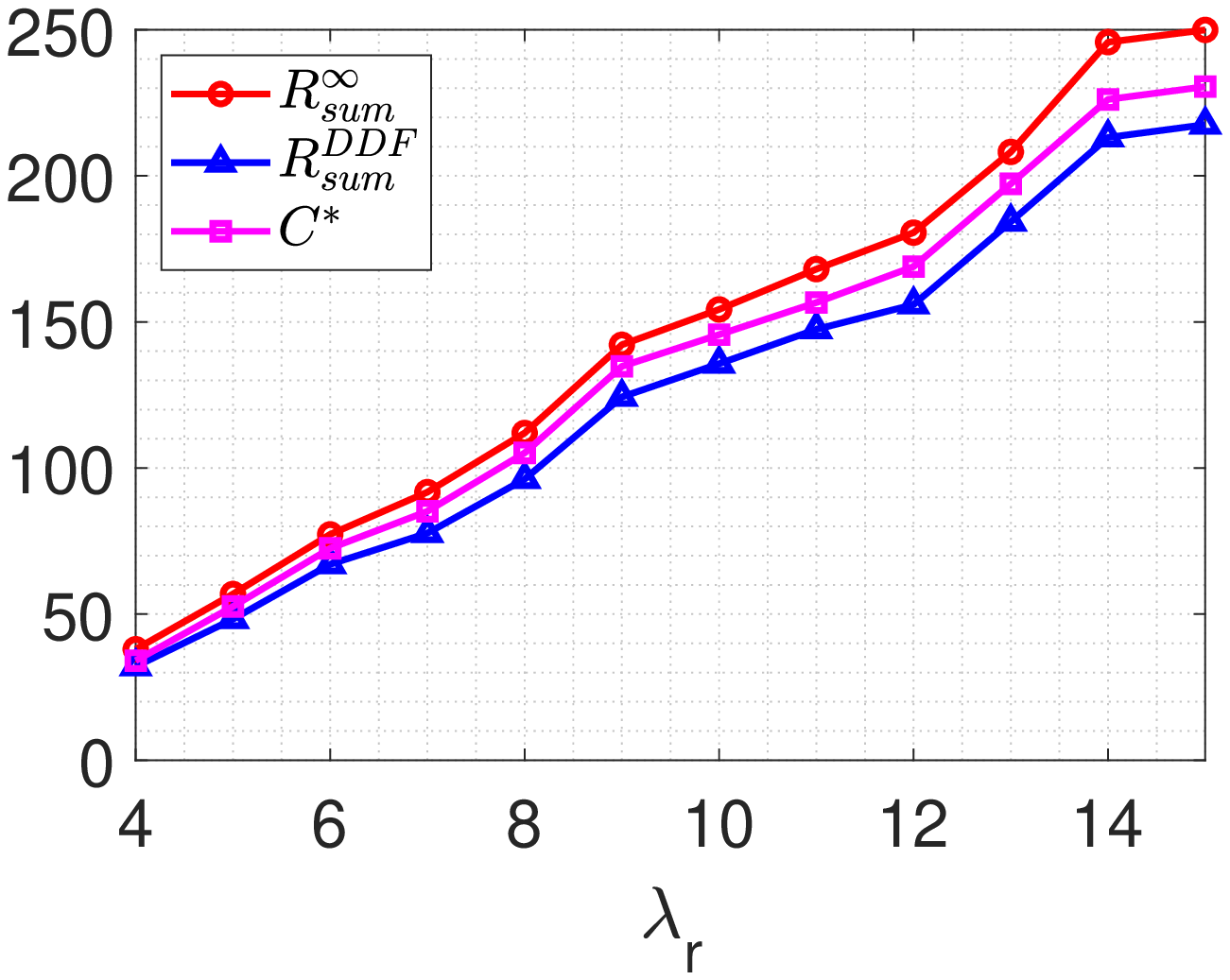}
\caption{Multipath, $\lambda_r = 2\lambda_u.$}
\label{fig:plot2c}
\end{subfigure}
\vskip\baselineskip
\begin{subfigure}[t]{0.33\textwidth}
\center
\includegraphics[scale=0.33]{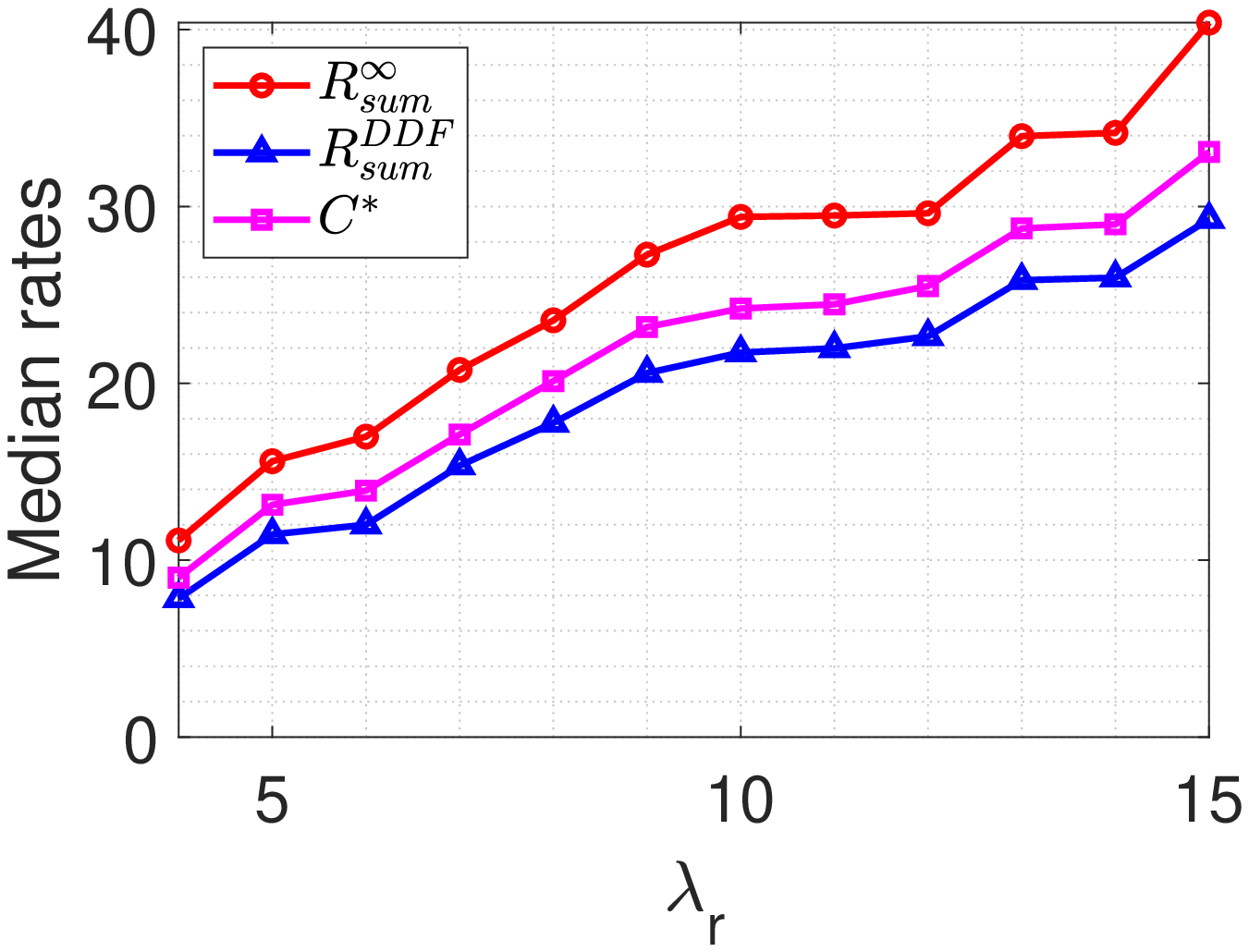}
\caption{$\beta = 2.5,$ $\lambda_r = \lambda_u^2.$}
\label{fig:plot2d}
\end{subfigure}%
\begin{subfigure}[t]{0.33\textwidth}
\center
\includegraphics[scale=0.33]{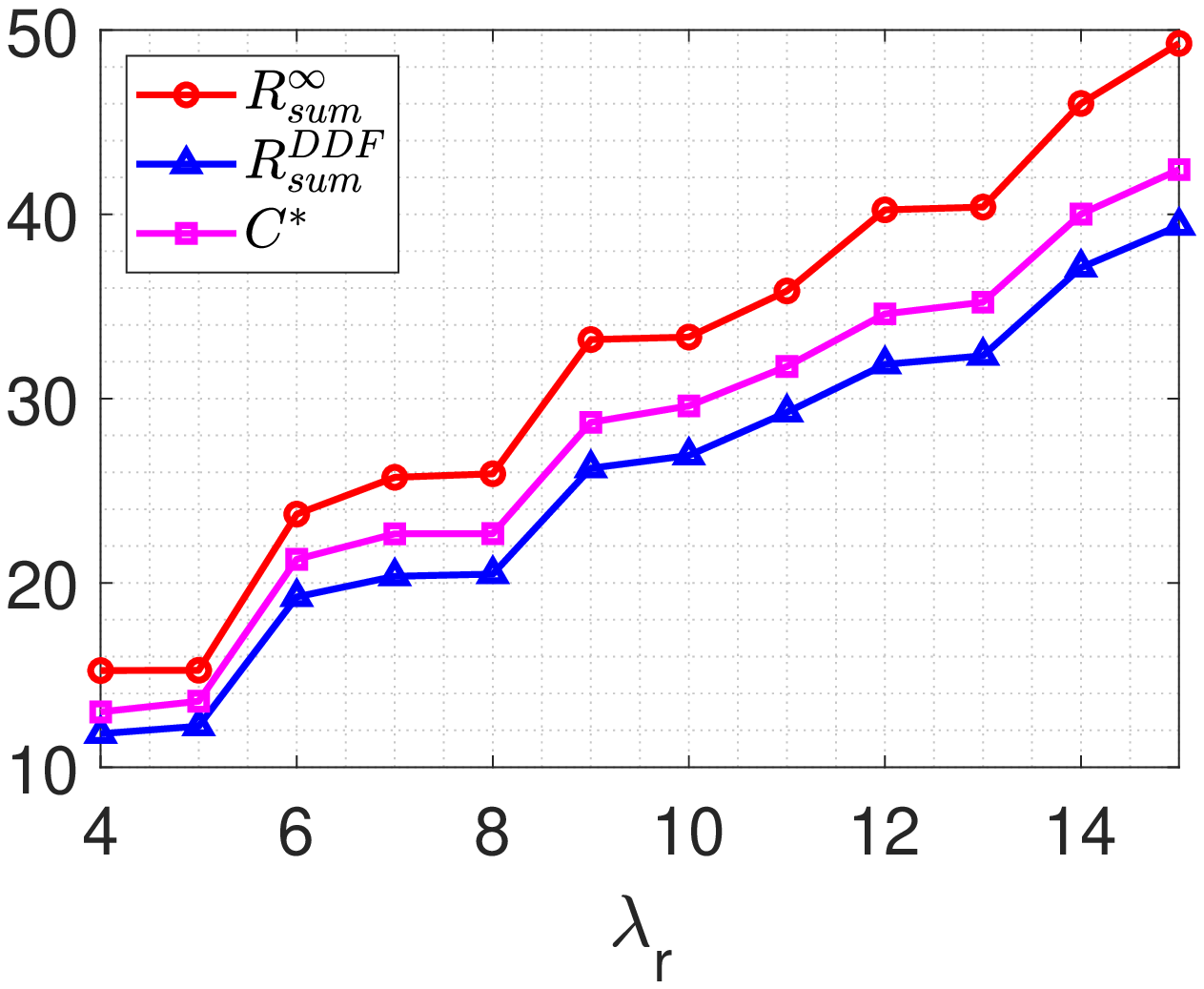}
\caption{$\beta = 3.5,$ $\lambda_r = \lambda_u^2.$}
\label{fig:plot2e}
\end{subfigure}
\begin{subfigure}[t]{0.33\textwidth}
\center
\includegraphics[scale=0.33]{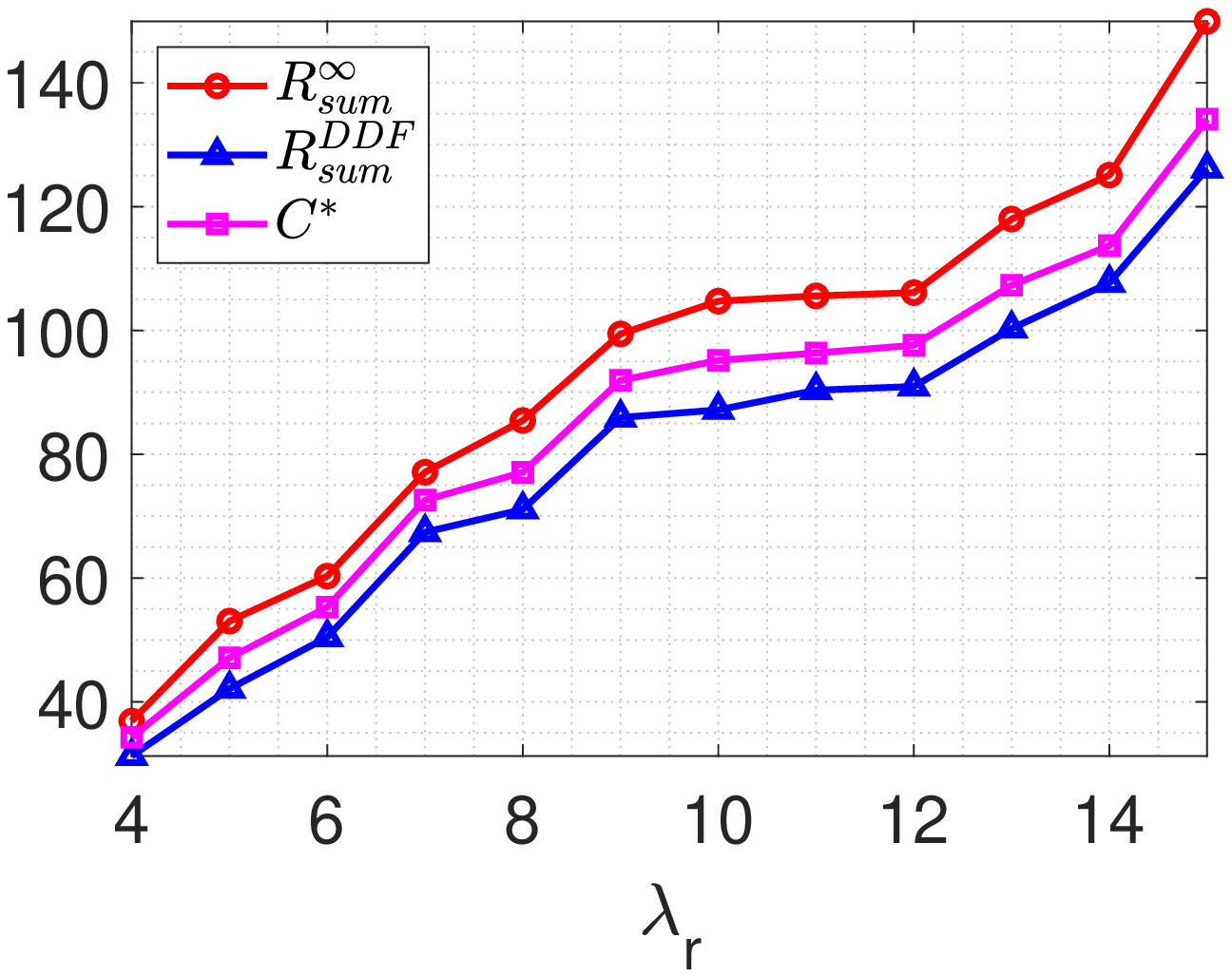}
\caption{Multipath, $\lambda_r = \lambda_u^2.$}
\label{fig:plot2f}
\end{subfigure}
\vskip\baselineskip
\begin{subfigure}[t]{0.33\textwidth}
\center
\includegraphics[scale=0.33]{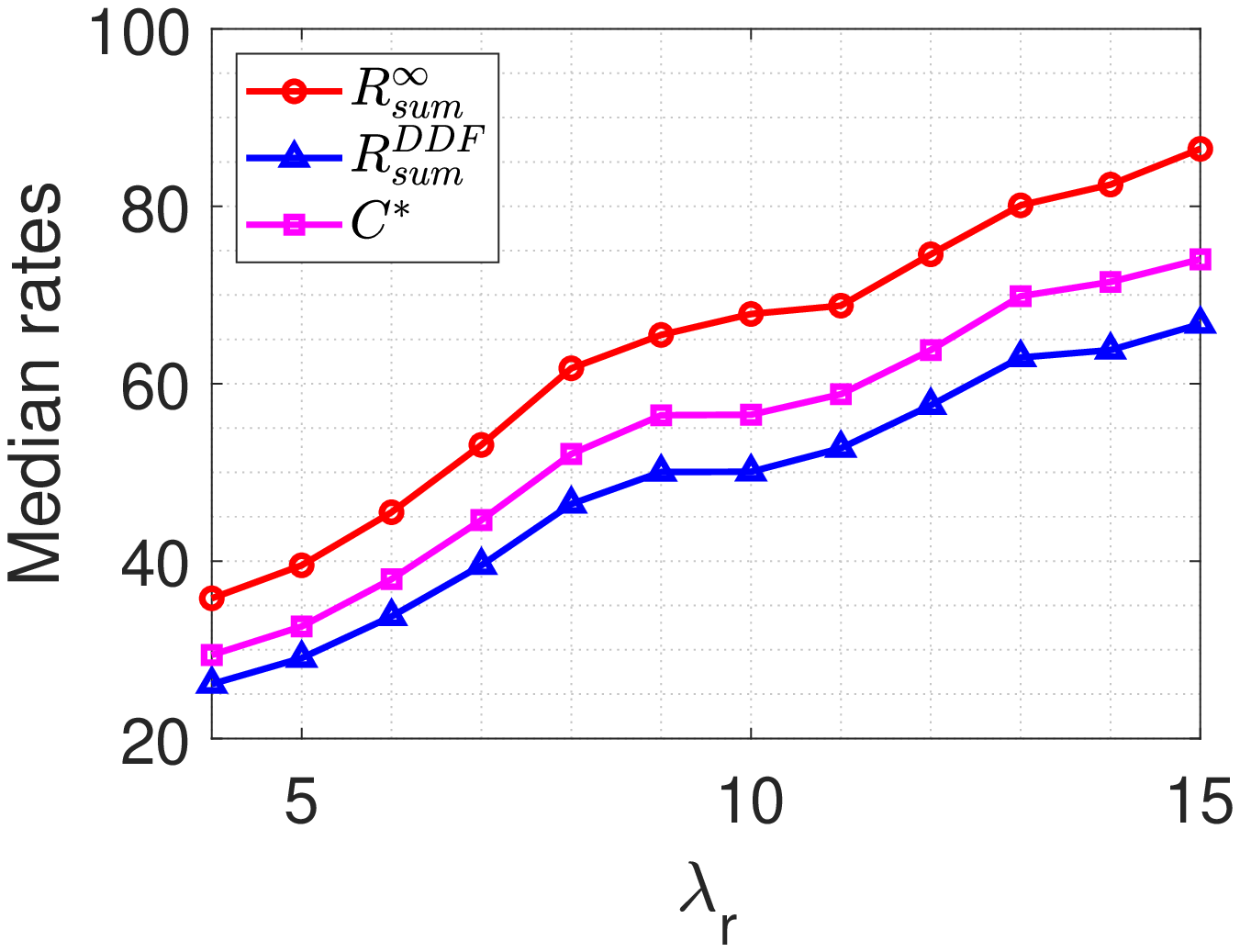}
\caption{$\beta = 2.5,$ $\lambda_u = 10.$}
\label{fig:plot2g}
\end{subfigure}%
\begin{subfigure}[t]{0.33\textwidth}
\center
\includegraphics[scale=0.33]{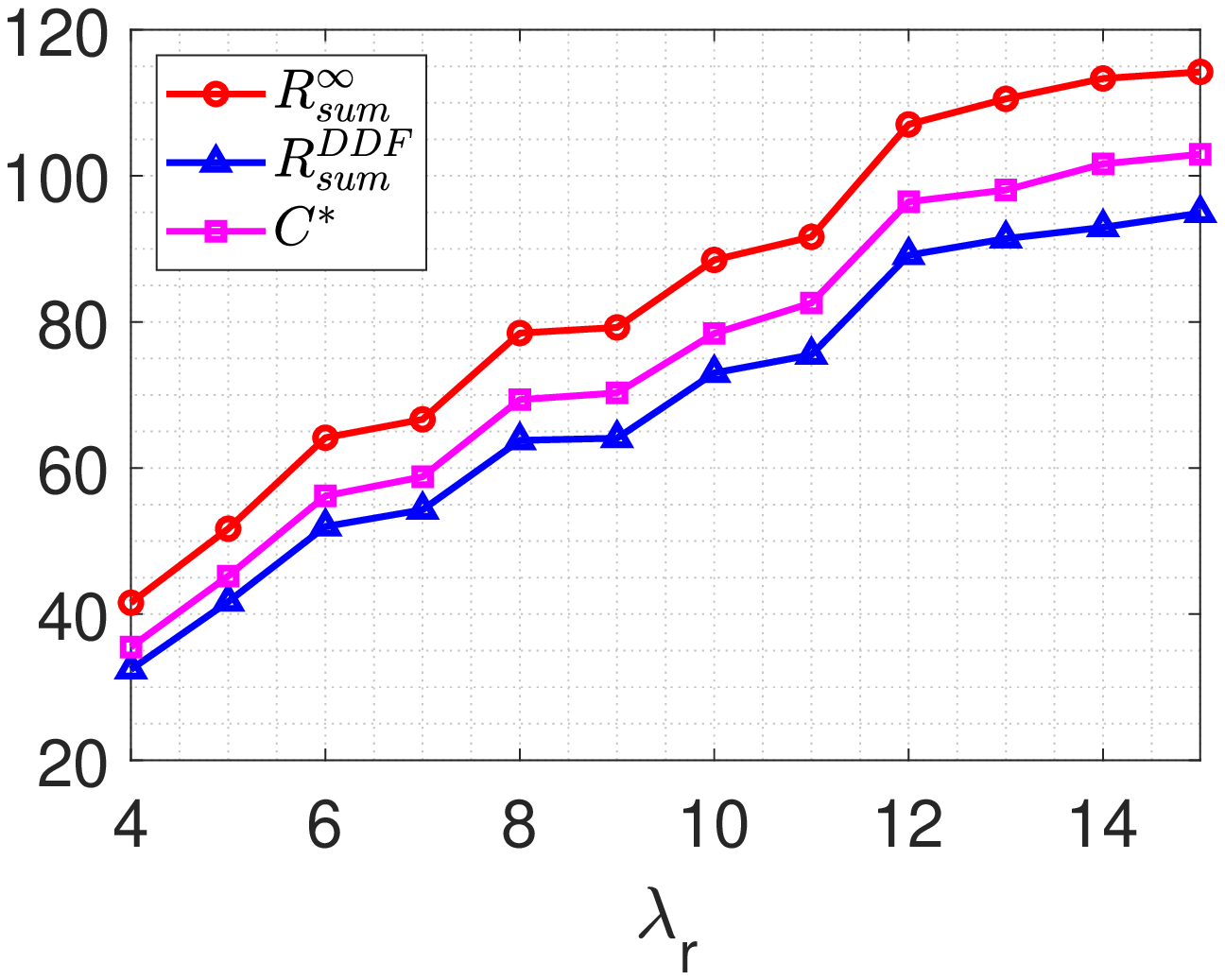}
\caption{$\beta = 3.5,$ $\lambda_u = 10.$}
\label{fig:plot2h}
\end{subfigure}
\begin{subfigure}[t]{0.33\textwidth}
\center
\includegraphics[scale=0.33]{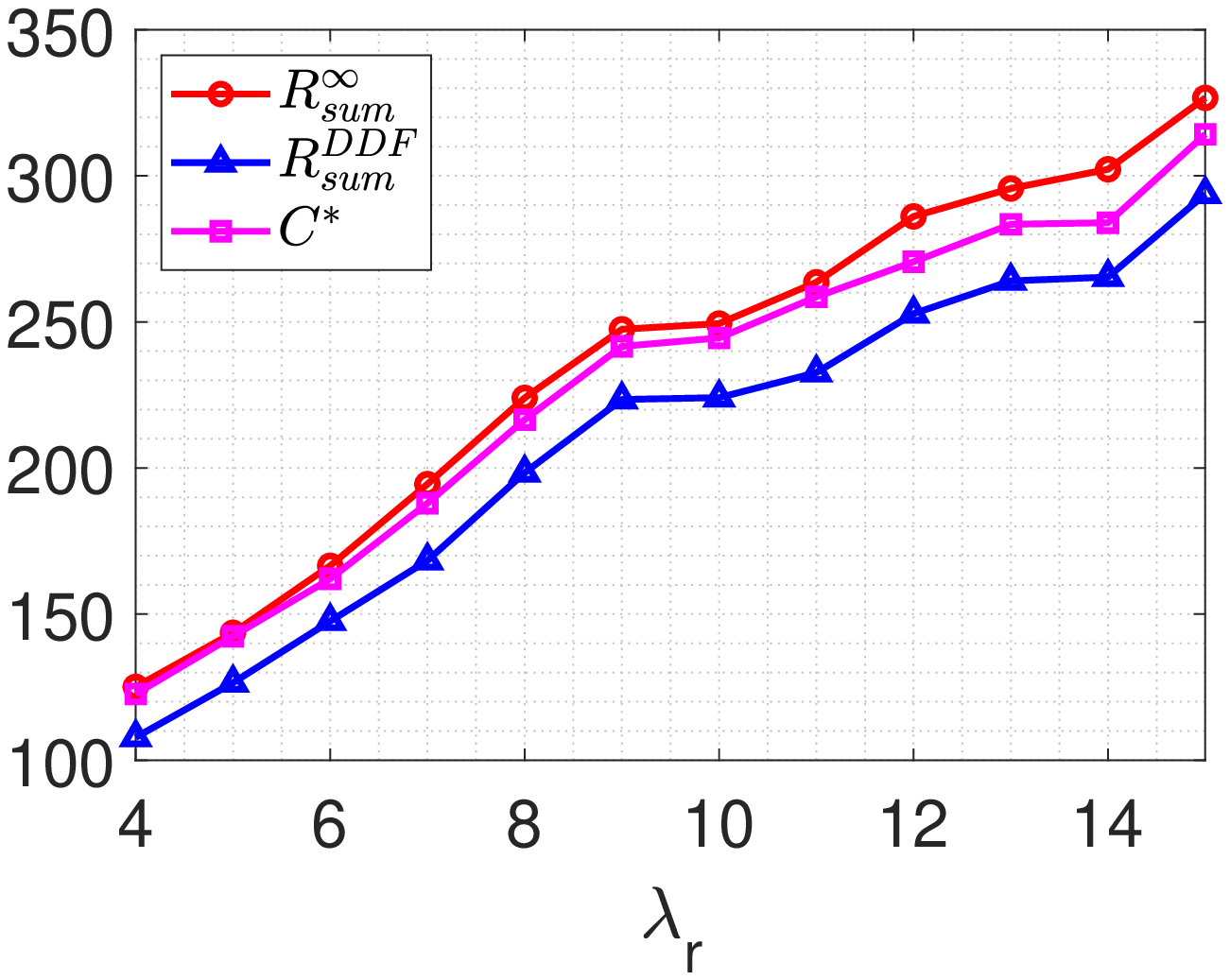}
\caption{Multipath, $\lambda_u = 10$.}
\label{fig:plot2i}
\end{subfigure}
\caption{MIMO downlink capacity scaling under stochastic geometry.}\label{fig:plotdlmimo}
\end{figure}

We have the following result on the large-network asymptotics of the MIMO C-RAN.
\begin{prop}\label{prop:mimoscaling}
Let the entries of the $N_rL\times N_uK$ channel gain matrix $G$ be distributed as i.i.d.\@ $\N(0,1),$ and let $\sigma^2 = \sigma^2(N_r, L, N_u, K) > 0.$ 
If $N_rL\to\infty$ such that $N_rL/N_uK\to\rho \in(1, \infty]$ and $N_rL/\sigma^2\to \infty,$ and if $N_u$ is kept fixed, then 
\[
R_\mathrm{sum}^\infty\sim\frac{N_uK}{2}\log(N_rL)
\]
and for every choice of $\Gamma_1, \ldots, \Gamma_K,$
\[
R_\mathrm{sum}^\text{NCF} = C^*- \frac{N_rL}{2}\log\left(1+\frac{1}{\sigma^2}\right)\sim \frac{N_uK}{2}\log(N_rL/\sigma^2),
\]
a.s.\@ in $G.$ 
Similarly, let the entries of the $N_uK\times N_rL$ channel gain matrix $H$ be distributed as i.i.d.\@ $\N(0,1),$ and let $\sigma^2 = \sigma^2(N_r, L, N_u, K) > 0.$ 
If $L\to\infty$ such that $N_rL/N_uK\to\rho \in(1, \infty]$ and $L/(N_u\sigma^2)\to \infty,$ and if $N_r$ is kept fixed, then for every choice of $\Gamma_1, \ldots, \Gamma_L,$
\[
R_\mathrm{sum}^\text{DDF} = C^*- \frac{N_uK}{2}\log\left(1+\frac{1}{\sigma^2}\right)\sim \frac{N_uK}{2}\log(L/\sigma^2) - \frac{N_uK}{2}\log\left(1+\frac{1}{\sigma^2}\right),
\]
a.s.\@ in $H.$
\end{prop}
\begin{remark}
Comparing Proposition~\ref{prop:mimoscaling} with Theorems~\ref{theorem:pggt} and \ref{theorem:phht} shows that under the rich scattering model, the large network asymptotics of the MIMO C-RAN is the same as if there were $N_uK$ users and $N_rL$ relays. Thus, Tables~\ref{table:uplinkscaling} and \ref{table:downlinkscaling} can be easily generalized to the MIMO case through appropriate choices of $\sigma^2(N_r, L, N_u, K).$
\end{remark}
\begin{remark}
In Proposition~\ref{prop:mimoscaling}, $N_u$ and $N_r$ are held fixed for uplink and downlink, respectively, so that the scaling results remain invariant to the power allocation across the local antennas at each user and at each relay, respectively.
\end{remark}
Similar to Sections~\ref{upscale} and \ref{downscale}, Figs.~\ref{fig:plotulmimo} and \ref{fig:plotdlmimo} plot $R_\mathrm{sum}^\text{NCF},$ $R_\mathrm{sum}^\text{DDF},$ $C^*,$ and $R_\mathrm{sum}^\infty$ under a stochastic geometry model. Both $N_r$ and $N_u$ are kept fixed ($N_r = N_u = 4$) for these simulations. For the downlink, as before, we plot an upper bound on $R_\mathrm{sum}^\infty,$ obtained by a grid search over eligible values of $Q$ in Remark~\ref{rem:mimodownlinksumrateduality}. At each node (user or relay), the shadowing effect is considered to be the same across all local antennas, while the small-scale fading is taken as i.i.d.  

\begin{figure}[h]
\begin{subfigure}[t]{0.33\textwidth}
\center
\includegraphics[scale=0.33]{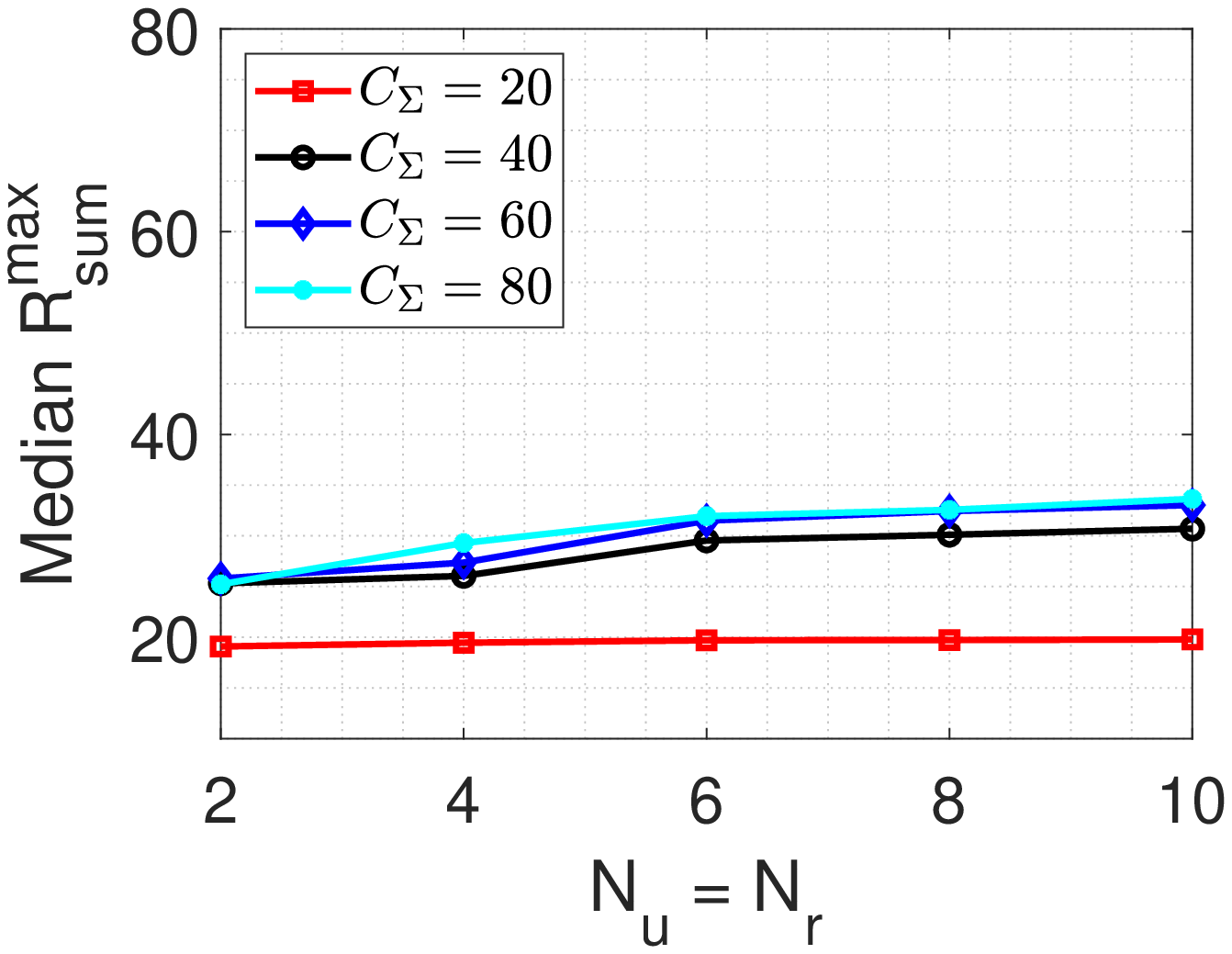}
\caption{$\beta = 2.5,$ $K = 4, L = 6.$}
\label{fig:plot2a}
\end{subfigure}%
\begin{subfigure}[t]{0.33\textwidth}
\center
\includegraphics[scale=0.33]{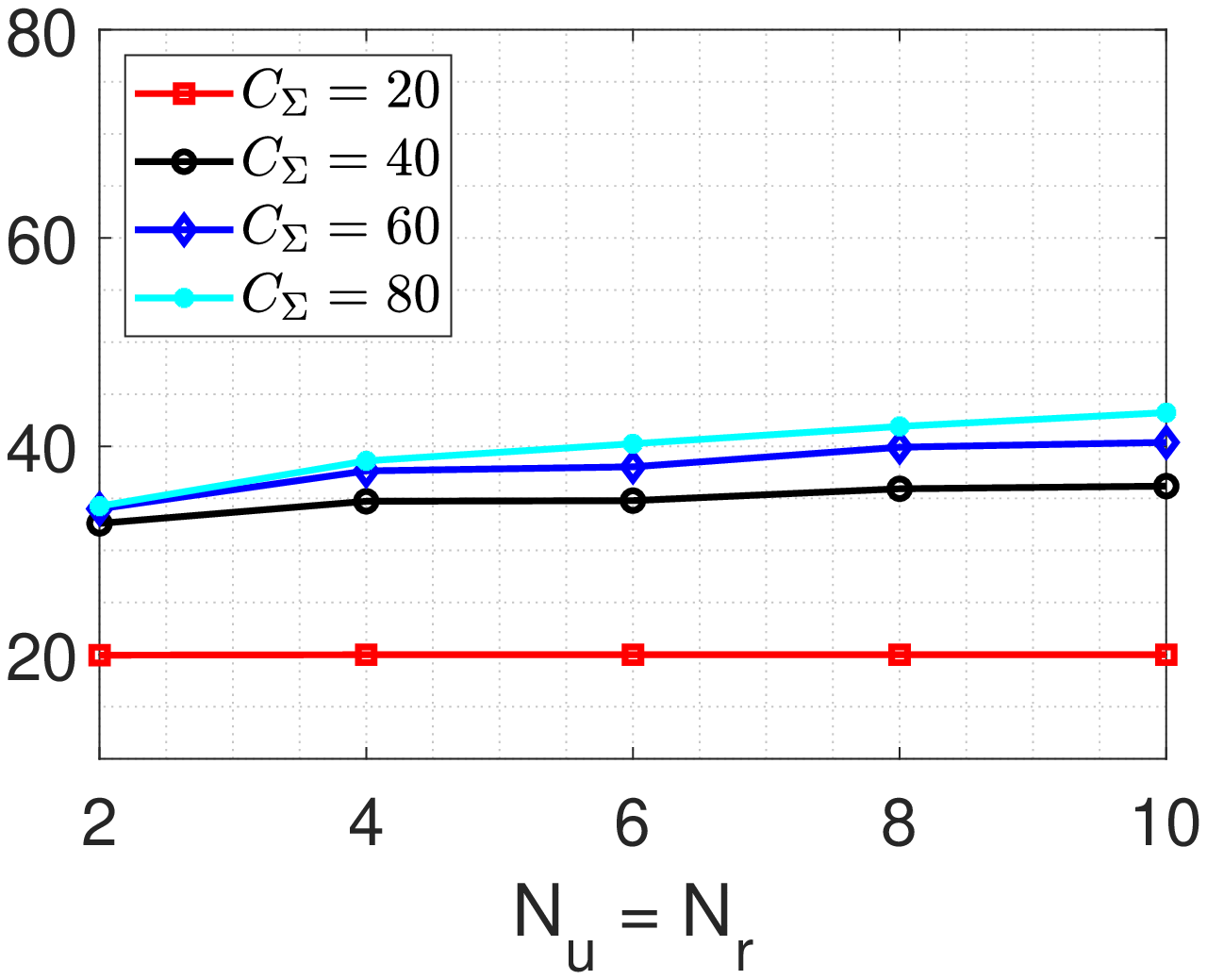}
\caption{$\beta = 3.5,$ $K = 4, L = 6.$}
\label{fig:plot2b}
\end{subfigure}
\begin{subfigure}[t]{0.33\textwidth}
\center
\includegraphics[scale=0.33]{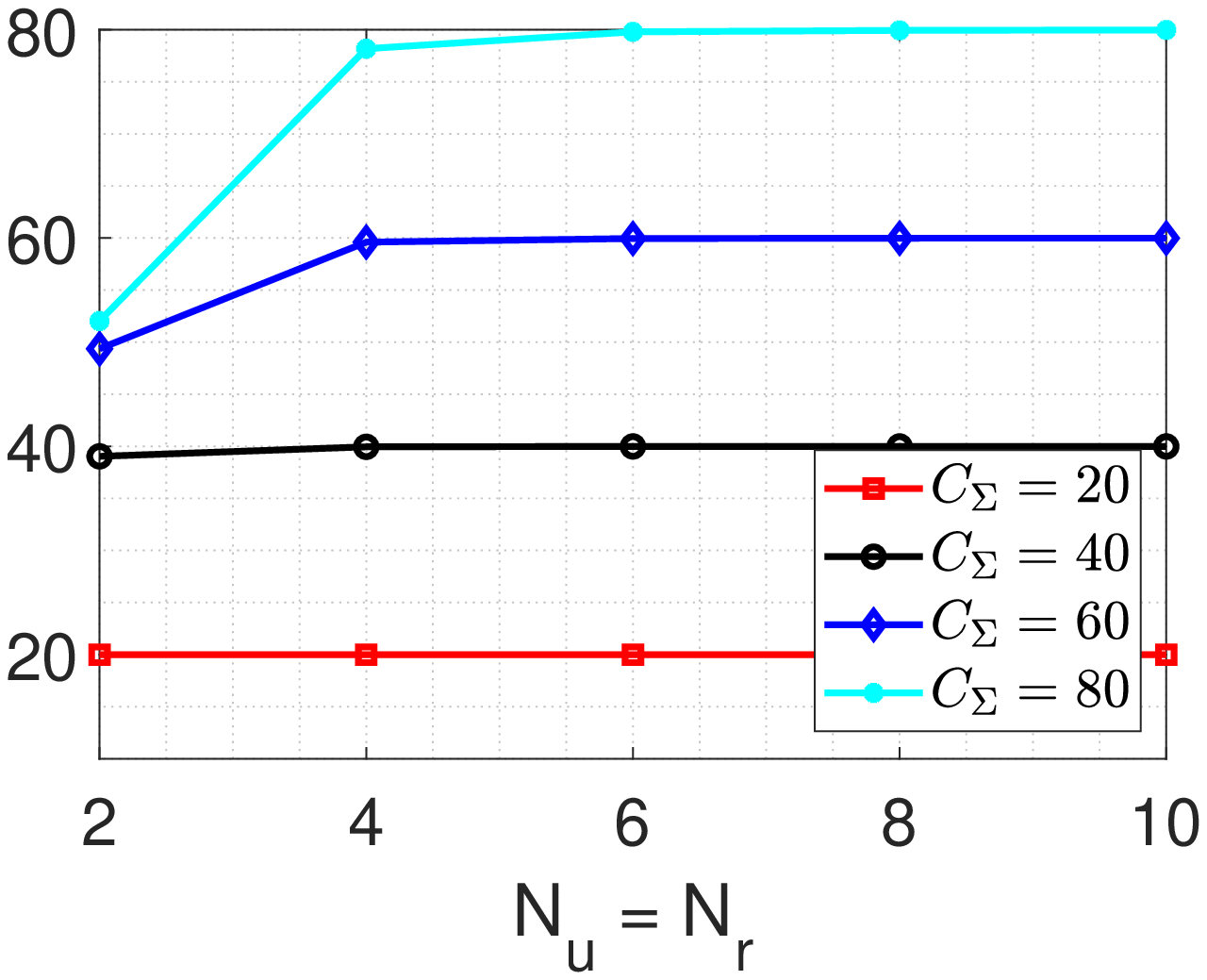}
\caption{Multipath, $K = 4, L = 6.$}
\label{fig:plot2c}
\end{subfigure}
%\vskip\baselineskip
\caption{MIMO uplink capacity scaling with antenna number under fixed sum-fronthaul.}\label{fig:plotulmimoCsum}
\end{figure}

\begin{figure}[h]
\begin{subfigure}[t]{0.33\textwidth}
\center
\includegraphics[scale=0.33]{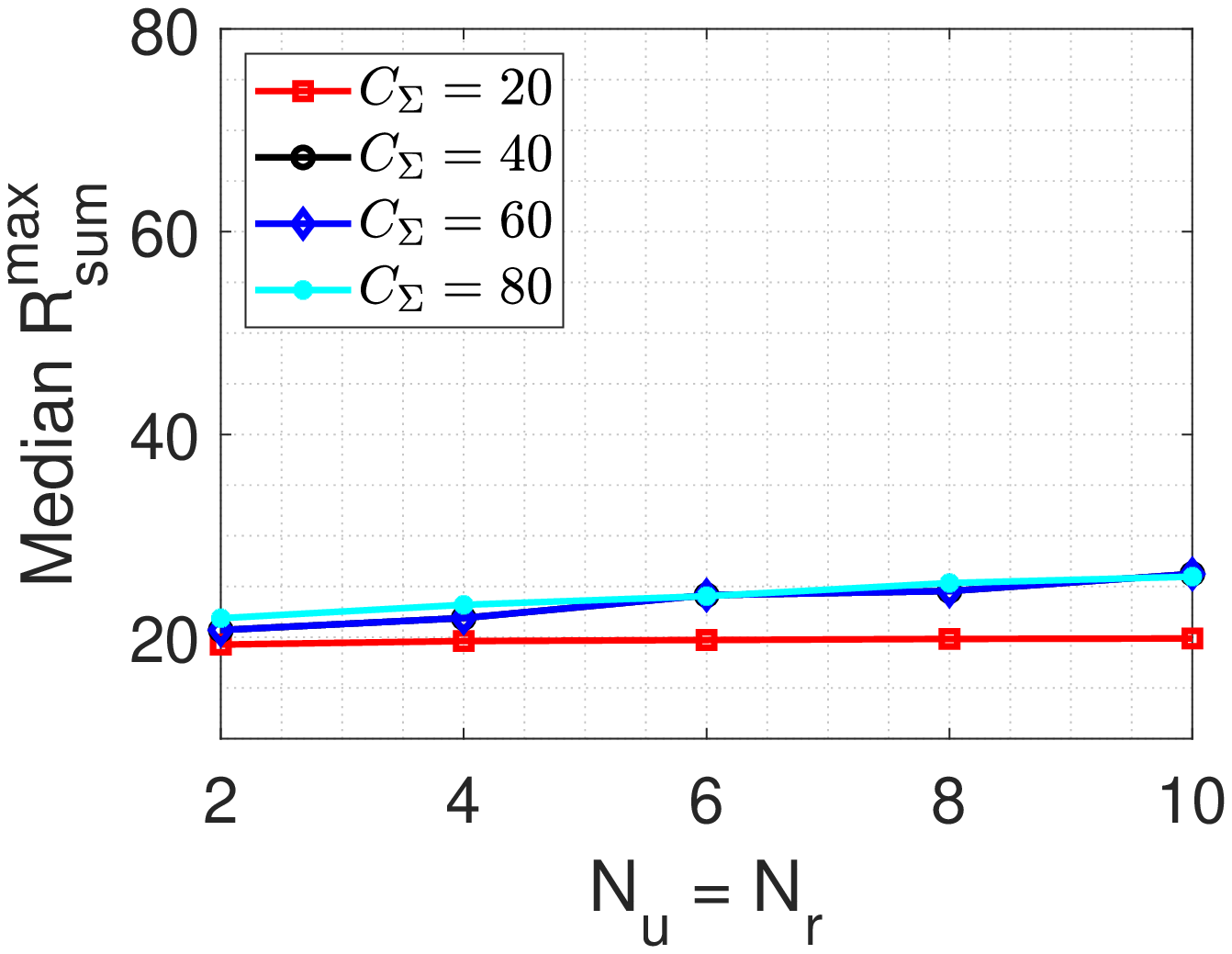}
\caption{$\beta = 2.5,$ $K = 4, L = 6.$}
\label{fig:plot2a}
\end{subfigure}%
\begin{subfigure}[t]{0.33\textwidth}
\center
\includegraphics[scale=0.33]{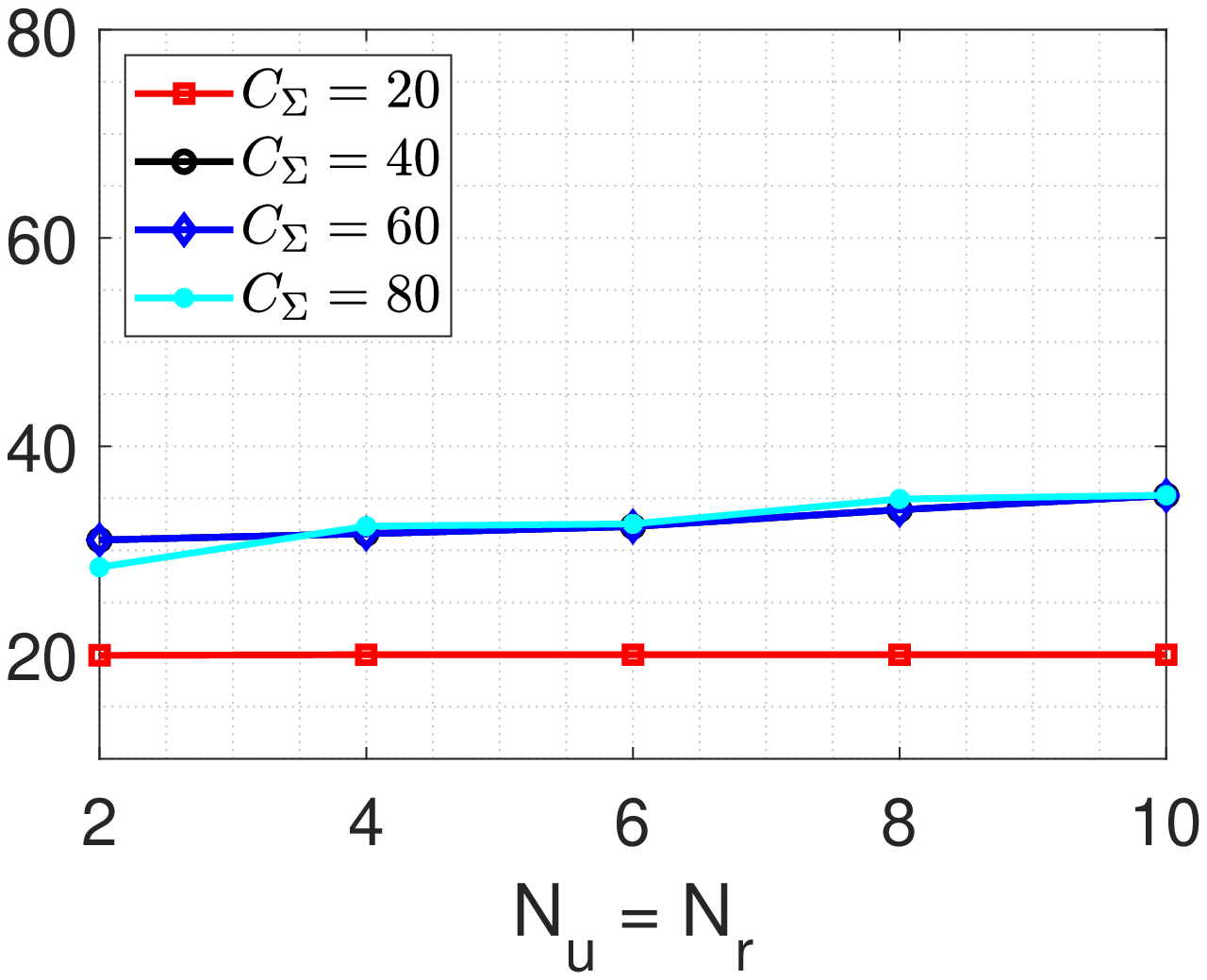}
\caption{$\beta = 3.5,$ $K = 4, L = 6.$}
\label{fig:plot2b}
\end{subfigure}
\begin{subfigure}[t]{0.33\textwidth}
\center
\includegraphics[scale=0.33]{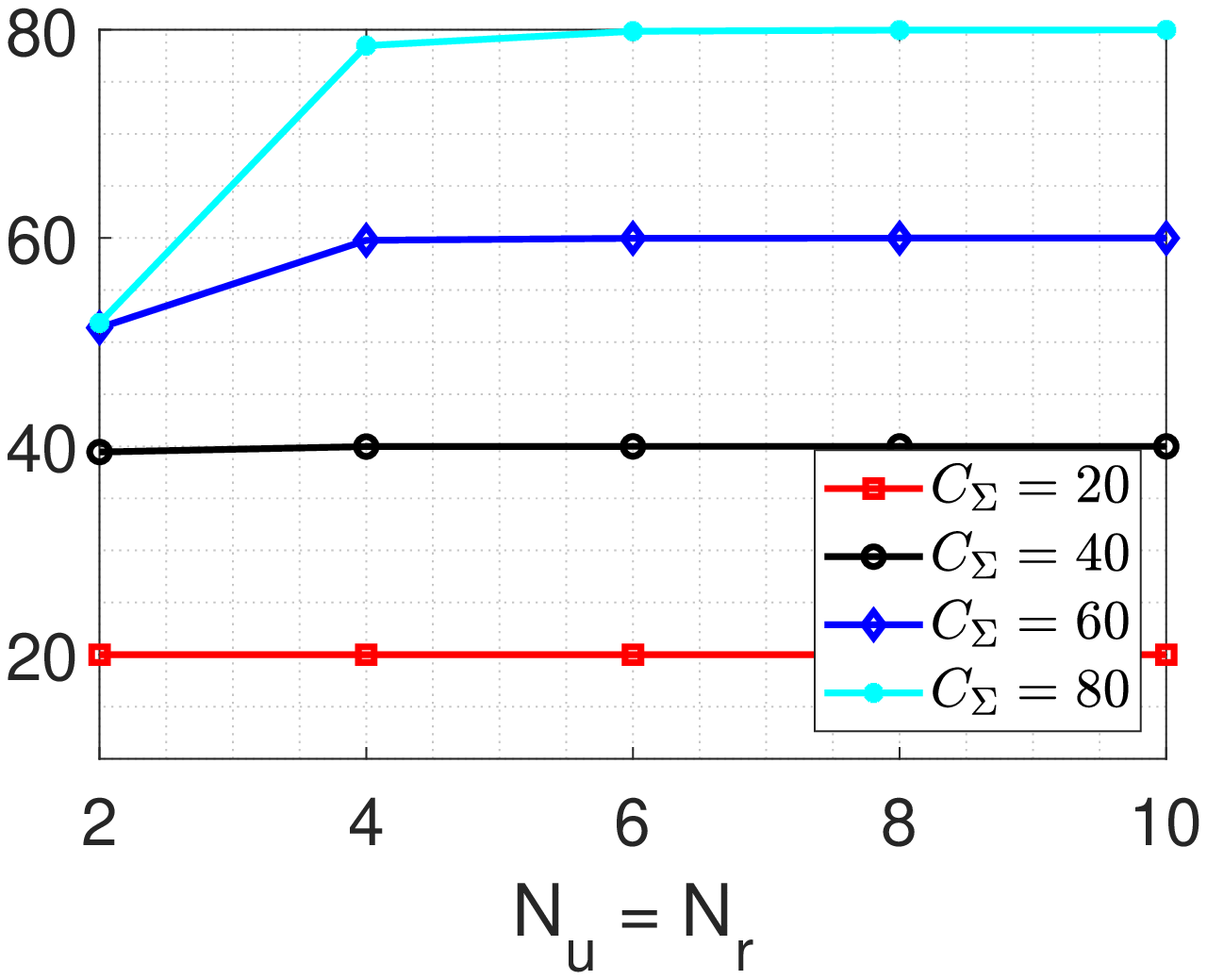}
\caption{Multipath, $K = 4, L = 6.$}
\label{fig:plot2c}
\end{subfigure}
%\vskip\baselineskip
\caption{MIMO downlink capacity scaling with antenna number under fixed sum-fronthaul.}\label{fig:plotdlmimoCsum}
\end{figure}

To quantify the advantages of having multiple local antennas at each user and each relay (i.e., the advantage of ``using MIMO''), Figs.~\ref{fig:plotulmimoCsum} and \ref{fig:plotdlmimoCsum} plot the best sum-rates achievable for a given $C_{\sum}$ as the number of local antennas grows, in accordance with Remark~\ref{rem:mimoCstartosumrate}. For these simulations, we take $K$ users and $L$ relays distributed uniformly over a $100\mathrm{m}\times 100\mathrm{m}$ area, where $K = 4$ and $L = 6.$ We consider the cases $C_{\sum} = 20,$ $40,$ $60,$ and $80$ bits per transmission. We take $N_u = N_r$ in all these simulations, for simplicity. The gains from MIMO are more significant at higher $C_{\sum}$ and under multipath models. Intuitively, a larger fronthaul provides a pipeline for the flow of the extra information available through the use of a larger number of antennas in the wireless hop of the network, while a smaller fronthaul is a bottleneck to achieving the full MIMO gains. In addition, for multipath models, i.i.d.\@ small-scale fading across different antennas at the local nodes leads to MIMO gains at higher $C_{\sum},$ while for the simple LOS models, the channel gains across different antennas at each node are almost identical and provide little diversity gain. 

\section{Concluding Remarks}\label{conclude}
We have approximated the capacities of the fronthaul-limited uplink and downlink C-RAN within a constant gap using noisy network coding (or network compress--forward) and distributed decode--forward, respectively. While the fronthaul-unlimited C-RAN capacities cannot be achieved exactly using a finite fronthaul, we have demonstrated that the former can be still \emph{approached} through judicious allocation of \emph{finite} fronthaul link capacities that are not too far off from the message rates. Practical implications of these findings to the design of network-layer protocols and architectures for C-RANs remain to be explored in more depth. 
%\useRomanappendicesfalse
\def\thesubsectiondis{\thesectiondis\arabic{subsection}.}
\appendices
\section{Uplink Propositions and Proofs}\label{upproofs}
\subsection*{Proof of Proposition~\ref{propo1}}
\subsubsection*{Coding scheme for uplink (specialization of noisy network coding)}
The noisy network coding scheme can be specialized to the uplink C-RAN model as follows.

\emph{Codebook generation.} Fix a pmf $p(q)\prod_{k=1}^Kp(x_k\cond q)\prod_{l=1}^Lp(\hat{y}_l\cond y_l, q).$ Randomly generate a time-sharing sequence $q^n\sim\prod_{i=1}^np_Q(q_i).$ For each message $m_k\in[2^{nR_k}],$ generate $x_k^n(m_k)\sim\prod_{i=1}^np_{X_k\cond Q}(x_{ki\cond q_i})$ conditionally independently. Define auxiliary indices $t_l\in[2^{n\hat{R}_l}],$ $l\in[L],$ for some auxiliary rates $\{\hat{R}_l, l\in[L]\}.$ For each $(w_l,t_l)\in[2^{nC_l}]\times[2^{n\hat{R}_l}]$ and $l\in[L]$, generate $\hat{y}_l^n(w_l,t_l)\sim\prod_{i=1}^np_{\hat{Y}_l\cond Q}(\hat{y}_{li}\cond q_i).$

\emph{Encoding.} For $k\in[K],$ to send message $m_k,$ encoder $k$ transmits $x_k^n(m_k).$

\emph{Relaying.} On receiving $y_l^n,$ relay $l$ finds $(w_l,t_l)$ such that $(q^n, y_l^n,\hat{y}_l^n(w_l,t_l))\in\aepvar$ and transmits $w_l$ to the central processor via the noiseless fronthaul.

\emph{Decoding.} Let $\epsilon>\epsilon^\prime$. Upon receiving the index tuple $w^L,$ the central processor finds message estimates $\hat{m}_1,\ldots,\hat{m}_K$ such that 
\[ 
(q^n, x_1^n(\hat{m}_1),\ldots,x_K^n(\hat{m}_K),\hat{y}_1^n(w_1,t_1),\ldots,\hat{y}_L^n(w_L,t_L))\in\mathcal{T}_\epsilon^{(n)}
\]
for some $t_1,\ldots,t_L.$ 
\subsubsection*{Analysis of the coding scheme}
Our analysis of the coding scheme follows that in \cite{nnc} but is considerably simpler because of the relative simplicity of our network model. We omit the time-sharing sequence $q^n$ for simplicity of notation.

Without loss of generality, let $m^K =(1, \ldots, 1)$ be the messages sent. Then the error events are:
\begin{align*}
\Ec_0  & = \Big\{\left(Y_l^n,\hat{Y}_l^n(w_l,t_l)\right)\notin\aepvar\text{ for all }(w_l,\ t_l,) \text{ for some }l\Big\}.\\
\Ec_1 & = \Big\{\Big(X_1^n(1),\ldots,X_K^n(1),\hat{Y}_1^n(W_1,t_1),\ldots, \hat{Y}_L^n(W_L,t_L)\Big)\notin\mathcal{T}_\epsilon^{(n)}\text{ for all }t^L\Big\}.\\
\Ec_2 & = \Big\{\Big(X_1^n(m_1),\ldots,X_K^n(m_K),\hat{Y}_1^n(W_1,t_1),\ldots, \hat{Y}_L^n(W_L,t_L)\Big)\in\mathcal{T}_\epsilon^{(n)}\text{ for some }t^L\text{ and }\\
& \quad\quad\quad\text{some }m^K\ne(1, \ldots, 1)\Big\}.
\end{align*}
Here, $(W_1, \ldots, W_L)$ represent the indices transmitted by the relays.
By the packing lemma and union of events, $\P(\Ec_0)\to 0$ as $n\to\infty$ if 
\[
C_l+\hat{R}_l  > I(Y_l;\hat{Y}_l)\numberthis\label{eq:uplinkcompressioncond}
\]
for all $l\in[L].$ 

By the Markov lemma \cite[Lemma 12.1]{nit} and union of events bound ($\hat{Y}_l\rightarrow Y_l\rightarrow X^K$ form a Markov chain), $\P(\Ec_1\cap\Ec_0^c)\to 0$ as $n\to\infty$.

To analyze $\P(\Ec_2)$, let $t^L = (1, \ldots, 1)$ be the $t$-indices chosen at the relays. Then, by the union of events bound,
\begin{align*}
\P\left(\Ec_2\right) & \le\sum_{\substack{m^K, t^L\\ m^K\ne(1, \ldots, 1)}}\P\left(\left(X_1^n(m_1),\ldots,X_K^n(m_K),\hat{Y}_1^n(W_1,t_1),\ldots,\hat{Y}_L^n(W_L,t_L)\right)\in\aep\right) \\
& =: \sum_{\substack{m^K, t^L:\\ m^K\ne (1, \ldots, 1)}}p_{m^K,t^L}.\numberthis\label{eq:unionbounduplink}
\end{align*}

In order to bound each term on the right-hand side of \eqref{eq:unionbounduplink}, we need the following generalization of the joint typicality Lemma. 

\begin{lemma}\cite[Lemma 2]{nnc}\label{jointtyp}
Let $(X^N,Y^N,Z)\sim p(x^N,y^N,z)$. Let the $n$-length random vector $\hat{\mathbf{Z}}$ be distributed according to some arbitrary pmf $p(\hat{\mathbf{z}})$ and let 
\[
(\hat{X}_1^n,\ldots, \hat{X}_N^n,\hat{Y}_1^n,\ldots, \hat{Y}_N^n)\sim\prod_{i=1}^np_{X^N}(\hat{x}_{1i},\ldots,\hat{x}_{Ni})\prod_{k=1}^N\prod_{i=1}^np_{Y_k|X_k}(\hat{y}_{ki}|\hat{x}_{ki})
\]
be distributed independently of $\hat{\mathbf{Z}}$. Then, there exists $\delta(\epsilon)$ that tends to zero as $\epsilon\to 0$, such that

\[ 
\P\left((\hat{\mathbf{Z}}, \hat{X}_1^n,\ldots, \hat{X}_N^n, \hat{Y}_1^n,\ldots, \hat{Y}_N^n)\in\mathcal{T}_\epsilon^{(n)}\right)
\le 2^{-n[I(Z;X^N)+\sum_{k=1}^NI(Y_k; X^N,Y^{k-1},Z|X_k)-\delta(\epsilon)]}. 
\]
\end{lemma}

For a given $t^L$ and $m^K\ne(1, \ldots, 1)$, let $\Sc_2(t^L) = \{l\in[L] : t_l\ne 1\}$ and $\Sc_1(m^K)=\{k\in[K] : m_k\ne 1\}.$ Then, $(X^n(\Sc_1(m^K)),\hat{Y}^n(\Sc_2(t^L)))$ is independent of $(X^n(\Sc_1^c(m^K)),\hat{Y}^n(\Sc_2^c(t^L))).$ Then, using $(X^n(\Sc_1^c(m^K)),\hat{Y}^n(\Sc_2^c(t^L)))$ as $\hat{\mathbf{Z}}^n$ in Lemma \ref{jointtyp}, we obtain
\begin{align*}
& p_{m^K, t^L} \\
& \le 2^{-n\Big[I(X(\Sc_1(m^K));X(\Sc_1^c(m^K)),\hat{Y}(\Sc_2^c(t^L)))+\sum_{l\in \Sc_2(t^L)}I(\hat{Y}_l; X(\Sc_1(m^K)),\hat{Y}([l-1]\cap \Sc_2(t^L)),X(\Sc_1^c(m^K)),\hat{Y}(\Sc_2^c(t^L)))-\delta(\epsilon)\Big]}.
\end{align*} 
The terms in the exponent of the right-hand expression (excluding the factor of $-n$ ) are given by
\begin{align*}
& I(X(\Sc_1(m^K));X(\Sc_1^c(\mv)),\hat{Y}(\Sc_2^c(t^L)))\\
& \quad\quad\quad + \sum_{l\in \Sc_2(t^L)}I(\hat{Y}_l; X(\Sc_1(m^K)),\hat{Y}([l-1]\cap \Sc_2(t^L)),X(\Sc_1^c(m^K)),\hat{Y}(\Sc_2^c(t^L)))- \delta(\epsilon)\\
& \stackrel{(a)}{=}  I(X(\Sc_1(m^K));\hat{Y}(\Sc_2^c(t^L))\cond X(\Sc_1^c(m^K)))+\sum_{l\in \Sc_2(t^L)}I(Y_l;\hat{Y}_l) \\
& \quad\quad\quad\quad\quad\quad -\sum_{l\in \Sc_2(t^L)}\left(I(Y_l;\hat{Y}_l)-I(\hat{Y}_l;\hat{Y}([l-1]\cap \Sc_2(t^L)),\hat{Y}(\Sc_2^c(t^L)),X^K)\right)-\delta(\epsilon)\\
\stackrel{(b)}{=} & I(X(\Sc_1(m^K));\hat{Y}(\Sc_2^c(t^L))\cond X(\Sc_1^c(m^K)))+\sum_{l\in \Sc_2(t^L)}I(Y_l;\hat{Y}_l)\\
& \quad\quad\quad\quad\quad\quad -\sum_{l\in \Sc_2(t^L)}I\left(Y_l;\hat{Y}_l\cond\hat{Y}([l-1]\cap \Sc_2(t^L)),\hat{Y}(\Sc_2^c(t^L)),X^K\right)-\delta(\epsilon).
\end{align*}
Here, $(b)$ follows from the fact that $\left(\hat{Y}([l-1]\cap \Sc_2(t^L)),\hat{Y}(\Sc_2^c(t^L)),X^K\right)\rightarrow Y_l\rightarrow \hat{Y}_l$ form a Markov chain and $(a)$ follows from the independence of $X(S_1(m^K))$ and $X(S_1^c(m^K)).$ 

Defining 
\[ 
J(\Sc_1,\Sc_2)
:=  I(X(\Sc_1);\hat{Y}(\Sc_2^c)|X(\Sc_1^c))+\sum_{l\in \Sc_2}I(Y_l;\hat{Y}_l)-\sum_{l\in \Sc_2}I\left(Y_l;\hat{Y}_l\Big|\hat{Y}([l-1]\cap \Sc_2),\hat{Y}(\Sc_2^c),X^K\right)
\]
and continuing \eqref{eq:unionbounduplink}, we have
\begin{align*}
\P(\Ec_2) & \le \sum_{\substack{m^K, t^L:\\ m^K\ne (1, \ldots, 1)}}2^{-n[J(\Sc_1(m^K),\Sc_2(t^L))-\delta(\epsilon)]}\\
& \le\sum_{\substack{\Sc_1\subseteq[K]\\ \Sc_2\subseteq[L]\\ \Sc_1\ne\emptyset}}2^{-n[J(\Sc_1,\Sc_2)-\sum_{k\in \Sc_1}R_k-\sum_{l\in \Sc_2}\hat{R}_l-\delta(\epsilon)]}.
\end{align*}
Therefore, $\P(\Ec_2)\to 0$ as $n\to\infty$ if
\begin{align}\sum_{k\in \Sc_1}R_k+\sum_{l\in \Sc_2}\hat{R}_l & < J(\Sc_1,\Sc_2)\label{eq:uplinkdecodingcond}\end{align}
for all $\Sc_1\subseteq[K]$ and $\Sc_2\subseteq[L]$ such that $\Sc_1\ne\emptyset.$ Combining \eqref{eq:uplinkdecodingcond} with \eqref{eq:uplinkcompressioncond} to eliminate the auxiliary rates $(\hat{R}_1, \ldots, \hat{R}_L),$ we obtain the inequalities
\begin{align*}
\sum_{k\in \Sc_1}R_k & < I(X(\Sc_1);\hat{Y}(\Sc_2^c)|X(\Sc_1^c))+\sum_{l\in \Sc_2}C_l-\sum_{l\in \Sc_2}I\left(Y_l;\hat{Y}_l\Big|\hat{Y}\left([l-1]\cap \Sc_2\right),\hat{Y}(\Sc_2^c),X^K\right)\numberthis
\end{align*}
for all $\Sc_1\subseteq[K]$ and $\Sc_2\subseteq[L]$ such that $\Sc_1\ne\emptyset.$

\subsection*{Proof of Proposition~\ref{upgenouter}}
For $k\in[K],$ let $M_k$ denote the message communicated by sender $k$ and let $W_l$ denote the index sent by relay $l$ to the central processor. Also, for $\Sc_1\subseteq[K]$ and $\Sc_2\subseteq[L],$ denote by $X_i(\Sc_1)$ the tuple $(X_{ki}, k\in \Sc_1)$ and by $Y_i(\Sc_2)$ the tuple $(Y_{li}, l\in \Sc_2).$ Similarly, $X^n(\Sc_1)$ stands for $(X_{ki}, k\in \Sc_1, i\in[n])$ and $Y^n(\Sc_2)$ stands for $(Y_{li}, l\in \Sc_2, i\in[n]).$ Then, for every $\Sc_1\subseteq[K]$ and $\Sc_2\subseteq[L],$ $X^n(\Sc_1)$ is a function of $M(\Sc_1)$ and $W(\Sc_2)$ is a function of $Y^n(\Sc_2).$ For every $\Sc_1\subseteq[K], \Sc_1\ne\emptyset$ and $\Sc_2\subseteq[L],$ we must have, by Fano's inequality, 
\[
H(M(\Sc_1)\cond M(\Sc_1^c), Y^n(\Sc_2^c), W^L)\le H(M(\Sc_1)\cond W^L)\le n\epsilon_n,
\]
where $\epsilon_n\to 0$ as $n\to\infty.$ Therefore, since $H(M(\Sc_1)\cond M(\Sc_1^c)) = n\sum_{k\in \Sc_1}R_k,$ we have
\begin{align*}
n\sum_{k\in \Sc_1}R_k & \le I(M(\Sc_1); Y^n(\Sc_2^c), W^L\cond M(\Sc_1^c)) + n\epsilon_n\\
& \stackrel{(a)}{=} I(M(\Sc_1); Y^n(\Sc_2^c), W(\Sc_2)\cond M(\Sc_1^c)) + n\epsilon_n\\
& = I(M(\Sc_1); Y^n(\Sc_2^c)\cond M(\Sc_1^c)) + I(M(\Sc_1); W(\Sc_2)\cond M(\Sc_1^c), Y^n(\Sc_2^c)) + n\epsilon_n\\
& \le \sum_{i=1}^nI(M(\Sc_1); Y_i(\Sc_2^c)\cond M(\Sc_1^c), Y^{i-1}(\Sc_2^c)) + H(W(\Sc_2)) + n\epsilon_n\\
& \stackrel{(b)}{=} \sum_{i=1}^nI(M(\Sc_1); Y_i(\Sc_2^c)\cond M(\Sc_1^c), X_i(\Sc_1^c), Y^{i-1}(\Sc_2^c)) + H(W(\Sc_2)) + n\epsilon_n\\
& \le \sum_{i=1}^nI(M(\Sc_1), M(\Sc_1^c), Y^{i-1}(\Sc_2^c); Y_i(\Sc_2^c)\cond X_i(\Sc_1^c)) + H(W(\Sc_2)) + n\epsilon_n\\
& \stackrel{(c)}{=}\sum_{i=1}^nI(M(\Sc_1), M(\Sc_1^c), X_i(\Sc_1), Y^{i-1}(\Sc_2^c); Y_i(\Sc_2^c)\cond X_i(\Sc_1^c)) + H(W(\Sc_2)) + n\epsilon_n\\
& \stackrel{(d)}{=}\sum_{i=1}^nI(X_i(\Sc_1); Y_i(\Sc_2^c)\cond X_i(\Sc_1^c)) + H(W(\Sc_2)) + n\epsilon_n\\
& \stackrel{(e)}{\le}\sum_{i=1}^nI(X_i(\Sc_1); Y_i(\Sc_2^c)\cond X_i(\Sc_1^c)) + n\sum_{l\in \Sc_2}C_l + n\epsilon_n.
\end{align*}
Here, $(a)$ follows since $W(\Sc_2^c)$ is a function of $Y^n(\Sc_2^c),$ $(b)$ follows since $X_i(\Sc_1^c)$ is a function of $M(\Sc_1^c),$ $(c)$ follows since $X_i(\Sc_1)$ is a function of $M(\Sc_1),$ $(d)$ follows since $(M(\Sc_1), M(\Sc_1^c),Y^{i-1}(\Sc_2^c))\rightarrow(X_i(\Sc_1), X_i(\Sc_1^c))\rightarrow(Y_i(\Sc_2^c))$ form a Markov chain (by the memorylessness of the first hop), and $(e)$ follows since $W(\Sc_2)$ is supported on a set of size $\prod_{l\in S_2}2^{nC_l}.$ Defining a random variable $Q\sim\mathrm{Unif}([n])$ independent of all other random variables, writing $X(\Sc_1) := X_Q(\Sc_1)$ and $Y(\Sc_2):= Y_Q(\Sc_2),$ and letting $n$ tend to infinity leads to
\[
\sum_{k\in \Sc_1}R_k \le I(X(\Sc_1); Y(\Sc_2^c)\cond X(\Sc_1^c), Q) + \sum_{l\in \Sc_2}C_l\numberthis\label{cutsetupbetappendix}
\]
for all $\Sc_1\subseteq[K], \Sc_1\ne\emptyset, \Sc_2\subseteq[L]$ for some pmf $p(q)\prod_{k\in[K]}p(x_k\cond q),$ and thus completes the proof.
\subsection*{Proof of Theorem~\ref{upgaptheo1}}
Recall from \eqref{upinngaussinformal} in Section~\ref{upgauss} that for the Gaussian uplink network model with channel gain matrix $G$ and average power constraint $P$ on each sender, a rate tuple $(R_1,\ldots,R_K)$ is achievable if for every $\Sc_1\subseteq[K]$ and $\Sc_2\subseteq[L],$
\[
\sum_{k\in \Sc_1}R_k < \frac{1}{2}\log\Big\vert\frac{P}{\sigma^2+1}G_{\Sc_2^c,\Sc_1}G_{\Sc_2^c,\Sc_1}^T+I\Big\vert +\sum_{l\in \Sc_2}C_l-\frac{\vert \Sc_2\vert}{2}\log\left(1+\frac{1}{\sigma^2}\right) =: f_{\mathrm{in}}(\Sc_1, \Sc_2).\numberthis\label{inngauss}
\]
We now establish the following useful property of the inner bound \eqref{inngauss}, which will be useful in developing some insight into the nature of the achievable region, as well as in proving Theorem~\ref{upgaptheo1}.

\begin{lemma}\label{monotonelemma}
For any $\Sc_2\subseteq[L]$ and $\Sc_1'\subseteq \Sc_1\subseteq[K],$ 
\begin{align*}
f_{\mathrm{in}}(\Sc_1^\prime,\Sc_2)\le f_{\mathrm{in}}(\Sc_1,\Sc_2).
\end{align*}
\end{lemma}
\begin{IEEEproof}
Letting $G_{\Sc_2^c,k}$ denote the column vector consisting of the elements of $G$ with row index in $\Sc_2^c$ and column index $k,$ we have
\[ 
G_{\Sc_2^c,\Sc_1}G_{\Sc_2^c,\Sc_1}^T = \sum_{k\in \Sc_1}G_{\Sc_2^c,k}G_{\Sc_2^c,k}^T \succeq\sum_{k'\in \Sc_1'}G_{\Sc_2^c,k'}G_{\Sc_2^c,k'}^T= G_{\Sc_2^c,\Sc_1'}G_{\Sc_2^c,\Sc_1'}^T,
\]
which implies that 
\[
\frac{1}{2}\log\Big\vert\frac{P}{\sigma^2+1}G_{\Sc_2^c,\Sc_1}G_{\Sc_2^c,\Sc_1}^T+I\Big\vert\ge\frac{1}{2}\log\Big\vert\frac{P}{\sigma^2+1}G_{\Sc_2^c,\Sc_1^\prime}G_{\Sc_2^c,\Sc_1^\prime}^T+I\Big\vert
\]
and hence, that $f_{\mathrm{in}}(\Sc_1',\Sc_2)\le f_{\mathrm{in}}(\Sc_1,\Sc_2)$, since the other terms remain the same.
\end{IEEEproof}
Lemma \ref{monotonelemma} immediately implies that 
\begin{align*}
\min_{\Sc_2}f_{\mathrm{in}}(\Sc_1^\prime,\Sc_2)\le \min_{\Sc_2}f_{\mathrm{in}}(\Sc_1,\Sc_2).\numberthis\label{monotone}
\end{align*}  
\begin{remark}\label{uplinkpolymatroid}
We can establish \eqref{monotone} directly from Proposition~\ref{propo1}, which implies that it continues to hold for the general inner bound. Moreover, we can show that the inner bound \eqref{inngauss} is a polymatroid for each fixed $\sigma^2>0.$
\end{remark}

We now specialize the cutset bound in Proposition~\ref{upgenouter} to the Gaussian uplink C-RAN model.
\begin{lemma}\label{cutsetuplink}
The cutset bound \eqref{cutsetupbetappendix} can be simplified and relaxed for the Gaussian model as
\begin{align*}
\sum_{k\in \Sc_1}R_k & \le \frac{1}{2}\log\Big\vert PG_{\Sc_2^c,\Sc_1}G_{\Sc_2^c,\Sc_1}^T+I\Big\vert+\sum_{l\in \Sc_2}C_l\\
& =: f_{\mathrm{out}}(\Sc_1,\Sc_2).\numberthis\label{cutsetupgauss1}
\end{align*}
\end{lemma}
\begin{IEEEproof}
Continuing from \eqref{cutsetupbetappendix}, we have
\begin{align*}
\sum_{k\in \Sc_1}R_k & \le I(X(\Sc_1);Y(\Sc_2^c)|X(\Sc_1^c),Q)+\sum_{l\in \Sc_2}C_l \\
& \stackrel{(a)}{=} h(Y(\Sc_2^c)|X(\Sc_1^c),Q)-\frac{|\Sc_2^c|}{2}\log(2\pi e)+\sum_{l\in \Sc_2}C_l\\
& = h\left(G_{\Sc_2^c,\Sc_1}X(\Sc_1)+G_{\Sc_2^c,\Sc_1^c}X(\Sc_1^c)+Z(\Sc_2^c)\cond X(\Sc_1^c),Q\right)-\frac{|\Sc_2^c|}{2}\log(2\pi e)+\sum_{l\in \Sc_2}C_l\\
& = h\left(G_{\Sc_2^c,\Sc_1}X(\Sc_1)+Z(\Sc_2^c)\cond X(\Sc_1^c),Q\right)-\frac{|\Sc_2^c|}{2}\log(2\pi e)+\sum_{l\in \Sc_2}C_l\\
& \stackrel{(b)}{=} h(G_{\Sc_2^c,\Sc_1}X(\Sc_1)+Z(\Sc_2^c)\cond Q)-\frac{|\Sc_2^c|}{2}\log(2\pi e)+\sum_{l\in \Sc_2}C_l\\
& =\sum_qh(G_{\Sc_2^c,\Sc_1}X(S_1)+Z(\Sc_2^c)\cond Q=q)p(q)-\frac{|\Sc_2^c|}{2}\log(2\pi e)+\sum_{l\in \Sc_2}C_l\\
& \stackrel{(c)}{\le}\sum_q\frac{1}{2}\log\left((2\pi e)^{|\Sc_2^c|}\left|G_{\Sc_2^c,\Sc_1}\E\left[X(\Sc_1)X(\Sc_1)^T\cond Q=q\right]G_{\Sc_2^c,\Sc_1}^T+I\right|\right)p(q)-\frac{|\Sc_2^c|}{2}\log(2\pi e)+\sum_{l\in \Sc_2}C_l\\
& \stackrel{(d)}{=}\sum_q\frac{1}{2}\log\left|G_{\Sc_2^c,\Sc_1}K_{\Sc_1}^\prime(q)G_{\Sc_2^c,\Sc_1}^T+I\right|p(q)+\sum_{l\in \Sc_2}C_l\\
& \stackrel{(e)}{\le}\frac{1}{2}\log\left|G_{\Sc_2^c,\Sc_1}K_{\Sc_1}G_{\Sc_2^c,\Sc_1}^T+I\right|+\sum_{l\in \Sc_2}C_l\\
& \le\frac{1}{2}\log\left|PG_{\Sc_2^c,\Sc_1}G_{\Sc_2^c,\Sc_1}^T+I\right|+\sum_{l\in \Sc_2}C_l.
\end{align*}
Here, $(a)$ follows since $Y(\Sc_2^c)$ is an i.i.d.\@ Gaussian vector given $X^K,$ $(b)$ follows since $X(\Sc_1)$ and $X(\Sc_1^c)$ are conditionally independent given $Q,$ $(c)$ follows from the (vector) maximum entropy principle, and in $(d),$ $K_{\Sc_1}^\prime(q)$ is a diagonal matrix consisting of $\{\E[X_k^2\cond Q=q], k\in \Sc_1\}.$ In $(e),$ $K_{\Sc_1}$ is a diagonal matrix consisting of $\{\E[X_k^2], k\in \Sc_1\},$ and $(e)$ follows from the concavity of the log--determinant function of a symmetric matrix. Finally, the last inequality follows since each diagonal entry of $K_{\Sc_1}$ is upper-bounded by $P.$ 
\end{IEEEproof}

These results finally lead to a proof of Theorem~\ref{upgaptheo1}.
%\begin{IEEEproof}[Proof of Theorem~\ref{upgaptheo1}]
Let
\begin{align}
\Delta & := \max_{\substack{\Sc_1\subseteq[K]\\ \Sc_1\ne\emptyset}}\frac{\min_{\Sc_2}f_{\mathrm{out}}(\Sc_1,\Sc_2)-\min_{\Sc_2}f_{\mathrm{in}}(\Sc_1,\Sc_2)}{|\Sc_1|}.\label{gengap}
\end{align} 
Suppose that $(R_1,\ldots, R_K)$ lies in the cutset bound, and let $\Ac=\{k: R_k>\Delta\}$. Then, for every nonempty $\Sc_1\subseteq[K]$,
\begin{align*}
 \sum_{k\in \Sc_1}(R_k-\Delta)^+ & =  \sum_{k\in \Sc_1\cap\Ac}(R_k-\Delta)\\
& =  \sum_{k\in \Sc_1\cap\mathcal{A}}R_k - |\Sc_1\cap\mathcal{A}|\Delta\\
& \stackrel{(a)}{\le} \min_{\Sc_2}\Big[f_{\mathrm{out}}(\Sc_1\cap\mathcal{A},\Sc_2)\Big]-\left(\min_{\Sc_2}f_{\mathrm{out}}(\Sc_1\cap\mathcal{A},\Sc_2)-\min_{\Sc_2}f_{\mathrm{in}}(\Sc_1\cap\mathcal{A},\Sc_2)\right)\\
& =   \min_{\Sc_2}f_{\mathrm{in}}(\Sc_1\cap\mathcal{A},\Sc_2)\\
& \stackrel{(b)}{\le}  \min_{\Sc_2}f_{\mathrm{in}}(\Sc_1,\Sc_2), 
\end{align*}
where $(a)$ follows from the cutset bound (\ref{cutsetupgauss1}) and the fact that 
\begin{align*}
\Delta & = \max_{\Sc_1}\frac{\min_{\Sc_2}f_{\mathrm{out}}(\Sc_1,\Sc_2)-\min_{\Sc_2}f_{\mathrm{in}}(\Sc_1,\Sc_2)}{|\Sc_1|}\\
& \ge \frac{\min_{\Sc_2}f_{\mathrm{out}}(\Sc_1\cap\mathcal{A},\Sc_2)-\min_{\Sc_2}f_{\mathrm{in}}(\Sc_1\cap\mathcal{A},\Sc_2)}{|\Sc_1\cap\mathcal{A}|},
\end{align*}
and $(b)$ follows from (\ref{monotone}). Hence, $\Delta$, as defined in \eqref{gengap}, satisfies the requirements of Theorem \ref{upgaptheo1}. Now, for every $\sigma^2>0$,
\begin{align*}
\Delta & = \max_{\Sc_1}\frac{\min_{\Sc_2}f_{\mathrm{out}}(\Sc_1,\Sc_2)-\min_{\Sc_2}f_{\mathrm{in}}(\Sc_1,\Sc_2)}{|\Sc_1|}\\
& \stackrel{(a)}{\le}\max_{\Sc_1,\Sc_2}\frac{f_{\mathrm{out}}(\Sc_1,\Sc_2)-f_{\mathrm{in}}(\Sc_1,\Sc_2)}{|\Sc_1|}\\
& \stackrel{(b)}{=}\max_{\Sc_1,\Sc_2}\left[\frac{1}{2|\Sc_1|}\log\frac{\Big\vert PG_{\Sc_2^c,\Sc_1}G_{\Sc_2^c,\Sc_1}^T+I\Big\vert}{\Big\vert \frac{P}{\sigma^2+1}G_{\Sc_2^c,\Sc_1}G_{\Sc_2^c,\Sc_1}^T+I\Big\vert}\right]\\
& \stackrel{(c)}{=}\max_{\Sc_1,\Sc_2}\Bigg[\frac{1}{2|\Sc_1|}\sum_{i=1}^{\mathrm{rank}(G_{\Sc_2^c,\Sc_1})}\log\frac{P\beta_i+1}{\frac{P}{\sigma^2+1}\beta_i+1}+\vert \Sc_2\vert\log\left(1+\frac{1}{\sigma^2}\right)\Bigg]\\
& \stackrel{(d)}{\le}\max_{\substack{k\in[K]\\l\in\{0, \ldots, L\}}}\Big[\frac{\min\{L-l,k\}}{2k}\log(1+\sigma^2)+\frac{l}{2k}\log\left(1+\frac{1}{\sigma^2}\right)\Big].\numberthis\label{rankexpuplink}
\end{align*}
Here, $(a)$ follows from the fact that for functions $f$ and $g$ defined over a finite set $\Xc$, such that $g\ge f$ everywhere on $\Xc$, $\min_{x\in\Xc}g(x)-\min_{x\in\Xc}f(x)\le\max_{x\in\Xc}[g(x)-f(x)]$, $(b)$ follows from \eqref{inngauss} and \eqref{cutsetupgauss1}, and in $(c)$, $\beta_1,\beta_2,\ldots$ are the (nonnegative) eigenvalues of $G_{\Sc_2^c,\Sc_1}G_{\Sc_2^c,\Sc_1}^T$. Finally, in $(d)$, we take $|\Sc_1|=k$, $|\Sc_2|=l,$ and upper-bound $\mathrm{rank}(G_{\Sc_2^c,\Sc_1})$ by $\min\{L-l,k\}.$ 
The maximization in \eqref{rankexpuplink} yields
\begin{align*}
\Delta & \le  \begin{cases}\frac{1}{2}\log(\sigma^2+1)+\frac{L-1}{2}\log(1+\frac{1}{\sigma^2}), & \sigma^2\ge 1,\\ \frac{L}{2}\log(1+\frac{1}{\sigma^2}), & \sigma^2\le1.\end{cases}
\end{align*}
Since this holds for every $\sigma^2>0$, we set $\sigma^2=L-1$ for $L\ge 2$ to obtain
\begin{align*}
 \Delta & \le \frac{1}{2}\log L+\frac{L-1}{2}\log\left(1+\frac{1}{L-1}\right)\\
& \stackrel{(a)}{\le}  \frac{1}{2}\log L+\frac{L-1}{2}\cdot\frac{1}{L-1}\log e\\
& \le  \frac{1}{2}\log(eL).\numberthis\label{finalex}
\end{align*} 
Here, $(a)$ follows since from elementary calculus, we know that for $x>0,$ $\log(1+x)\le x\log e.$

For $L=1,$ we can choose $\sigma^2=1$ to obtain $\Delta\le 1.$ This, together with \eqref{finalex}, establishes the first part of Theorem \ref{upgaptheo1}. For the sum-rate gap, we simply consider
\begin{align*}
\Delta_\text{sum} & \le\max_{\Sc_1, \Sc_2}\left(f_\mathrm{out}(\Sc_1, \Sc_2) - f_\mathrm{in}(\Sc_1, \Sc_2)\right)\\
& \le \max_{\substack{k\in[K]\\l\in\{0, \ldots, L\}}}\Big[\frac{\min\{L-l,k\}}{2}\log(1+\sigma^2)+\frac{l}{2}\log\left(1+\frac{1}{\sigma^2}\right)\Big].\numberthis\label{rankexpuplinksumrate}
\end{align*}
Maximization of \eqref{rankexpuplinksumrate} over $l$ and $k$ yields, for $\sigma^2\ge 1,$
\[
\Delta_\text{sum}\le\begin{cases}\frac{K}{2}\log(1+\sigma^2)+\frac{L-K}{2}\log(1+\frac{1}{\sigma^2}), & L\ge K,\\ \frac{L}{2}\log(1+\sigma^2), & L<K.\end{cases}
\]
For $L\le 2K,$ we can then choose $\sigma^2 = 1$ to obtain an upper bound $\Delta_\text{sum}\le L/2.$ For $L>2K,$ we can choose $\sigma^2 = L/K - 1\ge 1$ to obtain
%\begin{align*}
%\Delta_\text{sum} & \le \frac{K}{2}\log\left(\frac{L}{K}\right) + \frac{L-K}{2}\log\left(1 + \frac{K}{L-K}\right)\numberthis\label{uplinkbinentropy}\\
%& \le \frac{K}{2}\log\left(\frac{L}{K}\right) + \frac{L-K}{2}\cdot\frac{K}{L-K}\log e\\
%& = \frac{K}{2}\log\left(\frac{eL}{K}\right),
%\end{align*}
\begin{align*}
\Delta_\text{sum} & \le \frac{K}{2}\log\left(\frac{L}{K}\right) + \frac{L-K}{2}\log\left(1 + \frac{K}{L-K}\right)\numberthis\label{uplinkbinentropy}\\
& = \frac{L}{2}\left(\frac{K}{L}\log\left(\frac{L}{K}\right) + \left(1-\frac{K}{L}\right)\log\frac{1}{1-\frac{K}{L}}\right)\\
& = \frac{L}{2}H(K/L),
\end{align*}
completing the proof. 
%\end{IEEEproof}
\subsection*{Derivation of Table~\ref{table:uplinkscaling}}
First note that for all cases considered, 
\[
R_\mathrm{sum}^\mathrm{MIMO}\sim\frac{\min\{K, L\}}{2}\log\left(P\max\{K, L\}\right)\sim\frac{\min\{K, L\}}{2}\log \left(\max\{K, L\}\right).
\]
For $L = \gamma K$ with $\gamma>1,$ we can choose $\sigma^2 = 1$ to obtain, from Theorem~\ref{theorem:Cstaruplink} and Lemma~\ref{theorem:pggt},
\[
C^*(\sigma^2)\sim\frac{K}{2}\log(L/2) + \frac{L}{2}\sim\frac{K}{2}\log L,
\]
and
\[
R_\mathrm{sum}^\text{NCF}(\sigma^2)\sim\frac{K}{2}\log(L/2)\sim\frac{K}{2}\log L.\numberthis\label{eq:ncflinear}
\]
For $L = \gamma K$ with $\gamma\in(0,1),$ we can similarly choose $\sigma^2 = 1$ to obtain, from Theorem~\ref{theorem:Cstaruplink} and Lemma~\ref{theorem:pggt},
\[
C^*(\sigma^2)\sim\frac{L}{2}\log(K/2) + \frac{L}{2}\sim\frac{L}{2}\log K,
\]
and
\[
R_\mathrm{sum}^\text{NCF}(\sigma^2)\sim\frac{L}{2}\log(K/2)\sim\frac{L}{2}\log K.\numberthis\label{eq:ncflinearsmallell}
\]
For $L = K^\gamma$ with $\gamma > 1,$ we can choose $\sigma^2 = K^{\gamma - 1}$ to obtain
\[
C^*(\sigma^2)\sim\frac{K}{2}\log(L/K^{\gamma -1}) + \frac{L\log e}{2K^{\gamma - 1}}\sim\frac{K}{2}\log K,
\]
and
\[
R_\mathrm{sum}^\text{NCF}(\sigma^2))\sim\frac{K}{2}\log(L/K^{\gamma - 1})\sim\frac{K}{2}\log K.
\]
Alternatively, for the same scaling regime, we can choose $\sigma^2 = K^{\gamma - 1 - \delta}$ for some $\delta\in(0, \gamma-1)$ to obtain
\[
C^*(\sigma^2)\sim\frac{K}{2}\log(L/K^{\gamma -1-\delta}) + \frac{L\log e}{2K^{\gamma - 1 - \delta}}\sim\frac{K^{1+\delta}}{2}\log e,
\]
and
\[
R_\mathrm{sum}^\text{NCF}(\sigma^2)\sim\frac{K}{2}\log(L/K^{\gamma - 1 - \delta})\sim(1+\delta)\frac{K}{2}\log K.
\]
For $L = K^\gamma$ with $\gamma\in(0,1),$ we can choose $\sigma^2 = 1$ to obtain
\[
C^*(\sigma^2)\sim\frac{L}{2}\log(K/2) + \frac{L}{2}\sim\frac{L}{2}\log K,
\]
and
\[
R_\mathrm{sum}^\text{NCF}(\sigma^2)\sim\frac{L}{2}\log(K/2)\sim\frac{L}{2}\log K.
\]
For fixed $K$ with $L$ growing, we can choose $\sigma^2 = L^\epsilon$ for some $\epsilon\in(0, 1)$ to obtain
\[
C^*(\sigma^2)\sim\frac{K}{2}\log(L/L^\epsilon) + \frac{L\log e}{2L^\epsilon}\sim\frac{L^{1-\epsilon}}{2}\log e,
\]
and
\[
R_\mathrm{sum}^\text{NCF}(\sigma^2)\sim\frac{K}{2}\log(L/L^\epsilon)\sim(1-\epsilon)\frac{K}{2}\log L.
\]
Finally, for fixed $L$ with $K$ growing, we can choose $\sigma^2 = 1$ to obtain
\[
C^*(\sigma^2)\sim\frac{L}{2}\log(K/2) + \frac{L}{2}\sim\frac{L}{2}\log K,
\]
and
\[
R_\mathrm{sum}^\text{NCF}(\sigma^2)\sim\frac{L}{2}\log(K/2)\sim\frac{L}{2}\log K.
\]

\section{Downlink Propositions and Proofs}\label{downproofs}
\subsection*{Proof of Proposition~\ref{propo3}}
Throughout this proof, we use the following additional notation. For a function $f:{\mathbb N}\to [0, \infty)$ and a real number $r\ne 0,$ we say
\[
f(n)\overset{.}{=} 2^{nr}
\] 
if 
\[
r = \lim_{n\to\infty}\frac{\log f(n)}{n}.
\]
\subsubsection*{Coding scheme for downlink (specialization of distributed decode--forward)}
The distributed decode--forward coding scheme can be specialized to the downlink C-RAN model as follows.

\emph{Codebook generation.} Fix a pmf $p(x^L, u^K).$ For each $w_l\in[2^{nC_l}], l\in[L],$ generate $x_l^n(w_l)\sim \prod_{i=1}^np_{X_l}(x_{li}).$ Define auxiliary indices $s_k\in[2^{n\Rt_k}], k\in[K],$ for some auxiliary rates $(\Rt_k, k\in[k]).$ For each $(m_k,s_k)\in[2^{nR_k}]\times[2^{n\Rt_k}]$ and $k\in[K]$, generate $u_k^n(m_k,s_k)\sim\prod_{i=1}^np_{U_k}(u_{ki}).$

\emph{Encoding.} To transmit messages $m^K = (m_1, \ldots, m_K),$ the encoder transmits $w^L = (w_1, \ldots, w_L),$ such that
\[ 
(x_1^n(w_1),\ldots,x_L^n(w_L),u_1^n(m_1,s_1),\ldots,u_K^n(m_K,s_K))\in\aepvar
\]
for some $s^K\in[2^{n\Rt_1}]\times\cdots\times[2^{n\Rt_K}].$

\emph{Relaying.} On receiving the index $w_l,$ relay $l$ transmits $x_l^n(w_l).$

\emph{Decoding.} Let $\epsilon>\epsilon'.$ Upon receiving $y_k^n,$ receiver $k$ finds a message estimate $\mh_k$ such that 
\[
(u_k^n(\mh_k,s_k),y_k^n)\in\aep
\]
for some $s_k.$  

\subsubsection*{Analysis of the coding scheme}
For analyzing the coding scheme and proving Proposition~\ref{propo3}, we will need the Markov lemma \cite[Lemma 12.1]{nit} and the following additional elementary result.
\begin{lemma}\label{lemma:infochain}
Let $\Cc$, $\Tc_1$ and $\Tc_2$ be disjoint and finite index sets and fix a pmf $p(x(\Cc\cup\Tc_1\cup\Tc_2)).$ For each $k\in \Cc\cup\Tc_1\cup\Tc_2$, we independently generate $X_k^n$ according to the marginals $\prod_{i=1}^np_{X_k}(x_{ki})$. Then, as $n\to\infty$,
\[ 
\P\left(X^n(\Cc\cup \Tc_1\cup \Tc_2)\in\aep\right)\overset{.}{=}2^{-nI^*(X(\Cc\cup \Tc_1\cup \Tc_2))},\numberthis\label{eq:p1}
\]
and
\[
\P\left(X^n(\Cc\cup \Tc_1)\in\aep, X^n(\Cc\cup \Tc_2)\in\aep\right)\overset{.}{=}2^{-n[I^*(X(\Cc\cup \Tc_1))+I^*(X(\Tc_2))+I(X(\Cc);X(\Tc_2))]}.\numberthis\label{eq:p2}
\] 
\end{lemma}
Our analysis of the coding scheme follows that in \cite{ddf} but is considerably simpler because of the relative simplicity of our network model. 

Without loss of generality, let $m^K = (1, \ldots, 1)$ be the messages sent. Then the error events are:
\begin{align*}
\Ec_0 & = \Big\{\Big(X_1^n(w_1),\ldots,X_L^n(w_L),U_1^n(1,s_1),\ldots, U_K^n(1,s_K)\Big)\notin\aepvar\text{ for all }w^L\text{ and }s^K\Big\}.\\
\Ec_1 & = \Big\{\left(U_k^n(1,s_k),Y_k^n\right)\notin\aep\text{ for all }s_k,\text{ for some }k\Big\}.\\
\Ec_2 & = \Big\{\left(U_k^n(m_k,s_k),Y_k^n\right)\in\aep\text{ for some }k,\text{ some }s_k\text{, and some }m_k\ne 1\Big\}.
\end{align*}
By the packing lemma and union of events, $\P(\Ec_2\cap\Ec_1^c\cap\Ec_0^c)\to 0$ as $n\to\infty$ if 
\begin{align}R_k+\tilde{R}_k & < I(U_k;Y_k)\label{eq26}\end{align}
for all $k\in[K]$. By the Markov lemma and union of events bound ($U_k\rightarrow X^L\rightarrow Y_k$ form a Markov chain), $\P(\Ec_1\cap\Ec_0^c)\to 0$ as $n\to\infty$.

To analyze the error event $\Ec_0,$ we observe that by the manner in which the codebook is generated, $\P(\Ec_0)$ remains the same if we index the $U_k$s only by the indices $s_k$ and drop the $m_k.$ In the following analysis, we do this to simplify notation.

Let 
\[ 
\mathcal{A} := \Big\{(w^L, s^K):\left(X_1^n(w_1),\ldots,X_L^n(w_L),U_1^n(s_1),\ldots,U_K^n(s_K)\right)\in\aepvar\Big\}.
\] 
Then, $\P(\Ec_0) = \P(|\mathcal{A}|=0)$. We can write
\begin{align*}
|\mathcal{A}| & = \sum_{w^L, s^K}Z(w^L, s^K),
\end{align*}
where 
\[
Z(w^L,s^K) := \mathbbm{1}_{\left\{\left(X_1^n(w_1),\ldots,X_L^n(w_L),U_1^n(s_1),\ldots,U_K^n(s_K)\right)\in\aepvar\right\}}.
\] 
We have
\[ 
\E[Z(w^L,s^K)] = \P\left(\left(X_1^n(w_1),\ldots,X_L^n(w_L),U_1^n(s_1),\ldots,U_K^n(s_K)\right)\in\aepvar\right) =: p_1
\]
By \eqref{eq:p1}, 
\[ 
p_1\overset{.}{=}2^{-nI^*(X^L,U^K)}.\numberthis\label{eq:firstmoment}
\]
For $w^L, w'^L\in[2^{nC_1}]\times\cdots\times[2^{nC_l}]$ and $s^K, s'^K\in[2^{n\Rt_1}]\times\cdots\times[2^{n\Rt_K}],$ define
\begin{align*}
\Sc_1(w^L,w'^L) & := \left\{l\in[L]: w_l\ne w_l'\right\},\\
\Sc_2(s^K, s'^K) & := \left\{k\in[K]: s_k\ne s_k'\right\}. 
\end{align*}
Then, using \eqref{eq:p2} with index sets 
\begin{align*}
\Cc & := \left\{w_l: l\in S_1(w^L,w'^L)\right\}\cup\left\{s_k: k\in S_2(s^K,s'^K)\right\},\\
\Tc_1 & := \left\{w_l: l\in S_1^c(w^L, w'^L)\right\}\cup\left\{s_k: k\in S_2^c(s^K,s'^K)\right\}\text{, and}\\
\Tc_2 & := \left\{w_l^\prime: l\in S_1^c(w^L, w'^L)\right\}\cup\left\{s_k^\prime: k\in S_2^c(s^K, s'^K)\right\},
\end{align*}
we have
\begin{align*}
\E[Z(w^L, s^K)Z(w'^L, s'^K)]  & \overset{.}{=} 2^{-n[I^*(X^L,U^K)+I^*(X(\Sc_1),U(\Sc_2))+I(X(\Sc_1^c),U(\Sc_2^c);X(\Sc_1),U(\Sc_2))]}\\
 =: p_2(\Sc_1, \Sc_2),\numberthis\label{eq:secondmoment} 
\end{align*}
where in the definition of $p_2,$ we hide the dependence on $w^L, w'^L, s^K, s'^K.$ We then have
\begin{align*}
\E\left[|\Ac|^2\right] & = \sum_{w^L, s^K}\E[Z(w^L, s^K)] + \sum_{\substack{w^L, w'^L, s^K, s'^K:\\ w^L\ne w'^L\text{ or }s^K\ne s'^K}}\E[Z(w^L, s^K)Z(w'^L, s'^K)]\\
& = p_1\cdot 2^{n\left(\sum_{l=1}^LC_l+\sum_{k=1}^K\Rt_k\right)} \\
& \quad\quad + \sum_{\substack{\Sc_1\subseteq[L], \Sc_2\subseteq[K],\\ \Sc_1\ne\emptyset\text{ or }\Sc_2\ne\emptyset}}p_2(\Sc_1, \Sc_2)\cdot 2^{n\left(\sum_{l=1}^LC_l+\sum_{k=1}^K\Rt_k\right)}\cdot \left(2^{n\sum_{l\in \Sc_1}C_l}-1\right)\cdot\left(2^{n\sum_{k\in \Sc_2}\Rt_k}-1\right)\\
& \le p_1\cdot 2^{n\left(\sum_{l=1}^LC_l+\sum_{k=1}^K\Rt_k\right)} \\
& \quad\quad +\sum_{\substack{\Sc_1\subseteq[L], \Sc_2\subseteq[K],\\ \Sc_1\ne\emptyset\text{ or }\Sc_2\ne\emptyset}}p_2(\Sc_1, \Sc_2)\cdot 2^{n\left(\sum_{l=1}^LC_l+\sum_{k=1}^K\Rt_k+\sum_{l\in \Sc_1}C_l+\sum_{k\in \Sc_2}\Rt_k\right)}.
\end{align*}
Noting that $p_2([L], [K]) = p_1^2,$ we then have
\[ 
\Var(|\Ac|) \le p_1\cdot 2^{n\left(\sum_{l=1}^LC_l+\sum_{k=1}^K\Rt_k\right)} + \sum_{\substack{\Sc_1\subseteq[L], \Sc_2\subseteq[K],\\ \Sc_1\ne\emptyset\text{ or }\Sc_2\ne\emptyset,\\ \Sc_1\ne[L]\text{ or }\Sc_2\ne[K]}}p_2(\Sc_1, \Sc_2)\cdot 2^{n\left(\sum_{l=1}^LC_l+\sum_{k=1}^K\Rt_k+\sum_{l\in \Sc_1}C_l+\sum_{k\in \Sc_2}\Rt_k\right)}.\numberthis\label{eq:var}
\]
We also have
\[ 
\E[|\Ac|] = p_1\cdot 2^{n\left(\sum_{l=1}^LC_l+\sum_{k=1}^K\Rt_k\right)}.\numberthis\label{eq:expectation}
\] 
Using \eqref{eq:firstmoment}, \eqref{eq:secondmoment}, \eqref{eq:var}, and \eqref{eq:expectation} and manipulating exponents, we finally have, for some $\delta(\epsilon')$ that goes to zero as $\epsilon'\to 0,$ 
\[ 
\frac{\Var(|\Ac|)}{\E[|\Ac|]^2}\le\sum_{\substack{\Sc_1\subseteq[L], \Sc_2\subseteq[K],\\ \Sc_1\ne[L]\text{ or }\Sc_2\ne[K]}} 2^{-n\left(\sum_{l\in \Sc_1^c}C_l + \sum_{k\in \Sc_2^c}\Rt_k-I(X(\Sc_1^c), U(\Sc_2^c))-\delta(\epsilon^\prime)\right)}.
\]
Thus, using the inequality $\P(|\Ac| = 0)\le\Var(|\Ac|)/\E[|\Ac|^2],$ we conclude that $\P(\Ec_0)\to 0$ as $n\to\infty$ if
\begin{align}
\sum_{l\in\Sc_1^c}C_l+\sum_{k\in\Sc_2^c}\Rt_k & > I^*(X(\Sc_1^c),U(\Sc_2^c))\label{eq:basiccutsetcondition}
\end{align}
for all $\Sc_1\subseteq[L]$ and $\Sc_2\subseteq[K].$

Combining this with \eqref{eq26} to eliminate the auxiliary rates, the rates $R_k$ satisfy, for every $\Sc_1\subseteq[L]$ and $\Sc_2\subseteq[K]$,
\begin{align*}
& \sum_{k\in\Sc_2^c}R_k\\
 & < \sum_{l\in \Sc_1^c}C_l +\sum_{k\in \Sc_2^c}I(U_k;Y_k)-I^*(X(\Sc_1^c),U(\Sc_2^c))\\
& =  \sum_{l\in \Sc_1^c}C_l+\sum_{k\in \Sc_2^c}\left(I(U_k; X^L,Y_k)-I(U_k;X^L|Y_k)\right) -I^*(X(\Sc_1^c),U(\Sc_2^c))\\
& \stackrel{(a)}{=}  \sum_{l\in \Sc_1^c}C_l-\sum_{k\in \Sc_2^c}I(U_k;X^L|Y_k)+\sum_{k\in \Sc_2^c}I(U_k;X^L) -I^*(X(\Sc_1^c),U(\Sc_2^c))\\
& =  \sum_{l\in \Sc_1^c}C_l-\sum_{k\in \Sc_2^c}I(U_k;X^L|Y_k)-\sum_{k\in \Sc_2^c}h(U_k|X^L)-\sum_{l\in \Sc_1^c}h(X_l)+h(X(\Sc_1^c),U(\Sc_2^c))\\
& =  \sum_{l\in \Sc_1^c}C_l-\sum_{k\in \Sc_2^c}I(U_k;X^L|Y_k)-\sum_{k\in \Sc_2^c}h(U_k|X^L)-I^*(X(\Sc_1^c))+h(U(\Sc_2^c)|X(\Sc_1^c))\\
& =  \sum_{l\in \Sc_1^c}C_l-\sum_{k\in \Sc_2^c}I(U_k;X^L|Y_k)-h(U(\Sc_2^c)|X^L)-I^*(U(\Sc_2^c)|X^L)-I^*(X(\Sc_1^c))+h(U(\Sc_2^c)|X(\Sc_1^c))\\
 & =  \sum_{l\in \Sc_1^c}C_l-\sum_{k\in \Sc_2^c}I(U_k;X^L|Y_k) + I(X(\Sc_1);U(\Sc_2^c)|X(\Sc_1^c)) - I^*(X(\Sc_1^c))-I^*(U(\Sc_2^c)|X^L).\numberthis\label{eq:finalconditionsdownlink}
\end{align*}
Here, $(a)$ follows from the fact that $U_k\rightarrow X^L\rightarrow Y_k$ form a Markov chain. In addition, if $\Sc_2=[K]$ in \eqref{eq:basiccutsetcondition}, we obtain the additional conditions 
\begin{align*}
\sum_{l\in \Sc_1^c}C_l & > I^*(X(\Sc_1^c))\numberthis\label{eq:ddfadditionalconstraints}
\end{align*}
for every $\Sc_1\subsetneq[L].$ This completes the proof.
\begin{remark}
The constraints \eqref{eq:ddfadditionalconstraints} can be shown to be inactive using techniques similar to \cite[Appendix E]{ddf}.
\end{remark}
\subsection*{Proof of Proposition~\ref{downgenouter}}
We use $X_i(\Sc_1), X^n(\Sc_1), Y_i(\Sc_2), Y^n(\Sc_2)$ to convey similar meanings as in the proof of Proposition~\ref{upgenouter} in Appendix~\ref{upproofs}. For $k\in[K],$ let $M_k$ denote the message intended for receiver $k$ and let $W_l$ denote the index communicated by the central processor to relay $l.$ Then, for every $\Sc_1\subseteq[L],$ $X^n(\Sc_1)$ is a function of $W(\Sc_1),$ which is itself a function of $M^K.$ For every $\Sc_1\subseteq[L], \Sc_2\subsetneq[K],$ we must have, by Fano's inequality, 
\[
H(M(\Sc_2^c)\cond M(\Sc_2), Y^n(\Sc_2^c))\le H(M(\Sc_2^c)\cond Y^n(\Sc_2^c))\le n\epsilon_n,
\]
where $\epsilon_n\to 0$ as $n\to\infty.$ Therefore, since $H(M(\Sc_2^c)\cond M(\Sc_2)) = n\sum_{k\in \Sc_2^c}R_k,$ we have
\begin{align*}
n\sum_{k\in \Sc_2^c}R_k & \le I(M(\Sc_2^c); Y^n(\Sc_2^c)\cond M(\Sc_2)) + n\epsilon_n\\
& \le I(M(\Sc_2^c), W(\Sc_1^c); Y^n(\Sc_2^c)\cond M(\Sc_2)) + n\epsilon_n\\
&  = I(M(\Sc_2^c); Y^n(\Sc_2^c)\cond M(\Sc_2), W(\Sc_1^c)) + I(W(\Sc_1^c); Y^n(\Sc_2^c)\cond M(\Sc_2)) + n\epsilon_n\\
& \le I(M(\Sc_2^c); Y^n(\Sc_2^c)\cond M(\Sc_2), W(\Sc_1^c)) + H(W(\Sc_1^c)) + n\epsilon_n\\
& \stackrel{(a)}{=} I(M(\Sc_2^c), W(\Sc_1); Y^n(\Sc_2^c)\cond M(\Sc_2), W(\Sc_1^c)) + H(W(\Sc_1^c)) + n\epsilon_n\\
& = \sum_{i=1}^nI(M(\Sc_2^c), W(\Sc_1); Y_i(\Sc_2^c)\cond M(\Sc_2), W(\Sc_1^c), Y^{i-1}(\Sc_2^c)) + H(W(\Sc_1^c)) + n\epsilon_n\\
& \stackrel{(b)}{=}\sum_{i=1}^nI(M(\Sc_2^c), W(\Sc_1); Y_i(\Sc_2^c)\cond M(\Sc_2), W(\Sc_1^c), Y^{i-1}(\Sc_2^c), X_i(\Sc_1^c)) + H(W(\Sc_1^c)) + n\epsilon_n\\
& \le\sum_{i=1}^nI(M^K, W^L, Y^{i-1}(\Sc_2^c); Y_i(\Sc_2^c)\cond X_i(\Sc_1^c)) + H(W(\Sc_1^c)) + n\epsilon_n\\
& \stackrel{(c)}{=} \sum_{i=1}^nI(M^K, W^L, Y^{i-1}(\Sc_2^c), X_i(\Sc_1); Y_i(\Sc_2^c)\cond X_i(\Sc_1^c)) + H(W(\Sc_1^c)) + n\epsilon_n\\
& \stackrel{(d)}{=}\sum_{i=1}^nI(X_i(\Sc_1); Y_i(\Sc_2^c)\cond X_i(\Sc_1^c)) + H(W(\Sc_1^c)) + n\epsilon_n\\
& \stackrel{(e)}{\le}\sum_{i=1}^nI(X_i(\Sc_1); Y_i(\Sc_2^c)\cond X_i(\Sc_1^c)) + n\sum_{l\in \Sc_1^c}C_l + n\epsilon_n.
\end{align*}
Here, $(a)$ follows since conditioned on $M(\Sc_2),$ $W(\Sc_1)$ is a function of $M(\Sc_2^c);$ $(b)$ follows since $X_i(\Sc_1^c)$ is a function of $W(\Sc_1^c);$ $(c)$ follows since $X_i(\Sc_1)$ is a function of $W^L;$ $(d)$ follows since $(M^K, W^L, Y^{i-1}(\Sc_2^c))\rightarrow(X_i(\Sc_1), X_i(\Sc_1^c))\rightarrow(Y_i(\Sc_2^c))$ form a Markov chain (by the memorylessness of the second hop), and $(e)$ follows since $W(\Sc_1^c)$ is supported on a set of size $\prod_{l\in \Sc_1^c}2^{nC_l}.$ Defining a random variable $Q\sim\mathrm{Unif}([n])$ independent of all other random variables, writing $X(\Sc_1) := X_Q(\Sc_1)$ and $Y(\Sc_2):= Y_Q(\Sc_2),$ noting that $Q\rightarrow X^L\rightarrow Y^K$ form a Markov chain, and letting $n$ tend to infinity leads to
\begin{align*}
\sum_{k\in \Sc_1}R_k \le I(X(\Sc_1); Y(\Sc_2^c)\cond X(\Sc_1^c)) + \sum_{l\in \Sc_1^c}C_l\numberthis\label{cutsetdownappendix}
\end{align*}
for all $\Sc_1\subseteq[L], \Sc_2\subsetneq[K]$ for some pmf $p(x^L),$ and thus completes the proof.
%\end{IEEEproof}

\subsection*{Proof of Theorem~\ref{downgaptheo1}}
Recall from \eqref{downinngaussinformal} in Section~\ref{downgauss} that for the Guassian downlink network model with channel gain matrix $H$ and average power constraint $P$ on each relay, a rate tuple $(R_1,\ldots,R_K)$ is achievable if for every $\Sc_1\subseteq[L]$ and $\Sc_2\subsetneq[K],$   
\[
\sum_{k\in \Sc_2^c}R_k \le \frac{1}{2}\log\Big|\frac{P}{\sigma^2}H_{\Sc_2^c,\Sc_1}H_{\Sc_2^c,\Sc_1}^T+I\Big|+\sum_{l\in \Sc_1^c}C_l -\frac{|\Sc_2^c|}{2}\log\left(1+\frac{1}{\sigma^2}\right) =: F_{\mathrm{in}}(\Sc_1,\Sc_2).\numberthis\label{downinngaussappendix}
\]
The cutset bound \eqref{cutsetdownappendix} similarly simplifies (cf.\@ \eqref{cutsetdowngauss})to
\begin{align*}
\sum_{k\in \Sc_2^c}R_k & \le \frac{1}{2}\log\Big\vert H_{\Sc_2^c,\Sc_1}\Gamma_{\Sc_1|\Sc_1^c}H_{\Sc_2^c,\Sc_1}^T+I\Big\vert+\sum_{l\in \Sc_1^c}C_l\\
& =: F_\mathrm{out}(\Sc_1,\Sc_2)\numberthis\label{cutsetdowngaussappendix}
\end{align*}
for all $\Sc_1\subseteq[L]$ and $\Sc_2\subsetneq[K]$ for some covariance matrix $\Gamma\succeq 0$ satisfying $\Gamma_{ll}\le P$ for all $l\in[L].$ To prove Theorem~\ref{downgaptheo1} on the gap between the regions described by \eqref{downinngaussappendix} and \eqref{cutsetdowngaussappendix}, we need one more lemma, which is immediate from elementary calculus.
\begin{lemma}\label{littlelemma}
For $x>1,$ $x\log x-(x-1)\log(x-1)\le\log(ex).$
\end{lemma}
\begin{IEEEproof}
Let $f(x) = x\log x$ for $x>0.$ We then have $f'(x) = \log x+(1/\ln 2) = \log(ex),$ which is increasing in $x.$ Therefore, for $x>1,$
\begin{align*}
f(x) - f(x-1) & \le f'(x)\left(x-(x-1)\right)\\
& = \log(ex). 
\end{align*}
\end{IEEEproof}
%\begin{IEEEproof}[Proof of Theorem~\ref{downgaptheo1}] 
Note that unlike \eqref{monotone} in Appendix~\ref{upproofs}, $F_{\mathrm{in}}$ is not necessarily monotonic. We overcome this difficulty by rephrasing the inner bound \eqref{downinngaussappendix} as  
\begin{align*}
\sum_{k\in \Sc_2^c}R_k\le\min_{ \Tc_2\subseteq \Sc_2}F_{\mathrm{in}}(\Sc_1, \Tc_2).\numberthis\label{ddfmonotone}
\end{align*}   
We observe that the right-hand side of \eqref{ddfmonotone} is increasing with $\Sc_2^c$ for a fixed $\Sc_1$, so we can apply the technique developed in the proof of Theorem~\ref{upgaptheo1} to compute an upper bound on $\Delta$. We thus write
{\allowdisplaybreaks
\begin{align*}
\Delta & = \max_{\Sc_2\subsetneq[K]}\Bigg[\frac{\min_{\Sc_1}F_{\mathrm{out}}(\Sc_1,\Sc_2)}{|\Sc_2^c|}-\frac{\min_{\Sc_1}\min_{\Tc_2\subseteq\Sc_2}F_{\mathrm{in}}(\Sc_1, \Tc_2)}{|\Sc_2^c|}\Bigg]\\
& \le\max_{\substack{\Sc_1\subseteq[L]\\ \Sc_2\subsetneq[K]\\  \Tc_2\subseteq \Sc_2}}\frac{F_{\mathrm{out}}(\Sc_1,\Sc_2)-F_{\mathrm{in}}(\Sc_1, \Tc_2)}{|\Sc_2^c|}\\
& =  \max_{\substack{\Sc_1\subseteq[L]\\ \Sc_2\subsetneq[K]\\  \Tc_2\subseteq \Sc_2}}\frac{1}{2|\Sc_2^c|}\Bigg[\log\frac{\Big\vert H_{\Sc_2^c,\Sc_1}\Gamma_{\Sc_1|\Sc_1^c}H_{\Sc_2^c,\Sc_1}^T+I\Big\vert}{\Big\vert\frac{P}{\sigma^2}H_{\Tc_2^c,\Sc_1}H_{\Tc_2^c,\Sc_1}^T+I\Big\vert} +|\Tc_2^c|\log\left(1+\frac{1}{\sigma^2}\right)\Bigg]\\
& \stackrel{(a)}{\le}\max_{\substack{\Sc_1\subseteq[L]\\ \Sc_2\subsetneq[K]\\  \Tc_2\subseteq\Sc_2}}\frac{1}{2|\Sc_2^c|}\Bigg[\log\frac{\Big\vert H_{\Sc_2^c,\Sc_1}\Gamma_{\Sc_1}H_{\Sc_2^c,\Sc_1}^T+I\Big\vert}{\Big\vert\frac{P}{\sigma^2}H_{\Sc_2^c,\Sc_1}H_{\Sc_2^c,\Sc_1}^T+I\Big\vert}+| \Tc_2^c|\log\left(1+\frac{1}{\sigma^2}\right)\Bigg]s,
\numberthis\label{eq18}
\end{align*}}%
where $(a)$ follows since $\Gamma_{\Sc_1}\succeq\Gamma_{\Sc_1|\Sc_1^c}$ and for any matrix $A$ and $\alpha>0$, $|I+\alpha AA^T|$ increases when we add more rows to $A$. Writing $\Gamma_{\Sc_1}=U\Lambda U^T$ where $U$ is orthogonal and $\Lambda$ is diagonal, and letting $H_{\Sc_2^c,\Sc_1}U = [\begin{matrix}b_1 & b_2 & \cdots & b_{|\Sc_1|}\end{matrix}$], where $b_1,\ldots, b_{|\Sc_1|}$ are $|\Sc_2^c|\times 1$ vectors satisfying $\sum_{l=1}^{|\Sc_1|}\norm{b_l}^2 = \norm{H_{\Sc_2^c,\Sc_1}}_F^2$, we have
\begin{align*}
\log\frac{|H_{\Sc_2^c, \Sc_1}\Gamma_{\Sc_1}H_{\Sc_2^c,\Sc_1}^T+I|}{|\frac{P}{\sigma^2}H_{\Sc_2^c,\Sc_1}H_{\Sc_2^c,\Sc_1}^T+I|}
& = \log\frac{\Big|I+\sum_{l=1}^{|\Sc_1|}\lambda_lb_lb_l^T\Big|}{\Big|I+\frac{P}{\sigma^2}\sum_{l=1}^{|\Sc_1|}b_lb_l^T\Big|} \\
& \stackrel{(a)}{\le}   \log\frac{\Big|I+P|\Sc_1|\sum_{l=1}^{|\Sc_1|}b_lb_l^T\Big|}{\Big|I+\frac{P}{\sigma^2}\sum_{l=1}^{|\Sc_1|}b_lb_l^T\Big|}\\
& \stackrel{(b)}{=}   \sum_{k=1}^{|\Sc_2^c|}\log\frac{1+P|\Sc_1|\mu_k}{1+\frac{P}{\sigma^2}\mu_k}\\
& \le  |\Sc_2^c|\log\left(\sigma^2|\Sc_1|\right),
\end{align*}
provided $\sigma^2\ge\frac{1}{|\Sc_1|}$. Here, $(a)$ follows since the trace of $\Gamma_{\Sc_1}$ is upper bounded by $P|\Sc_1|$ and in $(b)$, $\mu_1,\ldots,\mu_{|\Sc_2^c|}$ are the (non-negative) eigenvalues of $\sum_{l=1}^{|\Sc_1|}b_lb_l^T$. Continuing from (\ref{eq18}), we thus have
\begin{align*}
\Delta & \le\max_{\substack{\Sc_1\subseteq[L]\\ \Sc_2\subsetneq[K]\\  \Tc_2\subseteq \Sc_2}}\Bigg[\frac{|\Tc_2^c|\log\left(1+\frac{1}{\sigma^2}\right)}{2|\Sc_2^c|}+\frac{1}{2}\log\left(\sigma^2|\Sc_1|\right)\Bigg]\\
&  = \frac{K}{2}\log\left(1+\frac{1}{\sigma^2}\right)+\frac{1}{2}\log(\sigma^2L).\numberthis\label{downlinkaftermaximization}
\end{align*}
This holds for every $\sigma^2\ge 1$, so we set $\sigma^2 = K-1$ (for $K\ge 2$) to obtain
\begin{align*}
\Delta & \le\frac{1}{2}\log L+\frac{1}{2}\left(K\log K-(K-1)\log(K-1)\right)\\
& \stackrel{(a)}{\le}\frac{1}{2}\left(\log L+\log K+\frac{1}{\ln 2}\right)\\
& = \frac{1}{2}\log(eKL).
\end{align*}
Here, $(a)$ follows from lemma~\ref{littlelemma}. For $K = 1,$ we can set $\sigma^2 = 1$ in \eqref{downlinkaftermaximization} to obtain 
\[
\Delta\le\frac{1}{2}\log(2L)\le\frac{1}{2}\log(eL).
\]
This establishes the first part of Theorem~\ref{downgaptheo1}. 

For the sum-rate gap, consider
\begin{align*}
\Delta_\text{sum} & \le\max_{\Sc_1, \Sc_2}\left(F_\mathrm{out}(\Sc_1, \Sc_2) - F_\mathrm{in}(\Sc_1, \Sc_2)\right)\\
& \le \max_{\Sc_1, \Sc_2}\left[\frac{1}{2}\log\frac{\left|H_{\Sc_2^c, \Sc_1}\Gamma_{\Sc_1}H_{\Sc_2^c, \Sc_1}^T + I\right|}{\left|\frac{P}{\sigma^2}H_{\Sc_2^c, \Sc_1}H_{\Sc_2^c, \Sc_1}^T + I\right|} + frac{|\Sc_2^c|}{2}\log\left(1+\frac{1}{\sigma^2}\right)\right]\\
& \le\max_{\Sc_1, \Sc_2}\left[\frac{1}{2}\log\frac{\left|I+\sum_{l=1}^{|\Sc_1|}\lambda_lb_lb_l^T\right|}{\left|I+\frac{P}{\sigma^2}\sum_{l=1}^{|\Sc_1|}b_lb_l^T\right|}+ frac{|\Sc_2^c|}{2}\log\left(1+\frac{1}{\sigma^2}\right)\right]\\
& \le \max_{\Sc_1, \Sc_2}\left[\frac{\min\{|\Sc_2^c|, |\Sc_1|\}}{2}\log(\sigma^2|\Sc_1|)+\frac{|\Sc_2^c|}{2}\log\left(1+\frac{1}{\sigma^2}\right)\right]\numberthis\label{rankexpdownlinksumrate}
\end{align*}
if $\sigma^2\ge 1/|\Sc_1|$ for each $\Sc_1\ne\emptyset.$ Maximization of \eqref{rankexpdownlinksumrate} over $|\Sc_1|$ and $|\Sc_2^c|$ yields, for $\sigma^2\ge 1,$
\[
\Delta_\text{sum}\le\frac{\min\{L, K\}}{2}\log(L\sigma^2) + \frac{K}{2}\log\left(1+\frac{1}{\sigma^2}\right).\numberthis\label{downlinksumrategapsigma2}
\]
We can then choose $\sigma^2 = 1$ in \eqref{downlinksumrategapsigma2} to obtain
\[
\Delta_\text{sum}\le\frac{\min\{L, K\}}{2}\log L + \frac{K}{2},
\]
completing the proof. 
\section{MIMO C-RAN Proof Sketches}\label{mimoproofs}
\subsection*{Proof Sketch of Proposition~\ref{mimogap}}
The proof is an extension of the proofs of Theorems~\ref{upgaptheo1} and \ref{downgaptheo1}, presented in Appendices~\ref{upproofs} and \ref{downproofs}, respectively. Let us focus on the uplink first. Recall (cf Section~\ref{mimocran}) that $\Rr_\mathrm{up}^\text{NCF}(\sigma^2,\Gamma_1,\ldots,\Gamma_K)$ is characterized by rate tuples $(R_1, \ldots, R_K)$ satisfying
\[
\sum_{k\in \Sc_1}R_k \le \frac{1}{2}\log\Big\vert\frac{\sum_{k\in \Sc_1}G_{\Sc_2^c,k}\Gamma_kG_{S_2^c,k}^T}{\sigma^2+1}+I\Big\vert +\sum_{l\in \Sc_2}C_l-\frac{N_r\vert \Sc_2\vert}{2}\log\left(1+\frac{1}{\sigma^2}\right) =: f_{\mathrm{in}}(\Sc_1,\Sc_2,\Gamma_1,\ldots,\Gamma_K)\numberthis\label{upinngaussinformalmimo1}
\]
for all $\Sc_1\subseteq[K]$ and $\Sc_2\subseteq[L].$ The cutset bound $\Rr_\mathrm{up}^\text{CS}$ is characterized by rate tuples $(R_1, \ldots, R_K)$ satisfying
\[
\sum_{k\in S_1}R_k \le \frac{1}{2}\log\Big\vert\sum_{k\in \Sc_1}G_{\Sc_2^c,k}\Gamma_kG_{\Sc_2^c,k}^T+I\Big\vert +\sum_{l\in \Sc_2}C_l =: f_{\mathrm{out}}(\Sc_1,\Sc_2,\Gamma_1,\ldots,\Gamma_K).\numberthis\label{upoutgaussinformalmimo1}
\] 
Similar to Lemma~\ref{monotonelemma} in Appendix~\ref{upproofs}, $f_{\mathrm{in}}(\Sc_1,\Sc_2,\Gamma_1,\ldots,\Gamma_K)$ satisfies the monotonicity property for each fixed $\Sc_2, \Gamma_1, \ldots, \Gamma_K.$ Therefore, similar to the line of argument in Appendix~\ref{upproofs}, the per-user rate gap from the cutset bound can be upper-bounded as 
\begin{align*}
\Delta^\mathrm{up} & \le \max_{\substack{\Sc_1\subseteq[K]\\ \Sc_1\ne\emptyset}}\frac{\max_{\Gamma_1, \ldots, \Gamma_K}\min_{\Sc_2}f_{\mathrm{out}}(\Sc_1,\Sc_2,\Gamma_1,\ldots, \Gamma_K)-\max_{\Gamma_1, \ldots, \Gamma_K}\min_{\Sc_2}f_{\mathrm{in}}(\Sc_1,\Sc_2,\Gamma_1,\ldots, \Gamma_K)}{|\Sc_1|}\\
& \le\max_{\substack{\Sc_1\subseteq[K]\\ \Sc_1\ne\emptyset}}\max_{\Gamma_1, \ldots, \Gamma_K}\max_{\Sc_2}\frac{1}{2|\Sc_1|}\log\frac{\Big\vert\sum_{k\in \Sc_1}G_{\Sc_2^c,k}\Gamma_kG_{\Sc_2^c,k}^T+I\Big\vert}{\Big\vert\frac{\sum_{k\in \Sc_1}G_{\Sc_2^c,k}\Gamma_kG_{\Sc_2^c,k}^T}{\sigma^2+1}+I\Big\vert}+\frac{N_r|\Sc_2|}{2}\log\left(1+\frac{1}{\sigma^2}\right)\\
& \stackrel{(a)}{\le}\max_{\substack{k\in[K]\\ l\in\{0,\ldots, l\}}}\left[\frac{\min\{N_r(L-l), N_uk\}}{2k}\log(1+\sigma^2) + \frac{N_rl}{2}\log\left(1+\frac{1}{\sigma^2}\right)\right],
\end{align*} 
where in $(a),$ we set $|\Sc_1| = k,$ $|\Sc_2| = l,$ and upper-bound $\rank\left(\sum_{k\in \Sc_1}G_{\Sc_2^c,k}\Gamma_kG_{\Sc_2^c,k}^T\right)$ by $\min\{N_r(L-l), N_uk\}.$  The maximization yields
\begin{align*}
\Delta^\mathrm{up} & \le \begin{cases}\frac{N_u}{2}\log(1+\sigma^2)+\frac{N_rL-N_u}{2}\log(1+\frac{1}{\sigma^2}), & \sigma^2\ge 1, N_rL\ge N_u,\\
 \frac{N_rL}{2}\log(1+\sigma^2), &  \sigma^2\ge 1, N_rL < N_u,\\
 \frac{N_rL}{2}\log(1+\frac{1}{\sigma^2}), & \sigma^2\le1.\end{cases}
\end{align*}
Since this holds for every $\sigma^2>0$, we set 
\[
\sigma^2=\frac{N_rL}{N_u}-1
\]
 for $L\ge 2N_u/N_r$ to obtain
\begin{align*}
 \Delta^\mathrm{up} & \le \frac{N_u}{2}\log\frac{N_rL}{N_u}+\frac{N_rL-N_u}{2}\log\left(1+\frac{N_u}{N_rL-N_u}\right)\\
& \stackrel{(a)}{\le}\frac{N_u}{2}\log\frac{eN_rL}{N_u}.\numberthis\label{finalexupmimo}
\end{align*} 
Here, $(a)$ follows since from elementary calculus, we know that for $x>0,$ $\log(1+x)\le x\log e.$
For $L<2N_u/N_r,$ we can choose $\sigma^2=1$ to obtain 
\[
\Delta^\mathrm{up}\le \frac{N_rL}{2}\stackrel{(a)}{\le}\frac{N_u}{2}\log\frac{eLN_r}{N_u}, 
\] 
where $(a)$ follows from the inequality $x\le\log(ex)$ for $x<1/2.$ This, together with \eqref{finalexupmimo}, establishes the per-user rate gap for the uplink MIMO C-RAN. For the sum-rate gap, we consider
\begin{align*}
\Delta_\mathrm{sum}^\mathrm{up} & \le\max_{\Sc_1, \Sc_2,\Gamma_1, \ldots, \Gamma_K}\left(f_\mathrm{out}(\Sc_1, \Sc_2,\Gamma_1,\ldots,\Gamma_K) - f_\mathrm{in}(\Sc_1, \Sc_2,\Gamma_1,\ldots,\Gamma_K)\right)\\
& \le \max_{\substack{k\in[K]\\l\in\{0, \ldots, L\}}}\Big[\frac{\min\{N_r(L-l),N_uk\}}{2}\log(1+\sigma^2)+\frac{N_rl}{2}\log\left(1+\frac{1}{\sigma^2}\right)\Big].\numberthis\label{rankexpuplinkmimosumrate}
\end{align*}
Maximization of \eqref{rankexpuplinkmimosumrate} over $l$ and $k$ yields, for $\sigma^2\ge 1,$
\[
\Delta_\mathrm{sum}^\mathrm{up}\le\begin{cases}\frac{N_uK}{2}\log(1+\sigma^2)+\frac{N_rL-N_uK}{2}\log(1+\frac{1}{\sigma^2}), & N_rL\ge N_uK,\\ \frac{N_rL}{2}\log(1+\sigma^2), & N_rL<N_uK.\end{cases}
\]
For $N_rL\le 2N_uK,$ we can then choose $\sigma^2 = 1$ to obtain an upper bound $\Delta_\mathrm{sum}^\mathrm{up}\le N_rL/2.$ For $N_rL>2N_uK,$ we can choose $\sigma^2 = N_rL/N_uK - 1\ge 1$ to obtain
\begin{align*}
\Delta_\mathrm{sum}^\mathrm{up} & \le \frac{N_uK}{2}\log\left(\frac{N_rL}{N_uK}\right) + \frac{N_rL-N_uK}{2}\log\left(1 + \frac{N_uK}{N_rL-N_uK}\right)\numberthis\label{uplinkbinentropy}\\
& = \frac{N_rL}{2}H(N_uK/N_rL).
\end{align*}

For the downlink MIMO C-RAN with channel gain matrix $H\in\Real^{N_uK\times N_rL},$ recall (cf Section~\ref{mimocran}) that $\Rr_\mathrm{down}^\text{DDF}(\sigma^2, \Gamma_1,\ldots,\Gamma_L)$ is characterized by rate tuples $(R_1, \ldots, R_K)$ satisfying
\[
\sum_{k\in \Sc_2^c}R_k \le \frac{1}{2}\log\Big\vert\frac{\sum_{l\in \Sc_1}H_{\Sc_2^c,l}\Gamma_lH_{\Sc_2^c,l}^T}{\sigma^2}+I\Big\vert +\sum_{l\in \Sc_1^c}C_l-\frac{N_u\vert \Sc_2^c\vert}{2}\log\left(1+\frac{1}{\sigma^2}\right) =: F_{\mathrm{in}}(\Sc_1,\Sc_2,\Gamma_1,\ldots,\Gamma_L)\numberthis\label{downinngaussinformalmimo1}
\]
for all $\Sc_1\subseteq[L]$ and $\Sc_2\subseteq[K].$ The cutset bound $\Rr_\mathrm{down}^\text{CS}$ is characterized by rate tuples $(R_1, \ldots, R_K)$ satisfying
\[
\sum_{k\in \Sc_2^c}R_k \le \frac{1}{2}\log\Big\vert H_{\Sc_2^c,\Sc_1}\tilde{\Gamma}_{\Sc_1\cond \Sc_1^c}H_{\Sc_2^c,\Sc_1}^T+I\Big\vert +\sum_{l\in \Sc_1^c}C_l =: F_{\mathrm{out}}(\Sc_1,\Sc_2,\tilde{\Gamma}),\numberthis\label{downoutgaussinformalmimo1}
\] 
where $\tilde{\Gamma}$ is a general $N_rL\times N_rL$ input covariance matrix satisfying the block trace constraints. 
Similar to Appendix~\ref{downproofs}, the per-user rate gap from the cutset bound can therefore be upper-bounded as 
{\allowdisplaybreaks
\begin{align*}
\Delta^\mathrm{down} & \le \max_{\Sc_2\subsetneq[K]}\Bigg[\frac{1}{|\Sc_2^c|}\left(\max_{\tilde{\Gamma}}\min_{\Sc_1}F_{\mathrm{out}}(\Sc_1,\Sc_2,\tilde{\Gamma})-\max_{\Gamma_1,\ldots,\Gamma_L}\min_{\Sc_1}\min_{\Tc_2\subseteq\Sc_2}F_{\mathrm{in}}(\Sc_1, \Tc_2,\Gamma_1,\ldots,\Gamma_L)\right)\Bigg]\\
& \le\max_{\substack{\Sc_1, \Sc_2, \Tc_2\subseteq\Sc_2}}\max_{\tilde{\Gamma}}\min_{\Gamma_1, \ldots, \Gamma_L}\frac{F_{\mathrm{out}}(\Sc_1,\Sc_2,\tilde{\Gamma})-F_{\mathrm{in}}(\Sc_1, \Tc_2,\Gamma_1, \ldots, \Gamma_L)}{|\Sc_2^c|}\\
& \le\max_{\substack{\Sc_1, \Sc_2, \Tc_2\subseteq \Sc_2}}\max_{\tilde{\Gamma}}\min_{\Gamma_1, \ldots, \Gamma_L}\frac{1}{2|\Sc_2^c|}\Bigg[\log\frac{\Big\vert H_{\Sc_2^c,\Sc_1}\tilde{\Gamma}_{\Sc_1}H_{\Sc_2^c,\Sc_1}^T+I\Big\vert}{\Big\vert\frac{\sum_{l\in \Sc_1}H_{\Sc_2^c,l}\Gamma_lH_{\Sc_2^c,l}^T}{\sigma^2}+I\Big\vert}+N_u|\Tc_2^c|\log\left(1+\frac{1}{\sigma^2}\right)\Bigg].
\numberthis\label{eq:beforeeigendecomposedowngap}
\end{align*}}%
Writing $\tilde{\Gamma}_{\Sc_1}=U\Lambda U^T$ where $U$ is orthogonal and $\Lambda$ is diagonal, letting $H_{\Sc_2^c,\Sc_1}U = [\begin{matrix}B_1 & B_2 & \cdots & B_{|\Sc_1|}\end{matrix}$], where $B_1,\ldots, B_{|\Sc_1|}$ are $N_u|S_2^c|\times N_r$ matrices satisfying $\sum_{l=1}^{|\Sc_1|}\tr(B_l^TB_l) = \norm{H_{\Sc_2^c,\Sc_1}}_F^2,$ and taking $\Gamma_l = (P/N_r)I$ for each $l,$ we have
\begin{align*}
\log\frac{\Big\vert H_{\Sc_2^c,\Sc_1}\tilde{\Gamma}_{\Sc_1}H_{\Sc_2^c,\Sc_1}^T+I\Big\vert}{\Big\vert\frac{\sum_{l\in \Sc_1}H_{\Sc_2^c,l}\Gamma_lH_{\Sc_2^c,l}^T}{\sigma^2}+I\Big\vert}
& = \log\frac{\Big|I+\sum_{l=1}^{|\Sc_1|}B_l\Lambda_lB_l^T\Big|}{\Big|I+\frac{P}{N_r\sigma^2}\sum_{l=1}^{|\Sc_1|}B_lB_l^T\Big|} \\
& \stackrel{(a)}{\le}   \log\frac{\Big|I+P|\Sc_1|\sum_{l=1}^{|\Sc_1|}B_lB_l^T\Big|}{\Big|I+\frac{P}{N_r\sigma^2}\sum_{l=1}^{|\Sc_1|}B_lB_l^T\Big|}\\
& \le  \min\{N_u|\Sc_2^c|, N_r|\Sc_1|\}\log\left(\sigma^2N_r|\Sc_1|\right),
\end{align*}
provided $\sigma^2\ge\frac{1}{N_r|\Sc_1|}$. Here, $(a)$ follows since the trace of $\tilde{\Gamma}_{\Sc_1}$ is upper bounded by $P|\Sc_1|.$ Continuing from \eqref{eq:beforeeigendecomposedowngap}, we thus have
\begin{align*}
\Delta^\mathrm{down} & \le\max_{\substack{\Sc_1, \Sc_2, \Tc_2\subseteq \Sc_2}}\Bigg[\frac{N_u|\Tc_2^c|\log\left(1+\frac{1}{\sigma^2}\right)}{2|\Sc_2^c|}+\frac{\min\{N_u|\Sc_2^c|, N_r|\Sc_1|\}}{2|\Sc_2^c|}\log\left(\sigma^2N_r|\Sc_1|\right)\Bigg]\\
&  = \frac{N_uK}{2}\log\left(1+\frac{1}{\sigma^2}\right)+\frac{\min\{N_u, N_rL\}}{2}\log(\sigma^2N_rL).\numberthis\label{downlinkmimoaftermaximization}
\end{align*}
This holds for every $\sigma^2\ge 1$, so we set $\sigma^2 = K-1$ for $K\ge 2$ and $N_rL\ge N_u$ to obtain
\begin{align*}
\Delta^\mathrm{down} & \le\frac{N_u}{2}\log L+\frac{N_u}{2}\left(\log N_r+K\log K-(K-1)\log(K-1)\right)\\
& \le \frac{N_u}{2}\log(eN_rLK).\numberthis\label{eq:dlcase1}
\end{align*}
For $K = 1$ and $N_rL\ge N_u,$ we can set $\sigma^2 = 1$ in \eqref{downlinkmimoaftermaximization} to obtain 
\[
\Delta^\mathrm{down}\le\frac{N_u}{2}\log(2N_rL)\le\frac{N_u}{2}\log(eN_rL).\numberthis\label{eq:dlcase2}
\]
For $N_rL<N_u$ and $N_uK\ge 2N_rL,$ set 
\[
\sigma^2 = \frac{N_uK}{N_rL}-1
\]
to obtain 
\[
\Delta^\mathrm{down}\le\frac{N_uK}{2}\log(N_uK) - \frac{N_uK-N_rL}{2}\log(N_uK-N_rL)\le\frac{N_rL}{2}\log(eN_uK),\numberthis\label{eq:dlcase3}
\]
and for $N_rL<N_u<2N_rL$ and $K = 1,$ set $\sigma^2 = 1$ to obtain
\begin{align*}
\Delta^\mathrm{down} & \le\frac{N_u}{2}+\frac{N_rL}{2}\log(N_rL)\\
& \le \frac{N_u}{2}\log(eN_rL).\numberthis\label{eq:dlcase4}
\end{align*}
The results \eqref{eq:dlcase1}--\eqref{eq:dlcase4} establish the per-user gap results for the downlink MIMO C-RAN.

For the sum-rate gap, we similarly obtain
\begin{align*}
\Delta_\mathrm{sum}^\mathrm{down} & \le\max_{\Sc_1, \Sc_2}\max_{\tilde{\Gamma}}\min_{\Gamma_1,\ldots,\Gamma_L}\left(F_\mathrm{out}(\Sc_1, \Sc_2,\tilde{\Gamma}) - F_\mathrm{in}(\Sc_1, \Sc_2,\Gamma_1,\ldots,\Gamma_L)\right)\\
& \le\max_{\Sc_1, \Sc_2}\left[\frac{\min\{N_u|\Sc_2^c|, N_r|\Sc_1|\}}{2}\log(\sigma^2N_r|\Sc_1|)+\frac{N_u|\Sc_2^c|}{2}\log\left(1+\frac{1}{\sigma^2}\right)\right]\numberthis\label{rankexpdownlinkmimosumrate}
\end{align*}
if $\sigma^2\ge 1/N_r|\Sc_1|$ for each $\Sc_1\ne\emptyset.$ Maximization of \eqref{rankexpdownlinkmimosumrate} over $|\Sc_1|$ and $|\Sc_2^c|$ yields, for $\sigma^2\ge 1,$
\[
\Delta_\mathrm{sum}^\mathrm{down}\le\frac{\min\{N_rL, N_uK\}}{2}\log(N_rL\sigma^2) + \frac{N_uK}{2}\log\left(1+\frac{1}{\sigma^2}\right).\numberthis\label{downlinkmimosumrategapsigma2}
\]
We can then choose $\sigma^2 = 1$ in \eqref{downlinkmimosumrategapsigma2} to obtain
\[
\Delta_\mathrm{sum}^\mathrm{down}\le\frac{\min\{N_rL, N_uK\}}{2}\log(N_rL) + \frac{N_uK}{2},
\]
completing the proof. 

%\end{IEEEproof}

\bibliography{refs}
\bibliographystyle{IEEEtran}
\end{document}